\documentclass[lettersize,journal]{IEEEtran}
\usepackage{amsmath,amsfonts}
\usepackage{algorithmic}
\usepackage{array}
\usepackage{hyperref}

\usepackage{textcomp}
\usepackage{amssymb,amsmath,color}
\usepackage{stfloats}
\usepackage{url}
\usepackage{verbatim}
\usepackage{graphicx}
\usepackage{tikz,pgfplots}
\usetikzlibrary{fit}
\usetikzlibrary{shapes}
\usetikzlibrary{calc,shapes.callouts,shapes.arrows}
\usepackage{tikzpeople}
\usetikzlibrary{shapes,positioning}
\usepackage{dsfont}
\usetikzlibrary{arrows,shapes}
\usepackage{etoolbox}
\usepackage{amsmath}
\newcommand{\mc}{\mathcal}
\usepackage{nccmath}
\usepackage{mathtools}
\usepackage{transparent}
\usepgfplotslibrary{colorbrewer}
\usepackage{subfigure}
\usepackage{makecell}
\usepackage{pbox}
\usepackage{multirow}

\newcommand{\R}{\mathds{R}}
\newcommand{\N}{\mathds{N}}
\newcommand{\dd}{\text{d}}
\newcommand{\defineas}{\coloneqq}
\newcommand{\map}[3]{#1: #2 \rightarrow #3}

\DeclareMathOperator*{\minimize}{minimize}
\definecolor{color0}{rgb}{0.847058823529412,0.749019607843137,0.847058823529412}
\definecolor{color1}{rgb}{0.576470588235294,0.43921568627451,0.858823529411765}
\definecolor{color2}{rgb}{0.294117647058824,0,0.509803921568627}

\def\BibTeX{{\rm B\kern-.05em{\sc i\kern-.025em b}\kern-.08em
    T\kern-.1667em\lower.7ex\hbox{E}\kern-.125emX}}

\newcommand{\subfigref}[2]{\hyperref[#1]{\ref{#1}(#2)}}

\begin{document}
\title{A Dynamic Macroscopic Framework for Pricing of Ride-hailing Services with an Optional Bus Lane Access for Pool Vehicles}
\author{Lynn Fayed$^{1}$ and Gustav Nilsson$^{1}$ and Nikolas Geroliminis$^{1}$
\thanks{$^{1}$The authors are with the Urban Transport Systems Laboratory (LUTS), École Polytechnique Fédérale de Lausanne (EPFL), Switzerland.
        {\tt\small \{lynn.fayed,gustav.nilsson, nikolas.geroliminis\}@epfl.ch}}%
\thanks{This work was supported by the Swiss National Science Foundation under NCCR Automation, grant agreement 51NF40\_180545.}
\thanks{A preliminary version of some of the work in this paper
was presented in~\cite{macroscopic_2023_fayed}.}%
}

\markboth{}%
{A Dynamic Macroscopic Modeling Framework for Pricing of Ride-hailing Services with Bus Lane Access}

\maketitle

\begin{abstract}
On-demand trip sharing is an efficient solution to mitigate the negative impact e-hailing has on congestion.
It motivates platform operators to reduce their fleet size, and serves the same demand level with a lower effective distance traveled. Users nevertheless prefer to travel solo and for shorter distances despite the fare discount they receive. 
By offering them the choice to pool and travel in high occupancy dedicated bus lanes, we provide them with a larger incentive to share their rides, yet this creates additional bus delays.
In this work, we develop dynamic feedback-based control schemes that adjust the price gap between solo and pool trips to improve multi-modal delays. First, we develop a modal- and space-dependent aggregate model for private vehicles, ride-pooling, and buses, and we use this model to test different control strategies.
To minimize the error between the target and actual speeds in the bus network, we design a PI controller and show that by adjusting pool trip fares, we can, with little input data, minimize this error. We also put forward a Model Predictive Control (MPC) framework to minimize the total Passenger Hours Traveled (PHT) and Waiting Times (WT) for the different travelers. Moreover, we show how the MPC framework can be utilized to impose a minimum speed in dedicated bus lanes to ensure that the buses operate on schedule.
The results mark the possibility of improving the overall network conditions by incentivizing or discouraging pooling in the vehicle or bus network.
\end{abstract}

\begin{IEEEkeywords}
Macroscopic fundamental diagrams, Model predictive control, Multi-modal networks, Ride-hailing, Space allocation, Pricing
\end{IEEEkeywords}

\section{Introduction}
The surge of on-demand mobility offers network commuters innovative transport alternatives for their trips. Characterized by their flexibility, convenience, and accessibility, on-demand modes have soon become widely popular, rooting their success in the fast-growing wireless communication technologies and the increasing interest in a more personalized mobility service. Ride-hailing, among many other similar modes, is nowadays a well-established transport alternative where a unique platform connects riders and drivers. Users request a ride, the request being most of the time instantaneous with no in-advance booking, and they are assigned to a nearby driver shortly after. The latter decides whether to carry users from their origins to their destinations based on their relative evaluation of the trip attractiveness. While being as convenient as private vehicles due to their door-to-door services, these services usually prevail over public transportation because of their shorter waiting times before pick-up~\cite{mobility_2017_shaheen}. 

The burgeoning number of ride-hailing services and their wide success required authorities to impose incentive- or enforcement-based regulation strategies. These strategies are introduced to contain the negative impact that ride-hailing has on multi-modal urban traffic and user mode choice distribution. 
A closer look at the operation of ride-hailing services holds drivers accountable for the increase in traffic congestion~\cite{transportation_2019_erhardt}. More specifically, a high number of idling ride-hailing drivers not only causes an increase in congestion but also has a counterproductive effect by extending the waiting for users due to longer dispatching time despite the high availability of empty vehicles~\cite{inefficiency_2021_beojone}.
Moreover, the vast majority of current ride-hailing users reported that they would use public transportation if such services were not available, hence creating a direct competition with buses~\cite{ride-hailing_2018_silva}. Clearly, the decline in the use of mass transit due to the shift in users' choice towards ride-hailing services raises multiple concerns about the undesired competition between these two modes.

To design well-informed and well-targeted policies, it is crucial to provide regulators with quantification studies that concretize the different impacts that ride-hailing has on the traffic externalities, the welfare of the drivers and riders, and the modal split between the different modes in a network~\cite{ridehailing_2019_tirachini}. This approach guides authorities to enforce appropriate actions to prevent further propagation of these services without proper regulations. 
A high fleet size, for instance, shortens the waiting time of users yet increases traffic congestion in urban spaces. One way to mitigate this is through sending empty vehicles with no assigned trip to available off-street parking locations~\cite{inefficiency_2021_beojone, offstreet_2020_li, optimal_2017_xu} or to enforce a cap on the fleet size or the maximum allowable VKT carried out by the ride-hailing fleet~\cite{balancing_2020_yu}.
Moreover, particularly when operating in a monopoly setting, ride-hailing platforms set a profit-maximizing fare without consideration of the rider's or driver's welfare. Many studies additionally investigated what the driver's wage and the rider's fare should be under a social welfare maximization framework. They argue that despite them not being sustainable from a revenue-maximization point of view, these pricing schemes are socially optimum when assessed in an equilibrium setting~\cite{economic_2016_zha, pricing_2020_ke}.

Promoting trip sharing is another strategy to reduce the total Vehicle Kilometers Travelled (VKT) by drivers to serve the same demand level~\cite{data_2021_ke}. It also allows the shrinking of the fleet size required to provide the same service level~\cite{real_2014_ma}. Generally referred to as ride-splitting or ride-pooling, this service prompts users to share their rides with other travelers in exchange for a fare discount to compensate for the extra detour incurred. Whether the passenger-to-passenger pool matching is successful depends on a myriad of factors, including the engagement levels in pooling, i.e., the willingness of users to share their rides with other users in the system, but also on their subsequent pick-up and drop-off locations~\cite{shared_2019_shaheen}. In the scope of this work, however, we ignore the possibility of a failed passenger-to-passenger pool matching, and we assume that all requests opting for pooling will eventually be pooled. We also limit the trip sharing to two passengers, even if high-capacity on-demand micro-transit services are gaining fast momentum~\cite{shared_2019_shaheen}.

Having highlighted the importance of trip-sharing as a way to alleviate the negative externalities of ride-hailing services~\cite{ridehailing_2020_tirachini}, these services are still attracting low to moderate demand levels. Therefore, we advance in this work an occupancy and space-dependent allocation strategy where pooling is motivated by allowing shared trips to use dedicated bus lanes. Although the majority of impact assessment and policy evaluation studies were formulated in a static equilibrium setting, we adopt in our framework a macroscopic dynamic approach with time-varying demand to capture the non-equilibrium and transient states of the network dynamics, and how the system evolves under different demand profiles during the day. More specifically, we tackle in this work the effect on multi-modal delays of incentivizing trip-sharing by entitling pool users to a spatial privilege or a higher fare discount.
In the remaining part of the introduction, we provide a detailed overview of the relevant research tackling different aspects of our work, including i) the static and dynamic modeling of ride-hailing/ride-splitting and their service optimization, and ii) the status of ride-hailing with respect to the other more traditional operating modes in the network.

The common ground in any study tackling ride-hailing is to have a representative model capable of capturing the main features and characteristics of these services and the different stakeholders involved.
The first insight into drawing the distinction between traditional taxi services and on-demand ride-hailing is to underline the appearance of a dispatching vehicle category that is non-existent in traditional taxi markets. It is the result of an online vehicle-passenger matching where the drivers' locations and the requests' origins are known to the platform~\cite{disruptive_2016_cramer}. Nevertheless, many studies have pointed out that, despite it sometimes being useful, online location access is a source of market inefficiency in the event of a demand surge where available vehicles are quickly depleted~\cite{surge_2017_castillo, xu_2020_supply}. A solution to this inefficiency is setting a surge pricing scheme that guarantees that the occurrence of these scenarios is avoided~\cite{role_2015_cachon}, even if this solution raises concerns about passengers' welfare. Service pricing is therefore an important element in ride-hailing modeling, thus justifying the large body of research investigating optimal full rider fare for ride-hailing~\cite{economic_2016_zha} and optimal discounted fare for ride-splitting~\cite{pricing_2020_ke, pool_2019_zhang}. Similarly, \cite{ridesourcing_2020_nourinejad} assessed service pricing but in a dynamic non-equilibrium setting with consideration of background traffic. Due to the spatial heterogeneity of demand and supply, it was also necessary to extend this framework to include a space-dependent pricing scheme balancing demand and supply in multi-regions, therefore guaranteeing a more efficient service level~\cite{geometric_2018_zha, spatial_2019_bimpikis}. Dynamic idle vehicle rebalancing strategies are also able to achieve the same outcomes but require having an accurate prediction of the demand in every region~\cite{dynamic_2023_ramezani, valadkhani_2023_dynamic}.

The research line we describe particularly focused on modeling ride-hailing services, and very few accounted for on-demand trip-sharing in their framework. This is mainly because microscopically modelling trip-sharing, which is one of the main contribution of this work, is complex to conduct, especially when the number of passengers participating in a trip exceeds two. The focus hence deviated towards finding some empirical and universal laws for driver and passenger detours~\cite{scaling_2016_tachet} or to assessing the different factors influencing the quality of a pooled trip~\cite{shareability_2023_soza_parra}. Moreover, the passenger-to-passenger matching and the vehicle dispatching require advanced exact algorithms or heuristics to solve them in real-time settings~\cite{dynamic_2016_jung, real_2014_ma, santos_2013_dynamic, mora_2017_ondemand}, even if in some work on two passenger-pooling, passenger-to-passenger matching probability prediction returned similar results to simulation settings~\cite{predicing_2021_wang}. Consequently, this computational effort makes it complex to integrate the matching with upper-level optimization problems like vehicle rebalancing or dynamic lane usage as in our case.

Positioning ride-hailing attractiveness relative to public transportation leads to questioning whether these two services are complimentary or substitutionary. In areas where public transit is well-developed, ride-hailing is rather viewed as a first/last-mile solution complementing bus or metro services~\cite{uber_2018_hall}.
However, this does not eliminate the potential competition between the two modes where, in many cases, ride-hailing substitutes transit, hence causing an inevitable increase in the total VKT~\cite{ridehailing_2019_tirachini}. This observation led many researchers to formulate user equilibrium under different available alternatives where users have the choice to use ride-hailing either for their full trips or for a subpart of their trips~\cite{equilibrium_2021_ke,competition_2021_zhu,economic_2023_ma}. They showed that subsidizing ride-hailing as a first/last-mile solution can indeed reinforce modal complementarity, despite reducing ride-hailing profit. 

The purpose of this work is to assess the importance of ride-hailing in a multi-modal context and to advance an adaptive pricing strategy which, combined with an adequate spatial allocation scheme, allows us to minimize total delays in the network. This occupancy- or modal-dependent space allocation framework has been previously studied in the context of High-Occupancy Toll (HOT) lanes~\cite{optimal_2017_toledo, impact_2022_cohen} or dedicated lanes for Autonomous Vehicles (AV) on the link level~\cite{use_2017_lamotte}, or for buses on the network level~\cite{modeling_2021_tsitsokas}. To the best of our knowledge, however, no work on modal- and occupancy-dependent allocation strategy to minimize overall network delays has considered ride-hailing services in its framework. The strategy we propose in this work was previously evaluated in a static setting, and network equilibrium solutions were computed. 
The foremost result in this direction pointed out the need for a dynamic control framework to regulate priority lane usage for varying demand \cite{utilization_2023_fayed}.

The contribution of this work is twofold. First, we build a dynamic macroscopic multi-modal dynamic model that utilizes Macroscopic Fundamental Diagrams (MFD)~\cite{existence_2008_geroliminis} to determine the dynamics of private vehicles, bus users, and ride-hailing services. We assume that ride-hailing users can choose to travel solo in the vehicle network or to pool in the bus network. In contrast to~\cite{macroscopic_2023_fayed}, we additionally consider the option of pooling in the vehicle network. Second, we introduce a regulatory pricing scheme for the pooling options and use the dynamic model developed within a control framework to determine what should the additional discount/fare that must be given/taken from pooling users be to steer the system toward its optimum. In other words, we propose a solution that minimizes the total delays and waiting times for all mode users in the network by adjusting the fares for pooled trips. Moreover, to guarantee minimal disturbances to buses, we impose a minimum speed in the bus network. This step ensures that buses continue to operate on schedule such that the utilization of their network by pool ride-hailing vehicles does not cause them significant delays.  

The remainder of the paper is organized as follows. In the following section, Section~\ref{sec:macro_model}, we describe the modal- and occupancy-dependent space allocation strategy we present in this work, and elaborate on the macroscopic state dynamics for the different transportation modes using the network under assessment. Next, in Section~\ref{sec:control}, we lay out the different control frameworks used for the purpose of improving network delays. To demonstrate the performance of the different controllers, we display the results of the simulations we run in Section~\ref{sec:results}. The paper finally concludes with the main findings and future directions in Section~\ref{sec:conclusion}.

\section{Macroscopic Multi-modal Framework}
\label{sec:macro_model}
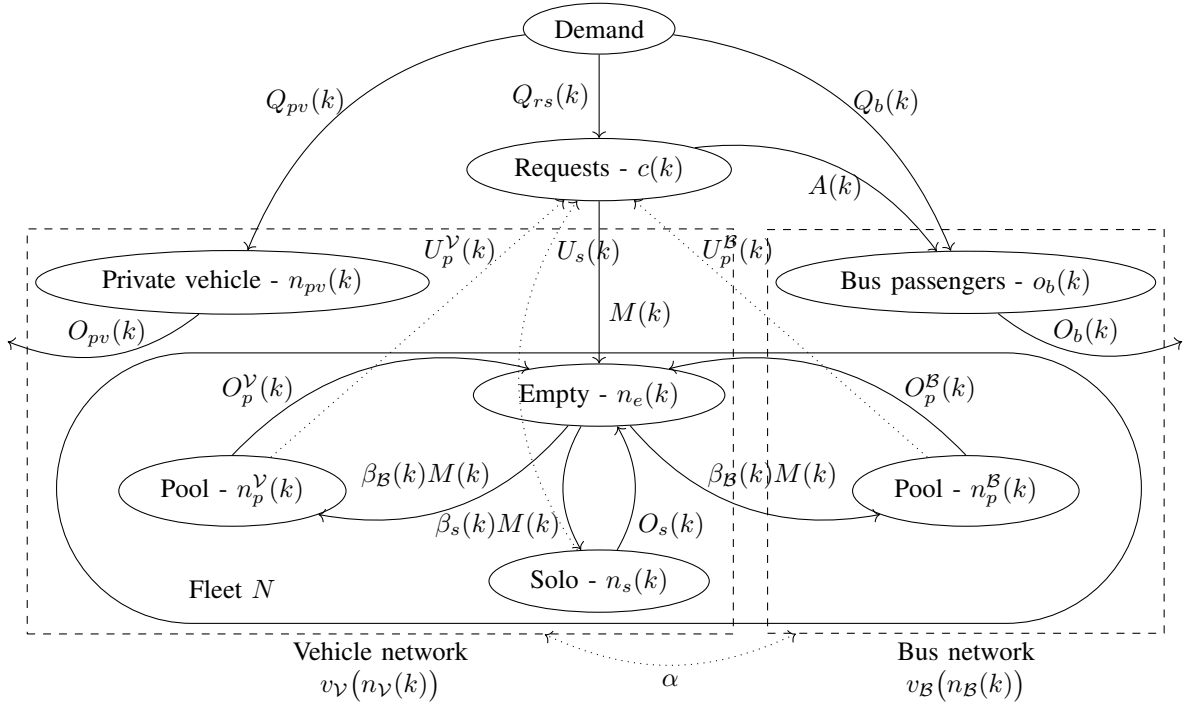
\begin{figure*}
    \centering
    \begin{tikzpicture}[scale=0.75]
        \node[draw, ellipse] (demand) at (0,0) {Demand};
        \node[draw, ellipse] (ride-hailing) at (0,-2.5) {Requests - $c(k)$};
        \node[draw, ellipse] (bus) at (6.5,-4.5) {Bus passengers - $o_{b}(k)$};
        \node[draw, ellipse] (pooled) at (6.5,-8.2) {Pool - $n_{p}^{\mc{B}}(k)$};
        \node[draw, ellipse] (solo) at (0,-9.8) {Solo -  $n_{s}(k)$};
         \node[] (dummy) at (6.5,-9.95) [opacity = 0]{Solo -  $n_{s}(k)$};
         \node[] (fleetsize) at (-6.5,-9.95) {Fleet $N$};
        \node[draw, ellipse] (poolV) at (-6.5,-8.2) {Pool -  $n_{p}^{\mc{V}}(k)$};
        \node[draw, ellipse] (empty) at (0,-6.5) {Empty -  $n_{e}(k)$};
        
        \node[draw, ellipse] (pv) at (-6.5,-4.5) {Private vehicle - $n_{pv}(k)$};
        \node[] (pvexit) at (-10.65,-5.5) {};
        \node[] (bexit) at (10.5,-5.5) {};
        \draw[->] (demand) to[bend left] node[right] {$Q_b(k)$} (bus);
        \draw[->] (demand) -- node[left] {$Q_{rs}(k)$} (ride-hailing);
        \draw[->] (demand) to[bend right] node[left] {$Q_{pv}(k)$} (pv);
        \draw[->] (ride-hailing) -- node[right,pos=0.7] {$M(k)$} (empty);
        \node[draw, dashed, rectangle, fit=(bus) (pooled) (dummy), inner xsep=3pt, inner ysep=8pt] (busnetwork) {};
            \node[draw, rounded rectangle, fit=(solo) (pooled) (empty) (poolV), inner xsep=-8pt, inner ysep=4pt] (rhfleet) {};
        \node[draw, dashed, rectangle, fit=(solo) (pv) (empty) (poolV), inner xsep=3pt, inner ysep=8pt] (vehiclenetwork) {};
        
        \draw[->] (empty) to[bend right] node[left = 0mm, pos = 0.8] {$\beta_s(k)M(k)$} (solo);
        \draw[->] (pv) to[bend left] node[above] {$O_{pv}(k)$} (pvexit);
        \draw[->] (bus) to[bend right] node[above] {$O_{b}(k)$} (bexit);
         \draw[->] (ride-hailing) to[bend left] node[left, pos = 0.68] {$A(k)$} (bus);
        \draw[ ->] (solo) to[bend right] node[right=0mm, pos = 0.2] {$O_{s}(k)$} (empty);
        \draw[->] (pooled.90) to[bend right] node[above=2mm, pos = 0.1] {$O_p^{\mc{B}}(k)$} (empty);
        \draw[->] (poolV.90) to[bend left] node[above=2mm, pos = 0.1] {$O_p^{\mc{V}}(k)$} (empty);
        \draw[->] (empty) to[bend right] node[right =2mm, pos = 0.25] {$\beta_{\mc{B}}(k) M(k)$} (pooled);
        \draw[->] (empty) to[bend left] node[left =2mm, pos = 0.25] {$\beta_{\mc{B}}(k) M(k)$} (poolV);
        \draw[->, dotted] (solo) to[bend left = 35] node[right,pos=0.85] {$U_s(k)$} (ride-hailing);
         \draw[->, dotted] (pooled) -- node[right,pos=0.8] {$U_p^{\mc{B}}(k)$} (ride-hailing);
        \draw[->, dotted] (poolV) -- node[left,pos=0.8] {$U_p^{\mc{V}}(k)$} (ride-hailing);
        \draw[dotted, <->, align=center] (busnetwork.230) to[bend left = 25] node[below] {$\alpha$} (vehiclenetwork);
        \node[below of= rhfleet, align=center] {};
        \node[below of= busnetwork, align=center] {\\ \\ \\ \\ \\ \\ \\ \\ \\ \\ \\ Bus network \\ $v_{\mathcal B}\bigl(n_{\mathcal B}(k)\bigr)$};
        \node[below of= vehiclenetwork,align=center] { \\ \\ \\ \\ \\ \\  \\ \\ \\ \\ \\ Vehicle network \\ $v_{\mathcal V}\bigl(n_{\mathcal{V}}(k)\bigr)$};
    \end{tikzpicture}
    \caption{A schematic sketch of the model under consideration. Private vehicles, empty ride-hailing vehicles, solo trips, and a portion of pool ride-hailing trips drive in the vehicle subnetwork $\mc{V}$, whereas buses and the remaining pool trip portion perform their trips in the bus subnetwork $\mc{B}$.} 
        \label{fig:summary_sketch}
\end{figure*}

In the following section, we describe our macroscopic modelling framework by delineating the modal-dependent allocation strategy we propose. Next, we put forward the aggregate traffic model according to the spatial allocation policy advanced, and use it to define the traffic dynamics for the different transportation modes under consideration. The entire modeling framework is summarized in Figure~\ref{fig:summary_sketch}.
\subsection{Modal-dependent Space Allocation}
In the network under consideration, travelers perform their trips using one of the available transportation alternatives in the set $\mathcal{M}$: private vehicles~$pv$, buses~$b$, and ride-hailing services~$rs$ such that $\mathcal{M} \defineas \{pv,b,rs\}$. 
The dynamics for each alternative are modeled in discrete time with time step $k \in \mathcal{K} \defineas \{0,\dots,k_\text{max}\}$, and the duration of each time step is $\tau>0$. Moreover, we let $\bar{\mc K} \defineas \mc K \setminus \{k_\text{max} \}$.
Every transportation mode under consideration has an exogenous and time-dependent demand $Q_j(k)$ for $j\in \mathcal{M}$ expressed in passengers per hour.
Commuters who opt for the ride-hailing alternative choose to either travel solo or to pool their trips with other users of the service. 
In terms of space utilization, private vehicles perform their trips in the subspace of the network occupying a fixed fraction $\alpha \in [0,1]$ of the total network space.
Buses from their sides solely travel in dedicated bus lanes spanning over a fraction $1-\alpha$ of the total space.  
We denote the vehicle and bus networks by $\mathcal{V}$ and $\mathcal{B}$ respectively, such that the set of available subnetworks is $\mc{N} \defineas \{\mc{V}, \mc{B}\}$.
Finally, if ride-hailing users opt for travelling solo, then the entirety of their trip is performed in the vehicle network $\mc{V}$. If however ride-hailing users choose to pool, they are granted the opportunity to either travel in the vehicle network $\mc{V}$ or the bus network $\mc{B}$.
This implies that the drivers of the ride-hailing fleet $N>0$, assumed to be time-independent in our framework, can exclusively be in one of the following states:

\begin{itemize}
\item[(i)] driving an empty vehicle in the vehicle network~$\mc{V}$, where the total number of drivers in this state is denoted by~$n_e$,
\item[(ii)] delivering a solo trip in the vehicle network~$\mc{V}$,  where the total number of drivers in this state is denoted by~$n_s$,
\item[(iii)] delivering a pool trip in the vehicle network~$\mc{V}$, where the total number of drivers in this state is denoted by~$n_p^{\mc{V}}$ and,
\item[(iv)] delivering a pool trip in the bus network~$\mc{B}$,  where the total number of drivers in this state is denoted by~$n_p^{\mc{B}}$.
\end{itemize}
It should be noted that we consider all of the aggregate states to be non-negative and continuous.

Moreover, let $n_{pv}$ be the number of private vehicles in the vehicle network~$\mc{V}$ and $n_b$ the number of buses in the bus network~$\mc{B}$. Having defined the different vehicle categories, we know that the ride-hailing fleet at any time step $k\in{\mc{K}}$, under the assumption of a fixed fleet size, has to satisfy $N=n_e(k) + n_s(k) + n_p^{\mc{V}}(k) + n_p^{\mc{B}}(k)$. The accumulation in the vehicle network~$n_{\mc{V}}$ at any time step $k\in{\mc{K}}$ is $n_{\mc{V}}(k) = n_{pv}(k) + n_e(k) + n_s(k) + n_p^{\mc{V}}(k)$, and the accumulation in the bus network~$n_{\mc{B}}$ at any time step $k\in{\mc{K}}$ is $n_{\mc{B}}(k) = n_b + n_p^{\mc{V}}(k)$. Note that the number of operating buses in $\mc{B}$ is assumed to be time-independent in our framework. 
\subsection{Aggregate traffic flow model}
\label{subsec:traffic}
In the following part, we elaborate on the aggregate traffic model we use to estimate the speed in the vehicle and bus networks, i.e., $\mathcal{V}$ and $\mathcal{B}$.
Let $\map{P}{\R_{\geq 0}}{\R_{\geq0}}$ denote the full network production function without any dedicated bus lanes. The production $P$ is dependent on the full network accumulation $n(k)$, for $k\in{\mc{K}}$, and can be calculated using the running network speed $\map{v}{\R_{\geq 0}}{\R_{\geq0}}$ such that $P\bigl( n(k) \bigr) = n(k)v\bigl(n(k)\bigr)$.
Following~\cite{city_2019_ni, sirmatel_2021_modeling}, we can compute the define $\map{P_{\mathcal{V}}}{\R_{\geq 0}}{\R_{\geq0}}$ in the vehicle network and the production $\map{P_{\mathcal{B}}}{\R_{\geq 0}}{\R_{\geq0}}$ in the bus network using the space allocation factor $\alpha$ such that $P_{\mathcal{V}}(\alpha n(k)) = \alpha P(n(k)) $ and $ P_{\mathcal{B}}(\bar{\alpha} n(k)) = \bar{\alpha} P(n(k))$ where $\bar{\alpha} = 1 - \alpha$.

The relationship between production and accumulation is valid as long as the vehicles commuting in the bus network are standard vehicles. 
However, since buses perform frequent stops to board and alight passengers, they slow down the remaining vehicles utilizing the same space.
To capture this interaction between buses and pooling vehicles utilizing the bus network, we partition the production in the bus network into pool vehicles production $\map{P_p}{\R_{\geq 0}\times \R_{\geq0}}{\R_{\geq0}}$ and bus production $\map{P_b}{\R_{\geq 0}\times \R_{\geq0}}{\R_{\geq0}}$, each dependent on both the number of pooling vehicles $n_p^{\mc{B}}$ and the number of buses $n_b$ to obtain a 3D-MFD~\cite{three_2014_geroliminis}.

Analogous to the reasoning behind the split of the production function, the same applies here to the aggregate estimation of the speed in the vehicle network $\map{v_{\mathcal{V}}}{\R_{\geq 0}}{\R_{\geq0}}$ and bus network $\map{v_{\mathcal{B}}}{\R_{\geq 0}}{\R_{\geq0}}$ where $v_{\mathcal{V}}\bigl(\alpha n(k)\bigr) = v\bigl(n(k)\bigr)$ and $v_{\mathcal{B}}\bigl(\bar{\alpha} n(k)\bigr) = v\bigl(n(k)\bigr)$.
Having estimated the individual speed function for every subnetwork, the production value in the vehicle network $\mathcal{V}$ at time step $k\in{\mc{K}}$ is hence given by  $P_{\mathcal{V}}(n_{\mathcal{V}}(k)) = n_{\mathcal{V}}(k)v_{\mathcal{V}}\bigl( n_{\mathcal{V}}(k)\bigr)$ and that in the bus network $\mathcal{B}$ is given by $P_{\mathcal{B}}(n_{\mathcal{B}}(k)) = n_{\mathcal{B}}(k)v_{\mathcal{B}}\bigl( n_{\mathcal{B}}(k)\bigr)$.
However, for the latter function, we account for the influence of buses on the vehicle speed by reducing $v_{\mathcal{B}}$ with a factor $r(n_b)$ dependent on the number of buses in the network where $r : \R _{\geq 0} \rightarrow (0, 1]$ and $\frac{\dd r}{\dd n_b}<0$.
Therefore, the actual speed of pool vehicles, that we denote by $\map{v_p}{\R_{\geq 0} \times \R_{\geq 0}}{\R_{\geq0}}$, is given by $v_p\bigr(n_{p}^{\mc{B}}(k), n_b\bigr) = v_{\mathcal{B}}\bigl( n_{\mathcal{B}}(k)\bigr)r\bigl(n_b\bigr)$.
Regarding the bus operating speed that we denote $\map{v_b}{\R_{\geq 0} \times \R_{\geq 0}}{\R_{\geq0}}$, it must account for the repetitive stops that buses perform at stations such that 
\begin{equation}
v_b\bigr(n^{\mc{B}}_{p}(k), n_b\bigr) = \left( \frac{1}{1 + v_p\bigr(n_{p}^{\mc{B}}(k), n_b\bigr)\frac{\bar{t}_{d}}{\bar{s}}}\right) v_p\bigr(n_p^{\mc{B}}{p}(k), n_b\bigr) \,, 
\label{eqn:bus_speed}
\end{equation}
where $\bar{t}_d$ and ${\bar{s}}$ are the average bus dwell time and the spacing between bus stops respectively.
It follows that the pool vehicle production in $\mathcal{B}$ at time step $k\in{\mc{K}}$ is $P_{p}\bigr(n_{p}^{\mc{B}}(k), n_b\bigr) = n_p^{\mc{B}}(k)v_p\bigr(n_p^{\mc{B}}(k), n_b\bigr)$ and the bus production is $P_{b}\bigr(n_p^{\mc{B}}(k), n_b(k)\bigr) = n_bv_b\bigr(n_p^{\mc{B}}(k), n_b\bigr)$.

The aggregate network-dependent production functions are used to estimate the trip completion rate --or outflow-- for every category of vehicles under consideration. While this approximation does not provide the same level of accuracy compared to trip-based models, it still allows a tractable analysis compared to the latter which is too complex for this sort of application \cite{sirmatel_2021_modeling}. 
As a result, in the following subsections, we utilize the production-based multi-modal macroscopic traffic model to estimate the changes as a function of time of the private vehicle accumulation, ride-hailing fleet assignment, and bus occupancy.
\subsection{Network dynamics}
Previously, we have defined the modal-dependent space allocation strategy and the subnetwork-dependent macroscopic traffic functions. In the following part, we elaborate on the aggregate dynamics of private vehicles, ride-hailing fleet, and bus average occupancies according to the proposed allocation strategy. 
Starting with the private vehicle category, the accumulation $n_{pv}$ is computed using
\begin{equation}
     n_{pv}(k+1) =  n_{pv}(k)   + \tau \left[\frac{Q_{pv}(k)}{\bar{o}_{pv}} - O_{pv}(k)\right] \,, \, \forall k \in \mc{\bar{K}} \,, \label{eq:npv}
\end{equation} 
where $\bar{o}_{pv} > 0$ is the average occupancy of a private vehicle and $O_{pv}(k)$ is the trip completion rate computed using
\begin{equation}
    O_{pv}(k) = \frac{n_{pv}(k)}{n_{\mathcal{V}}(k)} \frac{P_{\mathcal{V}}\bigl(n_{\mathcal{V}}(k)\bigr)}{\bar{l}_{pv}} \,,
\end{equation}
where $\bar{l}_{pv} > 0$ is the constant average trip distance between the origin-destination pairs of private vehicle users. Note that we assume a homogeneous mixture of private and ride-hailing vehicles. 

Moving to the ride-hailing mode, the various available options have different travel costs. Based on the relative cost of each option, the users choose to travel solo in $\mc{V}$, to pool in $\mc{V}$, or to pool in $\mc{B}$.
Therefore, we let $u_s$ denote the disutility for traveling solo, $u_p^{\mc{V}}$ the disutility for pooling in the vehicle network, and $u_p^{\mc{B}}$ the disutility for pooling in bus lanes. The expressions of the different utilities at time step $k\in\mc{K}$ are
\begin{subequations}
\begin{align}
U_s(k) & = \tilde{F}_s(k) + \kappa\frac{\bar{l}_s}{v_{\mathcal{V}}\bigl(n_{\mathcal{V}}(k)\bigr)} \,, \label{eq:gat1}\\
U_p^{\mc{V}}(k) & = \tilde{F}_p^{\mc{V}}(k) +  \kappa \frac{\bar{l}_s + \Delta l_p}{v_{\mathcal{V}}\bigl(n_{\mathcal{V}}(k)\bigr)}\,, \label{eq:gat2}\\
U_p^{\mc{B}}(k) & = \tilde{F}_p^{\mc{B}}(k) +  \kappa \frac{\bar{l}_s + \Delta l_p}{v_p\bigr(n_{p}^{\mc{B}}(k), n_b\bigr)}\,, \label{eq:gat3}
\end{align}
\end{subequations}
where $\kappa>0$ is the value of time, $\bar{l}_s>0$ is the average trip length for a solo trip, and $\Delta l_p \geq0$ is the pool detour distance that passengers incur in case they opt for pooling. The variable $\tilde{F}_s$ is the fare for travelling solo in~$\mc{V}$, $\tilde{F}_p^{\mc{V}}$ is the fare for pooling in~$\mc{V}$, and $\tilde{F}_p^{\mc{B}}$ is the fare for pooling in~$\mc{B}$. In this work, we assume that $\tilde{F}_s(k)$ is constant such that $\tilde{F}_s(k) = F_s$ for all $k\in \mc{K}$ where $F_s>0$ is the solo trip fare set by the platform operator. On the contrary, if $F_p>0$ is the pool trip fare set by operator, then $\tilde{F}_p^{\mc{V}}(k) = F_p + \phi_{\mc{V}}(k)$ and $\tilde{F}_p^{\mc{B}}(k) = F_p + \phi_{\mc{B}}(k)$ where  $\phi_{\mc{V}}(k)\in \R$ and $\phi_{\mc{B}}(k)\in \R$ are the control fares for pooling in the vehicle network~$\mc{V}$ and bus network~$\mc{B}$ respectively. The control fares are introduced to steer the total network towards different objectives that we expand on in Section~\ref{sec:control}. The relative values of each of the disutility functions are used to compute the modal share for every available ride-hailing alternative. Therefore, let $\beta_{\mc{V}} \in [0,1]$ and $\beta_{\mc{B}} \in [0,1]$ be the fraction of the total ride-hailing demand that will choose to pool in~$\mc{V}$ and $\mc{B}$ respectively at time step $k\in\mc{K}$, then using a multinomial logit model, we have that
\begin{subequations}
\begin{align}
    \beta_{\mc{V}}(k) & = \frac{e^{-\mu U_p^{\mc{V}}(k)}}{e^{-\mu U_s(k)} +  e^{-\mu U_p^{\mc{V}}(k)} +  e^{-\mu U_p^{\mc{B}}(k)} } \, , \\
    \beta_{\mc{B}}(k) & = \frac{e^{-\mu U_p^{\mc{B}}(k)}}{e^{-\mu U_s(k)} +  e^{-\mu U_p^{\mc{V}}(k)} +  e^{-\mu U_p^{\mc{B}}(k)} } \, ,
\end{align}
\end{subequations}
where $\mu>0$ is the scale parameter. The portion of users choosing to go solo at time step $k\in{\mc{K}}$ is $\beta_s(k)$ such that $\beta_s(k) = 1-\beta_{\mc{V}}(k) - \beta_{\mc{B}}(k)$.

For ease of implementation purposes, we reformulate the utility and mode choice functions to set apart the control and state variables. Consequently, if we redefine the disutilities of the three available ride-hailing alternatives, excluding the controllable price changes, by $u_s$, $u_p^{\mc{V}}$, and $u_p^{\mc{B}}$, we get that
\begin{subequations}
\begin{align}
u_s(k) & = F_s + \kappa\frac{\bar{l}_s}{v_{\mathcal{V}}\bigl(n_{\mathcal{V}}(k)\bigr)} \,, \\
u_p^{\mc{V}}(k) & = F_p +  \kappa \frac{\bar{l}_s + \Delta l_p}{v_{\mathcal{V}}\bigl(n_{\mathcal{V}}(k)\bigr)}\,, \\
u_p^{\mc{B}}(k) & = F_p +  \kappa \frac{\bar{l}_s + \Delta l_p}{v_p\bigr(n_{p}^{\mc{B}}(k), n_b\bigr)}\,, 
\end{align}
\end{subequations}
and if we set $\xi_{\mc{V}}(k)$ and $\xi_{\mc{B}}(k)$ as two variables that are function of $\phi_{\mc{V}}(k)$ and $\phi_{\mc{B}}(k)$ for all $k\in \mc{K}$, we get
\begin{align}
    \xi_{\mc{V}}(k) & \defineas e^{-\mu \phi_{\mc{V}}(k)}\, , \\
    \xi_{\mc{B}}(k) & \defineas  e^{-\mu \phi_{\mc{B}}(k)}\, .
\end{align}
Consequently, we can rewrite $\beta_{\mc{V}}(k)$ and $\beta_{\mc{B}}(k)$ as follows
\begin{align}
    \beta_{\mc{V}}(k) & = \frac{\xi_{\mc{V}}(k) e^{-\mu u_p^{\mc{V}}(k)}}{e^{-\mu u_s(k)} +  \xi_{\mc{V}}(k)e^{-\mu u_p^{\mc{V}}(k)} +  \xi_{\mc{B}}(k) e^{-\mu u_p^{\mc{B}}(k)}} \, , 
        \label{eqn:beta_v}\\
    \beta_{\mc{B}}(k) & = \frac{\xi_{\mc{B}}(k) e^{-\mu u_p^{\mc{B}}(k)}}{e^{-\mu u_s(k)} +  \xi_{\mc{V}}(k)e^{-\mu u_p^{\mc{V}}(k)} +  \xi_{\mc{B}}(k) e^{-\mu u_p^{\mc{B}}(k)}} \, .
        \label{eqn:beta_b}
\end{align}

We conclude that if $c(k)$ is the number of customers waiting to be matched at time $k$ for all $k\in \mc{K}$, then we know that the number of passengers choosing to travel solo is~$c_s(k) = \bigl(1-\beta_{\mc{V}}(k) - \beta_{\mc{B}}(k)\bigr)c(k)$ and the number of passengers choosing to pool is~$\bigl( \beta_{\mc{V}}(k) + \beta_{\mc{B}}(k)\bigr)c(k)$. By resorting to a Cobb-Douglas meeting function, we compute the matching rate~$M(k)$ at time step $k\in\mc{K}$ using
\begin{equation}
    M(k) = a_0 n_e(k)^{\alpha_e} \left(c_s(k) + \frac{1}{2} c_p(k)\right)^{\alpha_c}\,,
    \label{eqn:cobbdouglas}
\end{equation}
where $a_0 >0$, $\alpha_e>0$, and $\alpha_c>0$ are the Cobb-Douglas meeting function parameters. Note that a factor $\frac{1}{2}$ is added to $c_p$ to model that every single pool trip consists of two passengers. Subsequently, we can compute the dynamics of empty vehicles $n_e$ at any time step using
\begin{multline}
     n_e(k+1) =  n_e(k)  +  \tau \biggl[\frac{n_{s}(k)}{n_{\mathcal{V}}(k)} \frac{P_{\mathcal{V}}\bigl(n_{\mathcal{V}}(k)\bigr)}{\bar{l}_{s}}   + \\ \frac{n_{p}^{\mc{V}}(k)}{n_{\mathcal{V}}(k)} \frac{P_{\mathcal{V}}\bigl(n_{\mathcal{V}}(k)\bigr)}{\bar{l}_{s} + \Delta l_d}+ \frac{P_p\bigl(n_p^{\mc{B}}(k), n_b\bigr)}{\bar{l}_{s} + \Delta l_d} - M(k)\biggr] \,,  \quad \forall k \in \mathcal{\bar{K}}\,. 
    \label{eq:ne}
\end{multline}
where the first three elements of~\eqref{eq:ne} represent the completion rate of solo trips $O_s(k) =\frac{n_{s}(k)}{n_{\mathcal{V}}(k)} \frac{P_{\mathcal{V}}\bigl(n_{\mathcal{V}}(k)\bigr)}{\bar{l}_{s}}$, the completion rate of pool trips in~$\mc{V}$ $O_p^{\mc{V}}(k) = \frac{n_{p}^{\mc{V}}(k)}{n_{\mathcal{V}}(k)} \frac{P_{\mathcal{V}}\bigl(n_{\mathcal{V}}(k)\bigr)}{\bar{l}_{s} + \Delta l_d}$, and the completion rate of pool trips in $\mc{B}$ $O_p^{\mc{B}}(k) = \frac{P_p\bigl(n_p^{\mc{B}}(k), n_b\bigr)}{\bar{l}_{s} + \Delta l_d}$. The last element of~\eqref{eq:ne} denotes the number of empty vehicles that are matched and have therefore exited this category. The variable $\Delta l_d\geq0$ represents the driver detour, which is the additional distance traveled by drivers to perform a pool trip relative to a solo one.

Moving to the discretized dynamics of the solo vehicle category $n_s$, we get that
\begin{equation}
    \begin{split}
    \label{eq:ns}
       n_s(k+1) = n_s(k) + \tau \biggl[&\beta_s(k)  M(k) - \\ &\frac{n_{s}(k)}{n_{\mathcal{V}}(k)} \frac{P_{\mathcal{V}}\bigl(n_{\mathcal{V}}(k)\bigr)}{\bar{l}_{s}}\biggr] \,, \quad \forall k \in \mc{\bar{K}}\, .
    \end{split}
\end{equation}

Similarly, the discretized dynamics for the pool vehicle in~$\mc{V}$ category, $n_p^{\mc{V}}$, are given by
\begin{equation}
    \begin{split}
    \label{eq:np_v}
       n_p^{\mc{V}}(k+1) = n_p^{\mc{V}}(k) + \tau \biggl[&\beta_{\mc{V}}(k)  M(k) - \\ &\frac{n_{p}^{\mc{V}}(k)}{n_{\mathcal{V}}(k)} \frac{P_{\mathcal{V}}\bigl(n_{\mathcal{V}}(k)\bigr)}{\bar{l}_{s}+ \Delta l_d} \biggr] \,, \quad \forall k \in \mc{\bar{K}}\, .
    \end{split}
\end{equation}

Finally, the discretized dynamics for the pool vehicles in~$\mc{B}$, category, $n_p^{\mc{B}}$, are given by
\begin{equation}
\begin{split}\label{eq:np_b}
    n_p^{\mc{B}}(k+1) = n_p^{\mc{B}}(k) + \tau \biggl[&\beta_{\mc{B}}(k) M(k) - \\ & \frac{P_p\bigl( n_p^{\mc{B}}(k), n_b\bigr)}{\bar{l}_{s} + \Delta l_d} \biggr] \,, \quad \forall k \in \mc{\bar{K}} \,.
\end{split}
\end{equation}

So far, we have defined the changes in the ride-hailing vehicles category. In a similar manner, the changes in the number of passengers in the queue waiting to be assigned are
\begin{equation}
\begin{split}
    \label{eq:waiting_request}
    c(k+1) =  &c(k)  + \tau \biggl[Q_{rs}(k) - \\ & \bigl(1+\beta_{\mc{V}}(k)+\beta_{\mc{B}}(k)\bigr) M(k)\biggr] - A(k) \,, \quad \forall k \in \mathcal{\bar{K}} \,.
\end{split}
\end{equation}
Note here that the $\bigl(1+\beta_{\mc{V}}(k)+\beta_{\mc{B}}(k)\bigr) M(k)$ represents the outflow of the waiting requests category. It accounts for the fact that a pool trip match requires taking out two passengers of this category compared to a solo trip match. Finally, the variable $A(k) \geq 0$ represents the number of abandoning requests due to long waiting periods before pick-up. Therefore, if the waiting tolerance of ride-hailing customers is given by $w_{\text{max}}$, then we can estimate the number of abandoning requests using 
\begin{equation}
A(k) = \max\left(c(k) - \frac{1}{k} \sum_{\tilde{k}=1}^{k}M(\tilde{k})w_\text{max}, 0\right) \,, \quad \forall k \in \mathcal{K}.
\label{eq:abandonment}
\end{equation}
The abandonment approximation in~\eqref{eq:abandonment} first computes the theoretical queue length if the ride-hailing request waiting time is $w_{\text{max}}$ and the matching rate is the average over the previous time steps, and then subtracts this quantity from the current queue length $c(k)$. If the resulting quantity is less than $0$, this implies that the service level is rather satisfactory. The abandoning requests will not disappear from the system, but will rather perform their trips using buses.

The dynamics of the third transportation mode, buses, are formulated in terms of changes in occupancy rather than changes in the number of vehicles. This is because we assume that the number of buses $n_b$ traveling in the bus network~$\mc{B}$ is constant and their occupancy is time-dependent. Assuming a uniform bus occupancy $o_b$ over the available fleet of buses $n_b$, the discretized dynamics of $o_b$ are given by
\begin{equation}
    \begin{split}
    o_b(k+1) = o_b(k) + &\frac{\tau}{n_b}\biggl[Q_b (k) + A(k) - \\ &\frac{P_b\bigl(n_p^{\mc{B}}(k), n_b\bigr)}{\bar{l}_b}o_b(k)\biggr] \,, \quad  \forall k \in  \mathcal{\bar{K}}\,,
    \label{eq:bus_occupancy}
    \end{split}
\end{equation}
where $\bar{l}_b$ is the average trip length by bus. Note here that the abandoning ride-hailing requests at time step $k\in{\mc{K}}$, $A(k)$, are considered here as an additional demand for the bus occupancy category. The last term of~\eqref{eq:bus_occupancy} represents the bus passenger trip completion rate $O_{b}(k) = \frac{P_b\bigl(n_p^{\mc{B}}(k), n_b\bigr)}{\bar{l}_b}o_b(k)$. 

In the next section, Section~\ref{sec:control}, we elaborate on the ride-hailing pricing scheme that we set forth to reduce the multi-modal user delays by utilizing the dynamic model we previously described. 

\section{Control Framework}
\label{sec:control}
The objective of the space allocation model we advance in this work is to redistribute the network space over the available transportation modes. However, the choice of ride-hailing users to travel solo or to pool is not always aligned with the objective of improving the total delays in the network. Therefore, in this section, we develop a regulatory pricing scheme to ensure that allowing a fraction of pool trips in bus lanes does not worsen multi-modal user delays. To do so, we evaluate two different control strategies. The first one is a Proportional-Integral (PI) controller with the objective of keeping the bus network at a certain speed. The PI controller is myopic and only requires information about the speed in the bus network to compute the control prices. The second control strategy we evaluate is the Model Predictive Control (MPC) with the objective of minimizing the total travel time. 
\subsection{PI control}
Because our strategy moves pool vehicles to bus lanes, it is crucial to ensure that the disturbances to bus users are minimized, all while improving the travel time for the remaining travelers.
We do so by changing the values of the pool trip control fares $\phi_{\mc{B}}$ according to the difference between the actual bus speed in network $\mathcal{B}$ and the target bus speed, which we denote by $\bar{v}_b > 0$.
This difference is also referred to as the bus speed error. 
The choice of~$\phi_{\mc{B}}$ as our control variable in the PI framework is justified by the direct effect this quantity has on the amount of trip pooling in the bus network.
In addition, we keep track of the previous errors for the last $N_e \in \N$ time steps.
As a consequence, if $\epsilon(k)\defineas \bigl(\bar{v}_b - v_b(n_p^{\mc{B}}(k), n_b) \bigr)$ defines the error term at time step $k$, then the expression for $\phi_{\mc{B}}$ at time step $k\in{\mc{K}}$ is 
\begin{equation}
\phi_{\mc B}(k) = K_p\epsilon(k) + \frac{K_i }{N_e}  \sum_{\tilde{k} = \max({k -(N_e+1),0)}}^{k-1} \epsilon(\tilde{k}) \, ,
\end{equation}
where $K_p > 0$ and $K_i \geq 0$ are the constant proportional gain and integral gain.
In the result section, we show how the two different parameters, $K_p$ and $K_i$, affect the performance of the PI controller by bridging the gap between the actual and target bus speed. The drawback of the PI controller, however, is that it only takes into account the speed of buses irrespective of the other mode users. Next, we show how the MPC implementation is capable of circumventing this matter. 

\subsection{Model predictive control}
Unlike the PI controller, the aim of the MPC framework is to improve the total network delays using our proposed strategy. We quantify delays using the total Passenger Hours Travelled (PHT) of multi-modal users at any time step $k\in \mc{K}$, here equal to $\text{PHT}(k) = \tau{\left[n_{pv}(k) \bar{o}_{pv} + n_{b} o_{b}(k) + n_s(k) +\bigl(n_p^{\mc{V}}(k)+n_p^{\mc{B}}(k)\bigr)\bar{o}_p\right]}$ and the total Waiting Time (WT) of ride-hailing requests at time step $k\in{\mc{K}}$, which is given by $\text{WT}(k) = \tau c(k)$. Therefore, the formulation of the MPC framework is given by 
\begin{equation}
\begin{aligned}
    \minimize \quad &\sum_{k \in 
    \mc K} \text{PHT}(k) + \text{WT}(k)  &\\
  \textrm{subject to} \quad & \xi_i(k)\in[\xi_{\text{min}}, \xi_{\text{max}}]  &\forall k \in \mc K, \,  \forall i\in \mc N  \\
     & \xi_i(k) = \xi_i(k+1) \, & \hspace{-1.6cm}\forall k \in \bar{\mc{K}} \setminus \{n \cdot N_u \mid n \in \N \},\\
     & & \forall i\in \mc N  \\
     & \eqref{eq:npv}, \eqref{eq:ne}, \eqref{eq:ns}, \eqref{eq:np_v}, \eqref{eq:np_b}, \eqref{eq:bus_occupancy}
      \label{eq:of} &
\end{aligned}
\end{equation}
where $\xi_{\text{min}}$ and $\xi_{\text{max}}$ are the exogenous lower and upper bound of the control variable, $\bar{o}_p$ is the average occupancy of a pool trip such that $\bar{o}_p\in (1,2]$. The second constraint of~\eqref{eq:of} makes sure that the control actions are only updated every $N_u \in \N$ time steps. This avoids frequent fluctuations in the service pricing of ride-hailing, which is desirable from a user perspective. 

\section{Numerical study}
In the following section, we perform a numerical study to evaluate the impact of our proposed allocation scheme and we show the influence of different control strategies on the overall network performance. To do so, we describe in Subsection~\ref{subsec:param} the simulation parameters that we utilize for illustration purposes. Next, in Subsections~\ref{subsec:without} and~\ref{subsec:with}, we assess network delays for scenarios with and without abandonment respectively. We do so to observe the impact of adding a ride-hailing user waiting tolerance on the objective function value defined as the sum of overall user delays and on-demand user waiting time.  

\label{sec:results}
\subsection{Macro-simulation parameters}
\label{subsec:param}
Next, we describe the simulation environment that we implement to test the potential benefits of our proposed strategy along with the different control schemes. 
For this purpose, we consider a network characterized by an MFD that aggregates the microscopic traffic dynamics.
Its functional form is given by the production function $P(n) = A_0n^3 + B_0 n^2 + C_0n$, $A_0 = 5.74 \cdot 10^{-9}$, $B_0 = - 1.02 \cdot 10^{-3}$, and $C_0 = 36$ for $n \in [0, 58536]$.
The fractional split $\alpha$ partitioning the full network space into a vehicle network and a bus network has a value equal to $0.8$, and is usually an unalterable property of the infrastructure.
The value of $\alpha$ yields two production functions for every subnetwork according to the relationship described in Section~\ref{sec:macro_model}. 
Transforming the obtained vehicle MFD into a three-dimensional bus MFD requires the multiplication of the speed with a reduction factor $r(n_b)$ such that $r(n_b) = e^{-6.5\cdot{10^{-4}}n_b}$.
The resulting function described in Subsection~\ref{subsec:traffic} allows the computation of the pool vehicles running speed in the bus network $v_p$.
Similarly, the bus operating speed is straightforwardly obtained by factoring in the spacing $\bar{s}$ between bus stations and the dwell time $\bar{t}_d$ at stops where $\bar{s} = 0.8$ km, and  $\bar{t}_d = 30$ s as shown in~\eqref{eqn:bus_speed}.
With respect to the Cobb-Douglas meeting function that we adopt in~\eqref{eqn:cobbdouglas}, its constant parameters are equal to $0.025$, $0.93$, and $0.98$ for $a_0$, $\alpha_e$, and $\alpha_c$ respectively.
To estimate the outflow for every category of vehicle, we assume that the average trip length for private vehicles is equal to that of solo trips such that $\bar{l}_{pv} = \bar{l}_s = 3.86$ km. 
Since bus trips are generally longer than direct origin-destination trips, we consider that the average bus trip distance is $\bar{l}_b = 1.4\bar{l}_{pv}$. 
Similarly, pool trips are also longer than $\bar{l}_s$ due to the additional detour that passengers/drivers have to incur.
The value of this detour is generally dependent on the number of passengers willing to engage in pooling.
For the scope of this analysis, however, we will assume that the detour remains constant such that $\Delta l_d = 0.7 \bar{l}_s$ and $\Delta l_p = 0.15 \bar{l}_s$.
The integration of a demand-dependent detour value will be examined in future works.
The average occupancies of the different types of vehicles are $1$ and $1.5$ for $\bar{o}_s$ and $\bar{o}_p$ respectively, as the driver is excluded from the occupancy count.
With respect to the private vehicles' occupancy $o_{pv}$, we use a value of $1.2$.

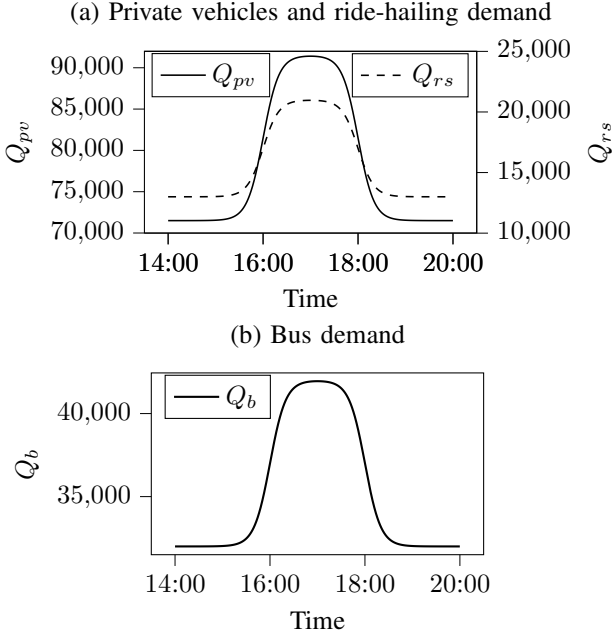
\begin{figure}
    \centering
\begin{tikzpicture}

\definecolor{color0}{rgb}{0.12156862745098,0.466666666666667,0.705882352941177}

\begin{axis}[
width=6cm,
height=4cm,
legend cell align={left},
legend style={at={(0.2,0.99)},anchor=north},
tick align=outside,
tick pos=left,
x grid style={white!69.0196078431373!black},
xlabel={Time},
xmin=13.5, xmax=20.5,
xtick style={color=black},
y grid style={white!69.0196078431373!black},
ylabel={$Q_{pv}$},
ymin=70000, ymax=92000,
ytick style={color=black},
scaled y ticks = false,
y tick label style={/pgf/number format/fixed},
xlabel near ticks,
ylabel near ticks,
scaled x ticks = false,
x tick label style={/pgf/number format/fixed},
x filter/.code={\pgfmathparse{#1/600+14}},
    xticklabel={ 
        \pgfmathsetmacro\hours{floor(\tick)}%
        \pgfmathsetmacro\minutes{(\tick-\hours)*0.6}%
        \pgfmathprintnumber{\hours}:\pgfmathprintnumber[fixed, fixed zerofill, skip 0.=true, dec sep={}]{\minutes}%
    },
title={(a) Private vehicles and ride-hailing demand}
]
\addplot [semithick, black]
table {%
0 71500.125
229 71501.2109375
337 71503.5703125
412 71507.5625
460 71512.21875
499 71518.0390625
533 71525.3359375
561 71533.5078125
587 71543.4375
610 71554.640625
631 71567.3671875
650 71581.3984375
669 71598.3515625
686 71616.46875
703 71637.90625
719 71661.640625
735 71689.421875
750 71719.734375
766 71757.375
781 71798.40625
796 71845.859375
811 71900.7109375
826 71964.0546875
841 72037.140625
856 72121.3671875
871 72218.3125
886 72329.734375
901 72457.5859375
916 72604
932 72783.265625
948 72989.3515625
964 73225.46875
980 73495
996 73801.3203125
1013 74170.8125
1030 74589.28125
1048 75089.203125
1066 75650.171875
1085 76309.7421875
1106 77117.9609375
1129 78091.9140625
1156 79334.7421875
1193 81150.03125
1252 84042.75
1280 85299.21875
1303 86237.9765625
1324 87010.859375
1344 87668.578125
1363 88222.765625
1381 88686.484375
1399 89093.9609375
1416 89430.921875
1433 89725.3671875
1450 89981.3359375
1466 90190.734375
1482 90372.8828125
1498 90530.828125
1514 90667.421875
1530 90785.2421875
1546 90886.6484375
1562 90973.734375
1578 91048.3515625
1594 91112.140625
1610 91166.5390625
1625 91210.1171875
1640 91247.4296875
1655 91279.2421875
1670 91306.234375
1685 91328.9609375
1700 91347.921875
1715 91363.515625
1730 91376.078125
1745 91385.8828125
1761 91393.5390625
1777 91398.484375
1792 91400.78125
1807 91400.8515625
1822 91398.7109375
1837 91394.296875
1852 91387.53125
1867 91378.25
1882 91366.2578125
1896 91352.390625
1911 91334.3671875
1925 91314.2578125
1940 91288.75
1955 91258.609375
1970 91223.2109375
1985 91181.8125
2000 91133.5703125
2015 91077.5
2030 91012.4921875
2045 90937.2734375
2060 90850.40625
2075 90750.296875
2090 90635.125
2105 90502.921875
2120 90351.4921875
2135 90178.484375
2151 89967.2734375
2167 89725.3671875
2183 89449.3828125
2199 89135.9453125
2216 88758.203125
2233 88330.8671875
2251 87821.0234375
2270 87216.25
2290 86504.8359375
2311 85677.5078125
2335 84639.9765625
2363 83329
2408 81100.1015625
2452 78956.96875
2480 77700.453125
2503 76761.640625
2524 75988.6875
2544 75330.875
2563 74776.5859375
2581 74312.7421875
2599 73905.1171875
2616 73567.9921875
2633 73273.359375
2650 73017.15625
2666 72807.5
2682 72625.0546875
2698 72466.75
2714 72329.734375
2730 72211.421875
2746 72109.4375
2762 72021.6796875
2778 71946.265625
2794 71881.5390625
2810 71826.046875
2826 71778.5078125
2842 71737.8203125
2858 71703.015625
2874 71673.2578125
2890 71647.828125
2907 71624.859375
2924 71605.4453125
2942 71588.1484375
2962 71572.234375
2982 71559.1796875
3002 71548.4765625
3025 71538.53125
3052 71529.4296875
3083 71521.59375
3117 71515.375
3158 71510.203125
3214 71505.828125
3291 71502.703125
3395 71500.953125
3599 71500.125
};
\addlegendentry{$Q_{pv}$}
\end{axis}

\begin{axis}[
width=6cm,
height=4cm,
axis y line=right,
legend cell align={left},
legend style={at={(0.8,0.99)},anchor=north},
tick align=outside,
x grid style={white!69.0196078431373!black},
xmin=13.5, xmax=20.5,
xtick pos=left,
xtick style={color=black},
y grid style={white!69.0196078431373!black},
ylabel={$Q_{rs}$},
ymin=10000, ymax=25000,
ytick pos=right,
ytick style={color=black},
yticklabel style={anchor=west},
scaled y ticks = false,
y tick label style={/pgf/number format/fixed},
xlabel near ticks,
ylabel near ticks,
scaled x ticks = false,
x tick label style={/pgf/number format/fixed},
axis line style={-},
x filter/.code={\pgfmathparse{#1/600+14}},
    xticklabel={ 
        \pgfmathsetmacro\hours{floor(\tick)}%
        \pgfmathsetmacro\minutes{(\tick-\hours)*0.6}%
        \pgfmathprintnumber{\hours}:\pgfmathprintnumber[fixed, fixed zerofill, skip 0.=true, dec sep={}]{\minutes}%
    }
]

\addplot [semithick, black, dashed]
table {%
0 13000.048828125
328 13001.3056640625
435 13003.806640625
503 13007.509765625
553 13012.375
593 13018.4462890625
626 13025.6357421875
655 13034.2236328125
680 13043.8896484375
703 13055.162109375
724 13067.9423828125
743 13082.013671875
761 13097.990234375
778 13115.884765625
795 13136.9912109375
811 13160.2841796875
827 13187.4443359375
843 13219.0771484375
859 13255.8740234375
875 13298.61328125
891 13348.1708984375
907 13405.51953125
923 13471.7333984375
939 13547.9775390625
955 13635.50390625
971 13735.6318359375
987 13849.7138671875
1004 13987.7294921875
1021 14144.5732421875
1038 14321.62890625
1056 14532.3515625
1075 14781.5869140625
1095 15073.783203125
1117 15429.138671875
1142 15871.43359375
1172 16443.59375
1270 18345.40234375
1295 18768.794921875
1317 19105.001953125
1337 19378.84765625
1356 19610.59375
1375 19815.33984375
1393 19985.65625
1410 20126.82421875
1427 20250.41796875
1444 20358.052734375
1461 20451.3515625
1477 20527.48046875
1493 20593.556640625
1509 20650.74609375
1525 20700.1171875
1541 20742.63671875
1557 20779.17578125
1573 20810.505859375
1589 20837.310546875
1606 20861.5
1623 20881.875
1640 20898.970703125
1658 20914.00390625
1677 20926.935546875
1697 20937.763671875
1717 20946.142578125
1739 20952.91015625
1762 20957.5703125
1786 20960.052734375
1810 20960.2421875
1834 20958.146484375
1857 20953.888671875
1879 20947.5546875
1900 20939.16796875
1919 20929.31640625
1938 20916.9609375
1956 20902.568359375
1973 20886.173828125
1990 20866.615234375
2006 20844.85546875
2022 20819.33984375
2038 20789.4921875
2054 20754.66015625
2070 20714.09765625
2085 20670.123046875
2101 20615.984375
2116 20557.5546875
2132 20485.97265625
2148 20403.646484375
2164 20309.28515625
2180 20201.552734375
2196 20079.087890625
2213 19931.34765625
2230 19764.009765625
2248 19564.083984375
2267 19326.5390625
2287 19046.55859375
2308 18720.212890625
2331 18327.63671875
2358 17827.791015625
2397 17059.9453125
2451 16001.5185546875
2479 15497.3271484375
2503 15104.6552734375
2524 14795.4736328125
2544 14532.3515625
2563 14310.6328125
2581 14125.0966796875
2599 13962.0478515625
2616 13827.1982421875
2633 13709.3447265625
2650 13606.861328125
2667 13518.1328125
2683 13445.7919921875
2699 13383.03515625
2715 13328.728515625
2731 13281.8359375
2747 13241.421875
2763 13206.6484375
2779 13176.76953125
2796 13149.6513671875
2813 13126.6259765625
2831 13106.04296875
2849 13088.7685546875
2868 13073.548828125
2888 13060.3173828125
2910 13048.478515625
2934 13038.18359375
2961 13029.181640625
2991 13021.638671875
3026 13015.2607421875
3068 13010.033203125
3120 13005.9677734375
3190 13002.96484375
3292 13001.0693359375
3478 13000.166015625
3599 13000.0498046875
};
\addlegendentry{$Q_{rs}$}
\end{axis}

\end{tikzpicture}
    
    \hspace{-1.5cm}
\begin{tikzpicture}

\begin{axis}[
width=6cm,
height=4cm,
legend cell align={left},
legend style={at={(0.2,0.99)},anchor=north},
tick align=outside,
tick pos=left,
x grid style={white!69.0196078431373!black},
xlabel={Time},
xmin=13.5, xmax=20.5,
xtick style={color=black},
y grid style={white!69.0196078431373!black},
ymin=31502.5371365936, ymax=42448.0718416422,
ytick style={color=black},
ylabel={$Q_{b}$},
scaled y ticks = false,
y tick label style={/pgf/number format/fixed},
xlabel near ticks,
ylabel near ticks,
scaled x ticks = false,
x tick label style={/pgf/number format/fixed},
x filter/.code={\pgfmathparse{#1/600+14}},
    xticklabel={ 
        \pgfmathsetmacro\hours{floor(\tick)}%
        \pgfmathsetmacro\minutes{(\tick-\hours)*0.6}%
        \pgfmathprintnumber{\hours}:\pgfmathprintnumber[fixed, fixed zerofill, skip 0.=true, dec sep={}]{\minutes}%
    },
title={(b) Bus demand}
]
\addplot [thick, black]
table {%
0 32000.060546875
317 32001.462890625
419 32004.0546875
485 32007.841796875
534 32012.794921875
573 32018.88671875
606 32026.251953125
634 32034.703125
659 32044.517578125
682 32055.96484375
703 32068.953125
722 32083.259765625
740 32099.517578125
757 32117.7421875
774 32139.255859375
790 32163.0234375
806 32190.771484375
822 32223.1328125
838 32260.83984375
853 32301.77734375
868 32348.912109375
883 32403.1015625
898 32465.30078125
913 32536.5625
929 32623.853515625
945 32724.259765625
961 32839.37890625
977 32970.87890625
993 33120.4609375
1010 33301.07421875
1027 33505.86328125
1045 33750.8515625
1063 34026.18359375
1082 34350.50390625
1103 34748.78125
1126 35230.01171875
1152 35822.484375
1187 36675.40234375
1254 38318.01953125
1281 38920.95703125
1304 39388.328125
1325 39772.78515625
1345 40099.72265625
1364 40375.03125
1382 40605.28125
1400 40807.515625
1417 40974.69140625
1434 41120.72265625
1451 41247.64453125
1467 41351.44140625
1483 41441.71484375
1499 41519.98046875
1515 41587.65234375
1531 41646.01953125
1547 41696.24609375
1563 41739.37109375
1579 41776.31640625
1595 41807.8984375
1611 41834.828125
1627 41857.71875
1644 41878.2109375
1661 41895.29296875
1678 41909.42578125
1696 41921.60546875
1715 41931.7578125
1734 41939.4765625
1755 41945.49609375
1776 41949.12890625
1798 41950.5390625
1820 41949.5625
1841 41946.3671875
1862 41940.81640625
1882 41933.12890625
1901 41923.3828125
1919 41911.64453125
1936 41897.98828125
1953 41881.45703125
1969 41862.875
1985 41840.90625
2001 41815.04296875
2016 41786.73046875
2031 41753.90234375
2046 41715.9296875
2061 41672.08203125
2076 41621.55078125
2091 41563.43359375
2106 41496.7265625
2122 41414.86328125
2138 41320.53125
2154 41212.17578125
2170 41088.15625
2186 40946.78515625
2203 40775.671875
2220 40581.1171875
2237 40361.3828125
2255 40099.72265625
2274 39790.04296875
2294 39426.7265625
2316 38984.51171875
2340 38456.44921875
2370 37744.34375
2472 35273.8984375
2496 34768.7578125
2518 34350.50390625
2538 34010.07421875
2557 33722.15234375
2575 33480.4609375
2593 33267.49609375
2610 33090.9609375
2627 32936.375
2644 32801.72265625
2661 32684.970703125
2677 32589.666015625
2693 32506.900390625
2709 32435.212890625
2725 32373.265625
2741 32319.841796875
2757 32273.845703125
2773 32234.3046875
2789 32200.35546875
2806 32169.564453125
2823 32143.435546875
2840 32121.283203125
2858 32101.5078125
2876 32084.927734375
2895 32070.3359375
2915 32057.66015625
2937 32046.326171875
2961 32036.4765625
2988 32027.869140625
3019 32020.45703125
3055 32014.28125
3098 32009.294921875
3152 32005.41796875
3224 32002.638671875
3331 32000.904296875
3547 32000.103515625
3599 32000.0625
};
\addlegendentry{$Q_{b}$}
\end{axis}

\end{tikzpicture}
    \caption{Time-dependent multi-modal demand profile in pax$/$hr.}
    \label{fig:demand}
\end{figure}

Moving to the multinomial logit model dictating the choice between solo and pooled trip, we consider a mode choice scale parameter $\mu = 1$, and a value of time $\kappa = 30$ CHF/hr.
The static trip fares for solo $F_s$ and for pool $F_p$ are $5$ and $4$ CHF, respectively.
We note here that these values usually change with the total ride-hailing demand.
Since the objective of this study is to determine the value of $\phi_{\mc{V}}$ and $\phi_{\mc{B}}$ irrespective of the values of $F_s$ and $F_p$, we discard the demand-dependent basic fare variations.
The fleet size $N$ remains constant over the full simulation framework and is set to $3500$ vehicles.
The simulation runs over six hours and is discretized such that the duration of every time step $\tau$ is equal to $6$ s.
The demand profile for private vehicles and ride-hailing is displayed in Figure~\subfigref{fig:demand}{a} where the increase in demand starts at around $16\text{:}00$ before it goes back to its original value at around $18\text{:}00$.
The demand for buses varies in a similar manner as also shown in Figure~\subfigref{fig:demand}{b}.

\subsection{Multi-modal network delays without abandonment}
\label{subsec:without}
\begin{figure}
\centering 
\begin{tabular}{cc}
\hspace{-0.2cm}
\begin{tikzpicture}

\definecolor{color0}{rgb}{0.12156862745098,0.466666666666667,0.705882352941177}

\begin{axis}[
width=4.3cm,
height=2.8cm,
tick align=outside,
tick pos=left,
x grid style={white!69.0196078431373!black},
xlabel={\footnotesize Time},
xmin=13.5, xmax=20.5,
xtick style={color=black},
y grid style={white!69.0196078431373!black},
ymin=9928.87765757532, ymax=16677.2933435358,
ytick style={color=black},
scaled y ticks = false,
y tick label style={/pgf/number format/fixed, font = \scriptsize},
xlabel near ticks,
ylabel near ticks,
scaled x ticks = false,
x tick label style={/pgf/number format/fixed, font = \scriptsize},
x filter/.code={\pgfmathparse{#1/600+14}},
    xticklabel={ 
        \pgfmathsetmacro\hours{floor(\tick)}%
        \pgfmathsetmacro\minutes{(\tick-\hours)*0.6}%
        \pgfmathprintnumber{\hours}:\pgfmathprintnumber[fixed, fixed zerofill, skip 0.=true, dec sep={}]{\minutes}%
    },
title={\footnotesize (a) Private vehicles -- $n_{pv}$}
]
\addplot [semithick, blue!50.1960784313725!black]
table {%
0 10500
26 10469.5419921875
52 10442.4423828125
78 10418.47265625
105 10396.6015625
133 10376.841796875
162 10359.1650390625
192 10343.50390625
223 10329.763671875
257 10317.1494140625
294 10305.896484375
334 10296.1474609375
378 10287.794921875
427 10280.8544921875
481 10275.525390625
539 10272.0576171875
598 10270.7548828125
652 10271.7451171875
699 10274.818359375
738 10279.5439453125
772 10285.9033203125
801 10293.55078125
827 10302.6826171875
851 10313.5419921875
872 10325.4169921875
892 10339.23046875
911 10355.0576171875
929 10372.921875
947 10394.0517578125
964 10417.4794921875
981 10444.7841796875
998 10476.5068359375
1015 10513.234375
1032 10555.587890625
1049 10604.2119140625
1066 10659.7607421875
1083 10722.869140625
1101 10798.587890625
1119 10884.056640625
1138 10985.37109375
1158 11104.677734375
1179 11243.9658203125
1201 11404.8349609375
1226 11604.8564453125
1254 11847.6923828125
1288 12162.9296875
1342 12687.5908203125
1407 13313.82421875
1446 13668.4912109375
1481 13967.0498046875
1513 14221.7119140625
1544 14451.0966796875
1574 14656.888671875
1603 14841.076171875
1632 15011.3701171875
1661 15168.4873046875
1689 15308.4169921875
1717 15437.509765625
1745 15556.46484375
1773 15665.947265625
1801 15766.5703125
1829 15858.87890625
1856 15940.4580078125
1883 16015.0849609375
1909 16080.61328125
1934 16137.890625
1958 16187.6201171875
1981 16230.388671875
2003 16266.685546875
2023 16295.5888671875
2042 16319.2216796875
2059 16336.98828125
2075 16350.5771484375
2089 16359.78515625
2102 16365.9228515625
2115 16369.5625
2127 16370.537109375
2139 16369.0478515625
2150 16365.361328125
2161 16359.2900390625
2172 16350.6669921875
2183 16339.30859375
2195 16323.5712890625
2207 16304.0810546875
2220 16278.40625
2233 16247.6240234375
2247 16208.3203125
2261 16162.1591796875
2276 16104.5439453125
2291 16037.9345703125
2307 15956.3828125
2323 15863.4609375
2340 15751.814453125
2358 15618.791015625
2377 15461.8369140625
2398 15269.1220703125
2421 15036.3759765625
2447 14749.3134765625
2478 14380.9970703125
2524 13804.7353515625
2587 13019.2236328125
2621 12622.8359375
2651 12298.51171875
2678 12030.1748046875
2703 11802.87109375
2727 11604.111328125
2750 11431.302734375
2773 11275.2490234375
2795 11140.953125
2816 11025.662109375
2837 10922.1806640625
2858 10829.6943359375
2878 10751.072265625
2898 10680.9599609375
2918 10618.6630859375
2938 10563.513671875
2958 10514.8779296875
2978 10472.154296875
2998 10434.779296875
3018 10402.2255859375
3038 10374.0048828125
3058 10349.6611328125
3078 10328.77734375
3098 10310.9677734375
3119 10295.1904296875
3141 10281.474609375
3164 10269.80078125
3189 10259.7490234375
3215 10251.7412109375
3244 10245.2255859375
3276 10240.3955078125
3313 10237.169921875
3357 10235.697265625
3413 10236.2099609375
3496 10239.451171875
3599 10244.583984375
};
\end{axis}

\end{tikzpicture} &
\hspace{-0.1cm}
\begin{tikzpicture}

\definecolor{color0}{rgb}{0.12156862745098,0.466666666666667,0.705882352941177}

\begin{axis}[
width=4.3cm,
height=2.8cm,
tick align=outside,
tick pos=left,
x grid style={white!69.0196078431373!black},
xlabel={\footnotesize Time},
xmin=13.5, xmax=20.5,
xtick style={color=black},
y grid style={white!69.0196078431373!black},
ymin=386.54033648655, ymax=682.652933782453,
ytick style={color=black},
scaled y ticks = false,
y tick label style={/pgf/number format/fixed, font = \scriptsize},
xlabel near ticks,
ylabel near ticks,
scaled x ticks = false,
x tick label style={/pgf/number format/fixed, font = \scriptsize},
x filter/.code={\pgfmathparse{#1/600+14}},
    xticklabel={ 
        \pgfmathsetmacro\hours{floor(\tick)}%
        \pgfmathsetmacro\minutes{(\tick-\hours)*0.6}%
        \pgfmathprintnumber{\hours}:\pgfmathprintnumber[fixed, fixed zerofill, skip 0.=true, dec sep={}]{\minutes}%
    },
title={\footnotesize (b) Solo trips -- $n_s$}
]
\addplot [semithick, blue!50.1960784313725!black]
table {%
0 400
9 400.670379638672
21 401.304595947266
38 401.930816650391
67 402.710296630859
126 404.014038085938
184 405.071716308594
239 405.826354980469
298 406.381958007812
366 406.767791748047
456 407.013549804688
705 407.552215576172
762 407.997100830078
806 408.579193115234
842 409.293609619141
873 410.150909423828
900 411.140228271484
924 412.261108398438
946 413.534423828125
966 414.936553955078
985 416.521697998047
1002 418.18017578125
1018 419.975036621094
1034 422.022094726562
1049 424.192596435547
1063 426.456237792969
1077 428.965545654297
1091 431.734588623047
1104 434.548919677734
1117 437.604858398438
1130 440.907012939453
1143 444.456787109375
1156 448.252014160156
1170 452.606384277344
1184 457.224792480469
1199 462.444244384766
1215 468.286712646484
1233 475.142425537109
1254 483.420349121094
1291 498.337219238281
1320 509.906890869141
1340 517.627685546875
1358 524.306640625
1374 529.98291015625
1389 535.05419921875
1404 539.864868164062
1419 544.403015136719
1433 548.387023925781
1447 552.126586914062
1461 555.623596191406
1476 559.106506347656
1491 562.324951171875
1506 565.289978027344
1521 568.014404296875
1537 570.671142578125
1554 573.231811523438
1571 575.544189453125
1589 577.746643066406
1608 579.824951171875
1628 581.770263671875
1650 583.660095214844
1674 585.466552734375
1700 587.173156738281
1729 588.827697753906
1762 590.461059570312
1801 592.138000488281
1849 593.946228027344
1918 596.271728515625
2055 600.841552734375
2102 602.687683105469
2139 604.377258300781
2171 606.084655761719
2199 607.829040527344
2224 609.63525390625
2247 611.546264648438
2268 613.532592773438
2288 615.66455078125
2308 618.051696777344
2327 620.569580078125
2346 623.337524414062
2365 626.352722167969
2385 629.778991699219
2407 633.810729980469
2433 638.845947265625
2511 654.165832519531
2530 657.531127929688
2547 660.287719726562
2562 662.47705078125
2576 664.28466796875
2590 665.839721679688
2603 667.038696289062
2616 667.985900878906
2628 668.625183105469
2640 669.029357910156
2652 669.190551757812
2664 669.101989746094
2676 668.758117675781
2688 668.154418945312
2700 667.287170410156
2712 666.153564453125
2724 664.751586914062
2736 663.080017089844
2748 661.138122558594
2760 658.926025390625
2772 656.444519042969
2784 653.695007324219
2796 650.679748535156
2808 647.401733398438
2820 643.864990234375
2832 640.074340820312
2844 636.035949707031
2857 631.3896484375
2870 626.471984863281
2883 621.295654296875
2897 615.449096679688
2911 609.34130859375
2926 602.534973144531
2943 594.53466796875
2962 585.297302246094
2984 574.316772460938
3023 554.512634277344
3054 538.900146484375
3075 528.594787597656
3093 520.024963378906
3110 512.205383300781
3126 505.122924804688
3141 498.749572753906
3156 492.649688720703
3170 487.212768554688
3184 482.028869628906
3198 477.100524902344
3212 472.427551269531
3226 468.007629394531
3240 463.83642578125
3254 459.908020019531
3269 455.960205078125
3284 452.272644042969
3299 448.834197998047
3315 445.427795410156
3331 442.276611328125
3348 439.191314697266
3365 436.359283447266
3383 433.617156982422
3402 430.986999511719
3421 428.605987548828
3441 426.34521484375
3462 424.21728515625
3484 422.231292724609
3508 420.318084716797
3533 418.573760986328
3560 416.939239501953
3589 415.433349609375
3599 414.96728515625
};
\end{axis}

\end{tikzpicture} \\
\begin{tikzpicture}

\definecolor{color0}{rgb}{0.12156862745098,0.466666666666667,0.705882352941177}

\begin{axis}[
width=4.3cm,
height=2.8cm,
tick align=outside,
tick pos=left,
x grid style={white!69.0196078431373!black},
xlabel={\footnotesize Time},
xmin=13.5, xmax=20.5,
xtick style={color=black},
y grid style={white!69.0196078431373!black},
ymin=845.692278208174, ymax=1320.58351452994,
ytick style={color=black},
scaled y ticks = false,
y tick label style={/pgf/number format/fixed, font = \scriptsize},
xlabel near ticks,
ylabel near ticks,
scaled x ticks = false,
x tick label style={/pgf/number format/fixed, font = \scriptsize},
x filter/.code={\pgfmathparse{#1/600+14}},
    xticklabel={ 
        \pgfmathsetmacro\hours{floor(\tick)}%
        \pgfmathsetmacro\minutes{(\tick-\hours)*0.6}%
        \pgfmathprintnumber{\hours}:\pgfmathprintnumber[fixed, fixed zerofill, skip 0.=true, dec sep={}]{\minutes}%
    },
title={\footnotesize (c) Pool in $\mc{V}$ -- $n_p^{\mc{V}}$}
]
\addplot [semithick, blue!50.1960784313725!black]
table {%
0 900
57 888.467407226562
75 885.507568359375
94 882.803161621094
114 880.376403808594
135 878.230590820312
158 876.279663085938
184 874.488342285156
213 872.905578613281
246 871.517883300781
284 870.327575683594
330 869.302490234375
386 868.466064453125
458 867.803283691406
551 867.365173339844
644 867.303588867188
718 867.625244140625
774 868.249267578125
818 869.12060546875
854 870.208740234375
885 871.523620605469
912 873.044799804688
937 874.849426269531
959 876.82421875
980 879.113403320312
999 881.581481933594
1017 884.317932128906
1034 887.301696777344
1050 890.501403808594
1066 894.113891601562
1081 897.902282714844
1096 902.101440429688
1111 906.727966308594
1126 911.79248046875
1141 917.297668457031
1156 923.237548828125
1171 929.595825195312
1187 936.808837890625
1204 944.908813476562
1223 954.406127929688
1246 966.359375
1323 1006.78363037109
1342 1016.09429931641
1359 1023.99548339844
1375 1031.00170898438
1390 1037.15368652344
1405 1042.87768554688
1420 1048.15771484375
1434 1052.67797851562
1448 1056.80346679688
1462 1060.53784179688
1477 1064.11389160156
1492 1067.26403808594
1507 1070.00573730469
1522 1072.35900878906
1538 1074.46643066406
1554 1076.18591308594
1571 1077.62084960938
1589 1078.73901367188
1608 1079.51708984375
1628 1079.94128417969
1650 1080.00219726562
1674 1079.65905761719
1701 1078.85583496094
1732 1077.51171875
1769 1075.48767089844
1819 1072.32690429688
1949 1063.93493652344
1995 1061.46411132812
2034 1059.76513671875
2068 1058.67333984375
2098 1058.08959960938
2126 1057.94140625
2151 1058.20239257812
2174 1058.83569335938
2195 1059.80236816406
2215 1061.12438964844
2233 1062.69848632812
2250 1064.56359863281
2266 1066.69372558594
2282 1069.2255859375
2297 1071.998046875
2312 1075.18884277344
2326 1078.57092285156
2340 1082.36694335938
2354 1086.59741210938
2367 1090.92956542969
2380 1095.66125488281
2393 1100.798828125
2406 1106.34411621094
2419 1112.29382324219
2433 1119.14318847656
2447 1126.43286132812
2462 1134.70178222656
2478 1143.99816894531
2495 1154.34411621094
2514 1166.36889648438
2539 1182.69641113281
2599 1222.10266113281
2618 1233.99133300781
2634 1243.5625
2649 1252.07922363281
2663 1259.56201171875
2676 1266.05615234375
2688 1271.62426757812
2700 1276.75073242188
2711 1281.03674316406
2722 1284.90625
2733 1288.33984375
2743 1291.06848144531
2753 1293.41064453125
2763 1295.35559082031
2773 1296.89404296875
2783 1298.017578125
2793 1298.71936035156
2803 1298.99377441406
2813 1298.83654785156
2823 1298.24462890625
2833 1297.21667480469
2843 1295.75244140625
2853 1293.85375976562
2863 1291.52331542969
2873 1288.76599121094
2883 1285.587890625
2893 1281.9970703125
2903 1278.00305175781
2914 1273.15747070312
2925 1267.85510253906
2936 1262.11572265625
2948 1255.3818359375
2960 1248.18566894531
2973 1239.90795898438
2987 1230.48559570312
3002 1219.875
3018 1208.05786132812
3036 1194.26965332031
3058 1176.90869140625
3095 1147.11071777344
3132 1117.49426269531
3156 1098.80810546875
3176 1083.7177734375
3195 1069.86999511719
3213 1057.24060058594
3230 1045.78210449219
3247 1034.79992675781
3264 1024.30627441406
3280 1014.88177490234
3296 1005.89575195312
3312 997.345092773438
3328 989.223510742188
3344 981.522277832031
3361 973.788330078125
3378 966.502136230469
3395 959.6474609375
3412 953.207214355469
3430 946.819946289062
3448 940.855590820312
3467 934.99462890625
3486 929.555236816406
3506 924.257446289062
3526 919.370666503906
3547 914.65185546875
3569 910.128662109375
3591 906.0029296875
3599 904.595031738281
};
\end{axis}

\end{tikzpicture} &
\hspace{-0.22cm}
\begin{tikzpicture}

\definecolor{color0}{rgb}{0.12156862745098,0.466666666666667,0.705882352941177}

\begin{axis}[
width=4.3cm,
height=2.8cm,
tick align=outside,
tick pos=left,
x grid style={white!69.0196078431373!black},
xlabel={\footnotesize Time},
xmin=13.5, xmax=20.5,
xtick style={color=black},
y grid style={white!69.0196078431373!black},
ymin=654.778948493802, ymax=1649.64208163015,
ytick style={color=black},
scaled y ticks = false,
y tick label style={/pgf/number format/fixed, font = \scriptsize},
xlabel near ticks,
ylabel near ticks,
scaled x ticks = false,
x tick label style={/pgf/number format/fixed, font = \scriptsize},
x filter/.code={\pgfmathparse{#1/600+14}},
    xticklabel={ 
        \pgfmathsetmacro\hours{floor(\tick)}%
        \pgfmathsetmacro\minutes{(\tick-\hours)*0.6}%
        \pgfmathprintnumber{\hours}:\pgfmathprintnumber[fixed, fixed zerofill, skip 0.=true, dec sep={}]{\minutes}%
    },
title={\footnotesize (d) Pool in $\mc{B}$ -- $n_p^{\mc{B}}$}
]
\addplot [semithick, blue!50.1960784313725!black]
table {%
0 700
12 703.834777832031
26 707.420349121094
43 710.864624023438
62 713.862548828125
84 716.504699707031
110 718.766784667969
140 720.5146484375
176 721.736267089844
220 722.359436035156
280 722.310852050781
385 721.241577148438
566 719.485900878906
675 719.263244628906
755 719.884460449219
814 721.129699707031
860 722.8857421875
898 725.130554199219
930 727.808776855469
958 730.940246582031
983 734.528747558594
1006 738.644226074219
1027 743.218566894531
1047 748.424865722656
1066 754.253784179688
1084 760.672607421875
1101 767.625
1117 775.031677246094
1133 783.340148925781
1149 792.607421875
1165 802.881652832031
1181 814.199157714844
1197 826.581848144531
1213 840.034973144531
1229 854.545776367188
1246 871.086791992188
1264 889.794921875
1283 910.766479492188
1304 935.234436035156
1327 963.320861816406
1355 998.863830566406
1397 1053.69201660156
1456 1130.54553222656
1488 1170.84033203125
1516 1204.81799316406
1542 1235.10998535156
1567 1262.99169921875
1591 1288.55639648438
1615 1312.92053222656
1639 1336.08129882812
1663 1358.05151367188
1687 1378.85546875
1711 1398.525390625
1735 1417.09899902344
1759 1434.61755371094
1783 1451.12341308594
1807 1466.65991210938
1832 1481.85778808594
1857 1496.09460449219
1883 1509.92639160156
1909 1522.80615234375
1935 1534.77026367188
1961 1545.84680175781
1987 1556.05529785156
2013 1565.40356445312
2039 1573.88671875
2065 1581.48327636719
2090 1587.91455078125
2114 1593.23120117188
2137 1597.47436523438
2159 1600.67883300781
2180 1602.8779296875
2199 1604.0673828125
2218 1604.41552734375
2236 1603.88903808594
2253 1602.55078125
2269 1600.47631835938
2285 1597.54516601562
2300 1593.96008300781
2315 1589.50964355469
2330 1584.142578125
2344 1578.26489257812
2358 1571.51477050781
2372 1563.86596679688
2386 1555.29992675781
2400 1545.80773925781
2415 1534.61193847656
2430 1522.37097167969
2446 1508.19787597656
2462 1492.9287109375
2479 1475.5888671875
2498 1454.98193359375
2519 1430.90771484375
2543 1402.05749511719
2573 1364.59936523438
2691 1215.74255371094
2720 1181.28466796875
2747 1150.44787597656
2774 1120.87854003906
2800 1093.60485839844
2827 1066.48828125
2854 1040.53149414062
2882 1014.75299072266
2911 989.190246582031
2940 964.719116210938
2969 941.313537597656
2997 919.74267578125
3024 899.946716308594
3050 881.879455566406
3075 865.491882324219
3099 850.721557617188
3123 836.934265136719
3146 824.667846679688
3169 813.3359375
3192 802.931823730469
3215 793.434875488281
3238 784.812561035156
3262 776.702331542969
3287 769.158020019531
3313 762.217102050781
3340 755.900573730469
3368 750.213562011719
3398 744.985900878906
3430 740.271728515625
3464 736.099731445312
3502 732.294799804688
3544 728.953552246094
3591 726.07080078125
3599 725.655212402344
};
\end{axis}

\end{tikzpicture} \\
\hspace{0.4cm}
\begin{tikzpicture}

\definecolor{color0}{rgb}{0.12156862745098,0.466666666666667,0.705882352941177}

\begin{axis}[
width=4.3cm,
height=2.8cm,
tick align=outside,
tick pos=left,
x grid style={white!69.0196078431373!black},
xlabel={\footnotesize Time},
xmin=13.5, xmax=20.5,
xtick style={color=black},
y grid style={white!69.0196078431373!black},
ymin=17.5003728696858, ymax=22.7380412734788,
ytick style={color=black},
scaled y ticks = false,
y tick label style={/pgf/number format/fixed, font = \scriptsize},
xlabel near ticks,
ylabel near ticks,
scaled x ticks = false,
x tick label style={/pgf/number format/fixed, font = \scriptsize},
x filter/.code={\pgfmathparse{#1/600+14}},
    xticklabel={ 
        \pgfmathsetmacro\hours{floor(\tick)}%
        \pgfmathsetmacro\minutes{(\tick-\hours)*0.6}%
        \pgfmathprintnumber{\hours}:\pgfmathprintnumber[fixed, fixed zerofill, skip 0.=true, dec sep={}]{\minutes}%
    },
title={\footnotesize (e) Speed in $\mc{V}$ -- $v_{\mc{V}}$}
]
\addplot [semithick, blue!50.1960784313725!black]
table {%
0 22.1764698028564
21 22.2054615020752
45 22.2340984344482
72 22.2617416381836
101 22.2869491577148
133 22.3102264404297
168 22.3311290740967
206 22.3493843078613
248 22.3652267456055
297 22.3791961669922
354 22.3908271789551
421 22.3998966217041
498 22.4059753417969
588 22.4086303710938
668 22.4069137573242
734 22.4012584686279
784 22.3927612304688
825 22.3815155029297
859 22.3679294586182
888 22.352165222168
914 22.3337783813477
937 22.3133106231689
958 22.2904319763184
978 22.2642288208008
996 22.2363605499268
1013 22.2057819366455
1029 22.1727638244629
1044 22.1376800537109
1059 22.0982284545898
1073 22.0571594238281
1087 22.0117111206055
1100 21.9653739929199
1113 21.9148826599121
1126 21.8601036071777
1139 21.8009490966797
1152 21.737377166748
1165 21.6694030761719
1178 21.597095489502
1191 21.5205936431885
1205 21.4337387084961
1219 21.3425674438477
1234 21.2405395507812
1250 21.1273822784424
1268 20.9956741333008
1290 20.8300094604492
1322 20.5840854644775
1364 20.2619400024414
1388 20.0824203491211
1409 19.9298210144043
1428 19.7961235046387
1446 19.6737403869629
1464 19.5558185577393
1481 19.4487133026123
1498 19.3458461761475
1515 19.2472534179688
1532 19.1529216766357
1549 19.062801361084
1566 18.9768142700195
1584 18.8901615142822
1602 18.8078861236572
1620 18.7298316955566
1639 18.6518402099609
1658 18.5781688690186
1678 18.5050678253174
1698 18.4362964630127
1719 18.368501663208
1741 18.3020629882812
1763 18.240047454834
1786 18.1796703338623
1809 18.123592376709
1833 18.0694141387939
1858 18.017448425293
1883 17.969841003418
1909 17.9247913360596
1935 17.8841972351074
1961 17.8480415344238
1986 17.8175411224365
2011 17.7913990020752
2035 17.7706718444824
2058 17.7551765441895
2080 17.7447776794434
2100 17.7395076751709
2119 17.7386379241943
2137 17.7419776916504
2154 17.7492961883545
2170 17.7603034973145
2186 17.7757453918457
2201 17.794677734375
2215 17.8166179656982
2229 17.8430576324463
2242 17.8719711303711
2255 17.9054069519043
2267 17.9405574798584
2279 17.9800701141357
2291 18.0241775512695
2302 18.0688323974609
2313 18.1176815032959
2324 18.1708564758301
2335 18.2284603118896
2346 18.2905712127686
2357 18.3572273254395
2368 18.4284343719482
2379 18.5041542053223
2390 18.5843086242676
2401 18.6687698364258
2412 18.7573661804199
2424 18.8584785461426
2436 18.9639186859131
2449 19.0825710296631
2463 19.2148609161377
2478 19.3609218597412
2496 19.5406761169434
2521 19.7951164245605
2561 20.2023906707764
2580 20.3912925720215
2596 20.5462913513184
2611 20.6873836517334
2625 20.8148288726807
2639 20.9377574920654
2652 21.0475635528564
2665 21.1529712677002
2678 21.2538223266602
2691 21.3500080108643
2703 21.4345970153809
2715 21.51513671875
2727 21.5916328430176
2740 21.6699771881104
2753 21.743673324585
2766 21.812816619873
2779 21.8775100708008
2792 21.9378814697266
2805 21.9940700531006
2819 22.0500736236572
2833 22.1015911102295
2847 22.1488285064697
2862 22.1949234008789
2877 22.2365989685059
2893 22.2764720916748
2909 22.311918258667
2926 22.3450736999512
2944 22.3755073547363
2962 22.4015426635742
2981 22.4246997833252
3001 22.4447612762451
3023 22.4623107910156
3046 22.4762630462646
3071 22.4870986938477
3099 22.4948253631592
3131 22.4991512298584
3169 22.499683380127
3217 22.4956359863281
3288 22.4846992492676
3490 22.4514083862305
3598 22.4391651153564
3599 22.4390716552734
};
\end{axis}

\end{tikzpicture} &
\hspace{0.2cm}
\begin{tikzpicture}

\definecolor{color0}{rgb}{0.12156862745098,0.466666666666667,0.705882352941177}

\begin{axis}[
width=4.3cm,
height=2.8cm,
tick align=outside,
tick pos=left,
x grid style={white!69.0196078431373!black},
xlabel={\footnotesize Time},
xmin=13.5, xmax=20.5,
xtick style={color=black},
y grid style={white!69.0196078431373!black},
ymin=18.8553944402523, ymax=21.7825384683567,
ytick style={color=black},
scaled y ticks = false,
y tick label style={/pgf/number format/fixed, font = \scriptsize},
xlabel near ticks,
ylabel near ticks,
scaled x ticks = false,
x tick label style={/pgf/number format/fixed, font = \scriptsize},
x filter/.code={\pgfmathparse{#1/600+14}},
    xticklabel={ 
        \pgfmathsetmacro\hours{floor(\tick)}%
        \pgfmathsetmacro\minutes{(\tick-\hours)*0.6}%
        \pgfmathprintnumber{\hours}:\pgfmathprintnumber[fixed, fixed zerofill, skip 0.=true, dec sep={}]{\minutes}%
    },
title={\footnotesize (f) Speed in $\mc{B}$ -- $v_{\mc{B}}$}
]
\addplot [semithick, blue!50.1960784313725!black]
table {%
0 21.649486541748
12 21.6378879547119
26 21.6270446777344
42 21.6171760559082
61 21.6079902648926
83 21.5998992919922
108 21.5931911468506
137 21.5878925323486
172 21.5840740203857
215 21.5820007324219
272 21.5819053649902
364 21.5845184326172
563 21.5905227661133
675 21.5912494659424
758 21.589241027832
817 21.5853385925293
862 21.5800094604492
899 21.5733089447021
931 21.5651397705078
959 21.5555934906006
984 21.5446586608887
1007 21.5321235656738
1028 21.5181980133057
1047 21.5032157897949
1065 21.4866333007812
1082 21.4685764312744
1098 21.4492359161377
1114 21.4274139404297
1129 21.4045333862305
1144 21.3791637420654
1158 21.3531322479248
1172 21.3247451782227
1186 21.2939472198486
1200 21.2607135772705
1214 21.2250461578369
1228 21.1869773864746
1242 21.1465816497803
1257 21.100830078125
1272 21.0526885986328
1288 20.9989280700684
1305 20.9393978118896
1324 20.8703842163086
1346 20.7879199981689
1374 20.6803665161133
1453 20.3753681182861
1478 20.2820472717285
1501 20.1986522674561
1522 20.1248970031738
1543 20.053638458252
1563 19.9882202148438
1583 19.9252700805664
1603 19.8648262023926
1623 19.8068981170654
1643 19.75146484375
1663 19.6984939575195
1683 19.6479358673096
1703 19.5997257232666
1724 19.5515594482422
1745 19.505823135376
1767 19.4604187011719
1789 19.41748046875
1811 19.3769054412842
1834 19.3369083404541
1858 19.2976951599121
1882 19.260950088501
1907 19.2251873016357
1932 19.1918907165527
1957 19.160982131958
1982 19.1324024200439
2008 19.1051177978516
2034 19.0803241729736
2059 19.0588760375977
2084 19.0398597717285
2108 19.0240230560303
2131 19.0112400054932
2153 19.0013980865479
2174 18.9943962097168
2194 18.9901294708252
2213 18.9884777069092
2231 18.9892959594727
2248 18.9924144744873
2264 18.9976196289062
2280 19.0052165985107
2295 19.0146827697754
2310 19.026575088501
2324 19.0400009155273
2338 19.0557880401611
2352 19.0740413665771
2366 19.0948486328125
2379 19.1165142059326
2392 19.140474319458
2405 19.1667404174805
2419 19.1975975036621
2433 19.2310752868652
2447 19.2670993804932
2461 19.3055591583252
2476 19.3493118286133
2491 19.3954944610596
2507 19.4471645355225
2525 19.5078773498535
2544 19.5744075775146
2567 19.6574745178223
2599 19.7758083343506
2662 20.0092887878418
2689 20.1065940856934
2713 20.1907253265381
2736 20.2689323425293
2758 20.3413581848145
2780 20.4113960266113
2802 20.4790573120117
2825 20.5473232269287
2848 20.6131839752197
2872 20.6794872283936
2897 20.7460861206055
2923 20.8128261566162
2949 20.8770847320557
2975 20.9388751983643
3000 20.9958953857422
3025 21.0504360198975
3049 21.1003036499023
3072 21.1456470489502
3094 21.1866474151611
3116 21.2252216339111
3137 21.2597122192383
3159 21.293363571167
3180 21.3231143951416
3202 21.3518295288086
3224 21.3780994415283
3246 21.4020175933838
3269 21.4246311187744
3293 21.4457836151123
3318 21.4653568267822
3344 21.4832763671875
3372 21.5000705718994
3401 21.5150299072266
3433 21.5290222167969
3467 21.5414028167725
3504 21.5524272918701
3545 21.5621891021729
3590 21.5705223083496
3599 21.5719394683838
};
\end{axis}

\end{tikzpicture} \\
\hspace{0.4cm}
\begin{tikzpicture}

\definecolor{color0}{rgb}{0.12156862745098,0.466666666666667,0.705882352941177}

\begin{axis}[
width=4.3cm,
height=2.8cm,
tick align=outside,
tick pos=left,
x grid style={white!69.0196078431373!black},
xlabel={\footnotesize Time},
xmin=13.5, xmax=20.5,
xtick style={color=black},
y grid style={white!69.0196078431373!black},
ymin=15.7621775749909, ymax=17.7562563730342,
ytick style={color=black},
scaled y ticks = false,
y tick label style={/pgf/number format/fixed, font = \scriptsize},
xlabel near ticks,
ylabel near ticks,
scaled x ticks = false,
x tick label style={/pgf/number format/fixed, font = \scriptsize},
x filter/.code={\pgfmathparse{#1/600+14}},
    xticklabel={ 
        \pgfmathsetmacro\hours{floor(\tick)}%
        \pgfmathsetmacro\minutes{(\tick-\hours)*0.6}%
        \pgfmathprintnumber{\hours}:\pgfmathprintnumber[fixed, fixed zerofill, skip 0.=true, dec sep={}]{\minutes}%
    },
title={\footnotesize (g) Bus speed -- $v_b$}
]
\addplot [semithick, blue!50.1960784313725!black]
table {%
0 17.6656169891357
12 17.6578922271729
26 17.650671005249
42 17.6440982818604
61 17.6379776000977
83 17.6325855255127
108 17.628116607666
137 17.624584197998
172 17.6220397949219
214 17.6206722259521
272 17.6205921173096
375 17.6225967407227
555 17.6262359619141
664 17.6268653869629
750 17.6257076263428
810 17.6232929229736
855 17.6200008392334
894 17.6155681610107
927 17.6101932525635
955 17.6040668487549
981 17.5967235565186
1004 17.588586807251
1025 17.5795383453369
1045 17.5692348480225
1064 17.5576915740967
1081 17.5457305908203
1097 17.5329113006592
1113 17.5184364318848
1128 17.5032501220703
1143 17.4863986968994
1157 17.4690933227539
1171 17.4502067565918
1185 17.4296989440918
1199 17.4075469970703
1213 17.3837509155273
1227 17.3583278656006
1242 17.3293304443359
1257 17.2985935211182
1272 17.2662258148193
1288 17.2300472259521
1305 17.1899490356445
1324 17.1434097290039
1345 17.0902938842773
1372 17.0202312469482
1465 16.7772769927979
1490 16.714542388916
1512 16.6609649658203
1533 16.6114368438721
1553 16.5658473968506
1573 16.5218753814697
1593 16.4795627593994
1613 16.4389324188232
1633 16.3999843597412
1653 16.3627052307129
1673 16.3270721435547
1693 16.2930488586426
1713 16.2605934143066
1734 16.2281551361084
1755 16.1973400115967
1777 16.1667366027832
1799 16.1377830505371
1822 16.1092071533203
1845 16.0822925567627
1868 16.0569667816162
1892 16.032169342041
1917 16.0080394744873
1942 15.9855842590332
1967 15.9647569656372
1993 15.9447927474976
2019 15.9265460968018
2044 15.9106359481812
2069 15.896372795105
2093 15.8843011856079
2116 15.8743181228638
2138 15.8663358688354
2160 15.8600282669067
2181 15.8557291030884
2200 15.8534421920776
2219 15.8528413772583
2237 15.8539876937866
2254 15.8567562103271
2270 15.8609933853149
2286 15.8669471740723
2301 15.8742055892944
2315 15.8825407028198
2329 15.8924674987793
2343 15.9040641784668
2357 15.9173974990845
2370 15.9313802719116
2383 15.9469375610352
2396 15.9640884399414
2409 15.982834815979
2422 16.0031623840332
2435 16.0250396728516
2449 16.0502796173096
2463 16.0771789550781
2478 16.1077251434326
2493 16.1399097442627
2509 16.1758518218994
2527 16.2179946899414
2547 16.2665424346924
2571 16.326530456543
2610 16.4260597229004
2655 16.5405464172363
2682 16.6075820922852
2706 16.6655559539795
2729 16.7194385528564
2751 16.7693157196045
2773 16.8175106048584
2795 16.8640270233154
2817 16.9089012145996
2840 16.9541358947754
2863 16.9977474212646
2887 17.0416316986084
2912 17.0856857299805
2938 17.1297874450684
2964 17.1721782684326
2989 17.2113094329834
3014 17.2487754821777
3038 17.2830867767334
3061 17.3143444061279
3084 17.3439159393311
3106 17.3705501556396
3127 17.3944129943848
3148 17.4167213439941
3170 17.4384098052979
3192 17.4583969116211
3214 17.476713180542
3237 17.494140625
3260 17.5098876953125
3284 17.52463722229
3309 17.5383033752441
3335 17.5508289337158
3362 17.5621891021729
3392 17.573055267334
3423 17.5825824737549
3457 17.5913181304932
3494 17.5991020202637
3534 17.6058521270752
3580 17.6118946075439
3599 17.6139488220215
};
\end{axis}
\end{tikzpicture} &
\hspace{0.1cm}
\begin{tikzpicture}

\definecolor{color0}{rgb}{0.12156862745098,0.466666666666667,0.705882352941177}
\definecolor{color1}{rgb}{1,0.498039215686275,0.0549019607843137}

\begin{axis}[
legend cell align={left},
legend style={at={(0.5,1.05)},anchor=north, fill opacity=0, draw opacity=1, text opacity=1, draw=white!80!black, legend columns=2, fill = none, legend style={draw=none}},
tick align=outside,
tick pos=left,
x grid style={white!69.0196078431373!black},
width=4.3cm,
height=2.8cm,
x grid style={white!69.0196078431373!black},
xlabel={\footnotesize Time},
xmin=13.5, xmax=20.5,
xtick style={color=black},
y grid style={white!69.0196078431373!black},
ymin=0.251238149184249, ymax=0.468025434430549,
ytick style={color=black},
scaled y ticks = false,
y tick label style={/pgf/number format/fixed, font = \scriptsize},
xlabel near ticks,
ylabel near ticks,
scaled x ticks = false,
x tick label style={/pgf/number format/fixed, font = \scriptsize},
x filter/.code={\pgfmathparse{#1/600+14}},
    xticklabel={ 
        \pgfmathsetmacro\hours{floor(\tick)}%
        \pgfmathsetmacro\minutes{(\tick-\hours)*0.6}%
        \pgfmathprintnumber{\hours}:\pgfmathprintnumber[fixed, fixed zerofill, skip 0.=true, dec sep={}]{\minutes}%
    },
title={\footnotesize (h) Choice -- $\beta_{\mc{V}}$ \& $\beta_{\mc{B}}$ }
]
\addplot [semithick, blue!50.1960784313725!black, dashed]
table {%
0 0.374650955200195
13 0.375757932662964
28 0.376857042312622
46 0.377986907958984
66 0.379054307937622
88 0.380045533180237
112 0.380948424339294
139 0.381781816482544
169 0.382523894309998
203 0.38317883014679
242 0.383741974830627
287 0.38420557975769
341 0.384575486183167
409 0.384850382804871
497 0.385014891624451
607 0.385037541389465
708 0.384888410568237
779 0.384613394737244
832 0.38423478603363
874 0.383760571479797
909 0.383191227912903
939 0.382530570030212
966 0.381759405136108
990 0.38089919090271
1012 0.379937052726746
1033 0.378837585449219
1052 0.377669215202332
1070 0.376392364501953
1088 0.374934554100037
1105 0.373379230499268
1122 0.371640920639038
1139 0.369712591171265
1155 0.367721557617188
1172 0.365419268608093
1189 0.362929940223694
1206 0.360263586044312
1224 0.357264041900635
1244 0.353745698928833
1267 0.34950578212738
1296 0.343960285186768
1379 0.327971696853638
1404 0.323397874832153
1427 0.319365501403809
1449 0.315687298774719
1470 0.312351226806641
1491 0.309191823005676
1512 0.306211113929749
1533 0.303407669067383
1555 0.300656914710999
1577 0.298090696334839
1599 0.295701265335083
1622 0.293383359909058
1646 0.291149854660034
1670 0.289093971252441
1695 0.287128925323486
1721 0.285263061523438
1748 0.283503532409668
1776 0.281856417655945
1805 0.280327796936035
1835 0.278924107551575
1866 0.277653574943542
1898 0.276527881622314
1929 0.275616765022278
1959 0.274909019470215
1988 0.274399757385254
2016 0.274089336395264
2042 0.273982763290405
2066 0.274064302444458
2088 0.274315595626831
2109 0.274738311767578
2128 0.275298714637756
2146 0.276008129119873
2163 0.276859045028687
2179 0.277840375900269
2194 0.278937220573425
2208 0.280130743980408
2222 0.281502962112427
2235 0.282950043678284
2248 0.28457510471344
2261 0.286389350891113
2273 0.288240551948547
2285 0.290268301963806
2297 0.292478322982788
2309 0.294874548912048
2321 0.297459125518799
2333 0.300231695175171
2345 0.303189635276794
2357 0.30632746219635
2369 0.309636831283569
2382 0.31340217590332
2396 0.317646622657776
2411 0.322381258010864
2428 0.327932715415955
2450 0.33530855178833
2500 0.352138876914978
2518 0.357977867126465
2534 0.362983107566833
2548 0.367189407348633
2562 0.371212840080261
2575 0.374771118164062
2588 0.378148555755615
2601 0.381338119506836
2614 0.38433575630188
2627 0.387139678001404
2640 0.389750003814697
2653 0.392168402671814
2666 0.394398212432861
2679 0.396443605422974
2692 0.39831018447876
2705 0.400003552436829
2719 0.401641011238098
2733 0.403093576431274
2747 0.404369831085205
2762 0.405551552772522
2777 0.406551480293274
2793 0.407429456710815
2809 0.408124685287476
2826 0.408676385879517
2843 0.409048676490784
2861 0.40926206111908
2880 0.409301519393921
2900 0.409155249595642
2921 0.408815264701843
2943 0.408277750015259
2967 0.407508850097656
2994 0.406455636024475
3025 0.405058622360229
3065 0.403064727783203
3186 0.396914839744568
3226 0.395127534866333
3264 0.393607974052429
3302 0.392269968986511
3341 0.391080737113953
3381 0.390041708946228
3424 0.389107704162598
3470 0.388291120529175
3521 0.38757061958313
3578 0.386951327323914
3599 0.386764049530029
};
\addlegendentry{\tiny $\beta_{\mathcal{V}}$}
\addplot [semithick, blue!50.1960784313725!black]
table {%
0 0.323701977729797
10 0.322291970252991
21 0.320918560028076
34 0.319484829902649
48 0.318125367164612
64 0.316763639450073
81 0.315504431724548
99 0.314349174499512
119 0.31324827671051
141 0.312225937843323
165 0.311299800872803
191 0.310480952262878
220 0.309752225875854
253 0.309111595153809
290 0.3085777759552
334 0.308131694793701
386 0.307791352272034
451 0.307554006576538
535 0.307438969612122
632 0.307488560676575
713 0.307697296142578
773 0.308019638061523
820 0.308444738388062
858 0.308961272239685
890 0.30956768989563
918 0.310271143913269
943 0.311073899269104
966 0.311992526054382
987 0.313012838363647
1006 0.314111232757568
1024 0.315328240394592
1041 0.316655158996582
1057 0.318078994750977
1073 0.319687962532043
1088 0.321377873420715
1103 0.323254704475403
1117 0.325183391571045
1131 0.327288866043091
1145 0.329575181007385
1159 0.332043647766113
1173 0.334692716598511
1187 0.337518215179443
1202 0.340732336044312
1217 0.344127178192139
1233 0.347928047180176
1251 0.35239565372467
1271 0.357549786567688
1296 0.364184617996216
1373 0.384760141372681
1395 0.390386343002319
1415 0.395322918891907
1434 0.399832010269165
1452 0.403927445411682
1470 0.407842636108398
1488 0.411572694778442
1506 0.41511607170105
1524 0.418473243713379
1542 0.421647071838379
1560 0.424641728401184
1578 0.427462577819824
1597 0.43025815486908
1616 0.43287467956543
1636 0.435444593429565
1656 0.437834978103638
1677 0.440162301063538
1698 0.442313551902771
1720 0.444389581680298
1743 0.446377992630005
1767 0.448266506195068
1791 0.449976682662964
1816 0.451579093933105
1842 0.453060984611511
1868 0.454361319541931
1894 0.455483198165894
1920 0.456425309181213
1945 0.457155704498291
1970 0.45770275592804
1993 0.458029747009277
2015 0.458166599273682
2036 0.458116412162781
2055 0.457898736000061
2073 0.457521200180054
2090 0.456991314888
2106 0.456319212913513
2121 0.455517768859863
2136 0.454531073570251
2150 0.453424692153931
2163 0.452220320701599
2176 0.450829148292542
2188 0.44936466217041
2200 0.447713136672974
2211 0.446022748947144
2222 0.444152355194092
2233 0.442091345787048
2243 0.440043330192566
2253 0.437821626663208
2263 0.435419321060181
2273 0.432830333709717
2283 0.430049657821655
2293 0.427073240280151
2303 0.423898696899414
2313 0.420524954795837
2323 0.416953206062317
2333 0.413186073303223
2343 0.409228920936584
2353 0.405088901519775
2363 0.400775671005249
2374 0.395845174789429
2385 0.390738487243652
2397 0.384992599487305
2411 0.378099918365479
2428 0.369534850120544
2486 0.340136051177979
2500 0.333323001861572
2512 0.327654600143433
2524 0.322172403335571
2535 0.317330121994019
2546 0.312677621841431
2556 0.30862283706665
2566 0.304741501808167
2576 0.30103874206543
2586 0.297517776489258
2596 0.294180870056152
2606 0.291028499603271
2616 0.288059949874878
2626 0.285273790359497
2636 0.282667398452759
2646 0.280237674713135
2656 0.277980446815491
2666 0.275891423225403
2677 0.273782014846802
2688 0.271863698959351
2699 0.270129203796387
2710 0.268571376800537
2721 0.26718282699585
2733 0.265852689743042
2745 0.264706254005432
2757 0.263734340667725
2770 0.262868165969849
2783 0.26218569278717
2796 0.261676669120789
2810 0.261311173439026
2824 0.261123776435852
2839 0.261107921600342
2854 0.261270880699158
2870 0.261627793312073
2887 0.262198209762573
2904 0.262949228286743
2922 0.263922929763794
2941 0.26512885093689
2962 0.266647219657898
2985 0.268497705459595
3011 0.270774602890015
3044 0.273856401443481
3157 0.284574508666992
3188 0.287244319915771
3217 0.28956139087677
3245 0.291618347167969
3273 0.293492913246155
3301 0.295186638832092
3330 0.296756863594055
3360 0.298195362091064
3391 0.299498319625854
3424 0.300699591636658
3459 0.301787376403809
3496 0.302755355834961
3536 0.303621768951416
3580 0.304394006729126
3599 0.304677724838257
};
\addlegendentry{\tiny $\beta_{\mathcal{B}}$}
\end{axis}

\end{tikzpicture}
\end{tabular}
\caption{Time-dependent model variables for the no control scenario without abandonment for (a) private vehicle accumulation, (b) solo trip ride-hailing vehicles, (c) pool trip ride-hailing accumulation in~$\mc{V}$, (d) pool trip ride-hailing accumulation in~$\mc{B}$, (e) speed in the vehicle network~$\mc{V}$, (f) vehicle speed in the bus network~$\mc{B}$, (g) bus speed in the bus network~$\mc{B}$, and (h) fraction of pool trip in $\mc{V}$ and $\mc{B}$ respectively.}
\label{fig:results_no_control_no_ab}
\end{figure}
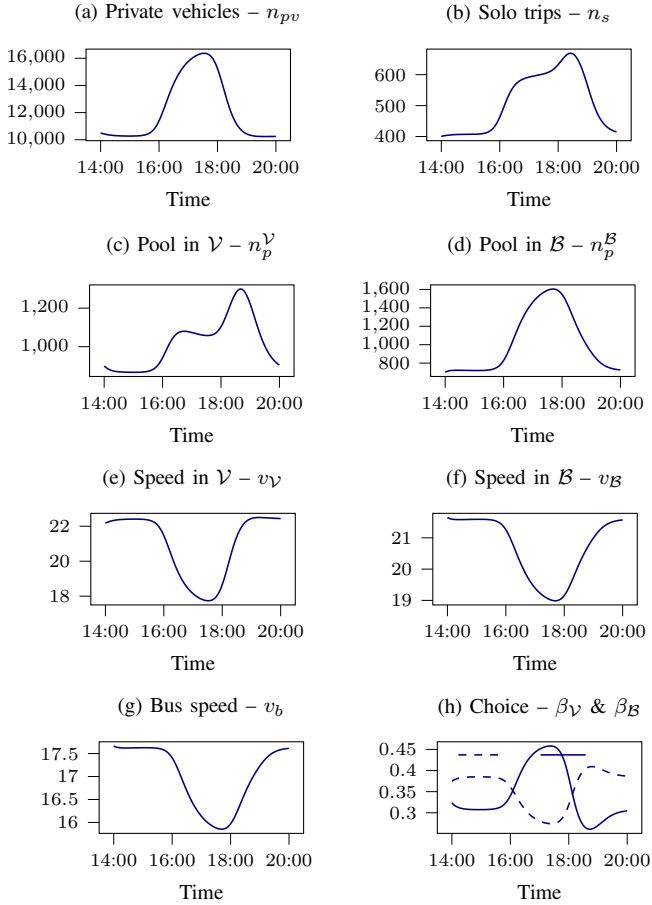

The changes of the state variables for the network under consideration without any regulatory interventions are shown in Figure~\ref{fig:results_no_control_no_ab}. Under this scenario, which we refer to as the no control scenario, the choice of users is solely dictated by the platform-set fares and the subnetworks' travel time. Note that we display here the simulation results with no abandonment, i.e., when $w_{\text{max}}$ is infinitely large or equivalently $A(k)=0$ for all $k \in \mc K$. During peak hours, the accumulation of private vehicles $n_{pv}$ spiraled up in Figure~\subfigref{fig:results_no_control_no_ab}{a}. With respect to the different states of the ride-hailing fleet size, it can be observed from Figures~\subfigref{fig:results_no_control_no_ab}{b}-\subfigref{fig:results_no_control_no_ab}{d}, that the number of solo trip vehicles $n_s$ and pool trip in $\mc{B}$ vehicles $n_{p}^{\mc{B}}$ increased during peak hours whereas the number of pool trip in $\mc{V}$ vehicles $n_{p}^{\mc{V}}$ remains almost constant after some initial transient behavior. The deterioration of the condition in the vehicle network, reflected in the reduction in speed $v_{\mc{V}}$ in Figure~\subfigref{fig:results_no_control_no_ab}{e},  prompts the user to pool in the bus network. This is confirmed by looking at the increase in the share of pool users opting to travel in the bus network $\beta_{\mc{B}}$ and the decrease in the share of pool users opting to travel in the bus network $\beta_{\mc{V}}$ as seen in \subfigref{fig:results_no_control_no_ab}{h}. The total delays for all users in the network for this specific scenario, i.e., when $\beta_{\mc{V}}\in[0,1]$ and $\beta_{\mc{B}}\in[0,1]$ are determined through~\eqref{eqn:beta_v} and~\eqref{eqn:beta_b} respectively, and are displayed in Table~\ref{tab:results_no_ab}, where, for $\beta_{\mc{V}}\in[0,1]$ and $\beta_{\mc{B}}\in[0,1]$, the sum of PHT and WT is equal to $188954$ pax.hr.
\begin{table}
    \centering
    \caption{Macro-simulation results without abandonment}
    \begin{tabular}{cccc}
    \hline
         \multirow{2}{*}{\textbf{Scenario}}& \textbf{PHT+WT} & \textbf{PHT}& \textbf{WT}  \\
          & \textbf{[pax.hr]}& \textbf{[pax.hr]} & \textbf{[pax.hr]} \\ \hline
        $\beta_{\mc{V}}=0$ \& $\beta_{\mc{B}}= 0$& $246980$ & $193913$& $53067$\\
         $\beta_{\mc{V}}\in[0,1]$ \& $\beta_{\mc{B}}= 0$& $239819$ &$198116$ & $41703$\\
         $\beta_{\mc{V}}=0$ \& $\beta_{\mc{B}}\in[0,1] $ & $190872$ &$176442$&$14430$\\
         $\beta_{\mc{V}}=0$ \& $\beta_{\mc{B}} = 1$ & $223210$ &$188097$&$35113$\\
         $\beta_{\mc{V}}\in[0,1]$ \& $\beta_{\mc{B}} \in[0,1]$ & $188954$ &$177389$&$11565$ \\
         PI control -- $\phi_{\mc{B}}$ &$191383$ &$178709$&$12674$\\
        MPC -- $\phi_{\mc{B}}$& $188316$ &$176721$&$11595$\\
        MPC ($\underbar{v}_b$) -- $\phi_{\mc{B}}$ & $191464$& $178515$&$12949$\\
         MPC -- $\phi_{\mc{V}}$ and $\phi_{\mc{B}}$& $187030$ &$176348$ & $10682$\\
        MPC ($\underbar{v}_b$) -- $\phi_{\mc{V}}$ and $\phi_{\mc{B}}$ & $189725$ & $178352$& $11373$ \\
         \hline
    \end{tabular}
    \label{tab:results_no_ab}
\end{table}

\begin{table}
    \centering
    \caption{Multi-modal Delays Without abandonment}
    \begin{tabular}{ccc}
    \hline
         \multirow{2}{*}{\textbf{Scenario}}& $\text{\textbf{PHT}}_\mathbf{pv}$ & $\text{\textbf{PHT}}_\mathbf{rs}$\\
          & \textbf{[pax.hr]} &  \textbf{[pax.hr]} \\ \hline
        $\beta_{\mc{V}}=0$ \& $\beta_{\mc{B}}= 0$& $114856$ & $18518$\\
         $\beta_{\mc{V}}\in[0,1]$ \& $\beta_{\mc{B}}= 0$& $114856$ & $22721$ \\
         $\beta_{\mc{V}}=0$ \& $\beta_{\mc{B}}\in[0,1] $ & $84623$ & $20770$\\
         $\beta_{\mc{V}}=0$ \& $\beta_{\mc{B}} = 1$ & $75462$ & $26713$\\
         $\beta_{\mc{V}}\in[0,1]$ \& $\beta_{\mc{B}} \in[0,1]$ & $87991$ & $21572$\\
         PI control -- $\phi_{\mc{B}}$ &$89527$  & $21798$\\
        MPC -- $\phi_{\mc{B}}$& $87355$ & $21564$ \\
        MPC ($\underbar{v}_b$) -- $\phi_{\mc{B}}$ &$91091$ & $21631$\\
         MPC -- $\phi_{\mc{V}}$ and $\phi_{\mc{B}}$& $87669$ &$21186$ \\
        MPC ($\underbar{v}_b$) -- $\phi_{\mc{V}}$ and $\phi_{\mc{B}}$ & $91308$ & $21459$\\
         \hline
    \end{tabular}
    \label{tab:results_multi_delay}
\end{table}

\begin{table}
    \centering
    \caption{Bus delays}
    \begin{tabular}{ccc}
    \hline
         \multirow{2}{*}{\textbf{Scenario}}& $\text{\textbf{PHT}}_\mathbf{b}$& $\mathbf{n_b\boldsymbol{\int} \boldsymbol{\max}(\bar{v}_b - v_b, 0)} dt$ \\
          & \textbf{[pax.hr]} &  \textbf{[km]}\\ \hline
        $\beta_{\mc{V}}=0$ \& $\beta_{\mc{B}}= 0$& $60539$ & $0$\\
         $\beta_{\mc{V}}\in[0,1]$ \& $\beta_{\mc{B}}= 0$& $60539$ & $0$ \\
         $\beta_{\mc{V}}=0$ \& $\beta_{\mc{B}}\in[0,1] $ & $71049$ & $2634$\\
         $\beta_{\mc{V}}=0$ \& $\beta_{\mc{B}} = 1$ & $85922$ & $11869$\\
         $\beta_{\mc{V}}\in[0,1]$ \& $\beta_{\mc{B}} \in[0,1]$ & $67825$ & $958$\\
         PI control -- $\phi_{\mc{B}}$ &$67384$  & $90$\\
        MPC -- $\phi_{\mc{B}}$& $67802$ & $1430$ \\
        MPC ($\underbar{v}_b$) -- $\phi_{\mc{B}}$ &$65792$ & $0$\\
         MPC -- $\phi_{\mc{V}}$ and $\phi_{\mc{B}}$& $67493$ &$1354$ \\
        MPC ($\underbar{v}_b$) -- $\phi_{\mc{V}}$ and $\phi_{\mc{B}}$ & $65584$ & $0$\\
         \hline
    \end{tabular}
    \label{tab:results_bus_delay}
\end{table}

To put the result above into context, we compare it with settings where the choices for the pooling users are limited. Table~\ref{tab:results_no_ab} shows the user delays for such scenarios. The worst performing scenario is when no pooling is involved, i.e., when both $\beta_{\mc{V}}$ and $\beta_{\mc{B}}$ are equal to zero. This is because for constant fleet sizes, solo travel results in longer queues and longer waiting times, especially when all ride-hailing vehicles are occupied and few vehicles are available for pick-up. When pooling is only allowed in the vehicle network, i.e., $\beta_{\mc{B}}=0$ and $\beta_{\mc V}\in [0,1]$, the total delays are much greater than the scenario where pooling is only allowed in the bus network with $\beta_{\mc{V}}=0$. This is due to users opting for pooling in the bus network, causing ride-hailing vehicles to travel at a larger speed, and making them available soon after to perform a new trip. We note here that we also provide the simulation results for when $\beta_{\mc{B}}=1$ which implies $\beta_{\mc{V}}=0$, meaning that all ride-hailing users are pooling in the bus network. The reason for that is to show that this extreme solution causes significant delays for bus users and long waiting times, and is therefore not attractive at the system level. Motivated by these observations, the need to regulate the share of each ride-hailing alternative becomes more substantiated. For this reason, we resort to different controllers to find a proper pricing scheme that minimizes the observed delays. 

\subsubsection{PI controller framework}

\begin{figure}
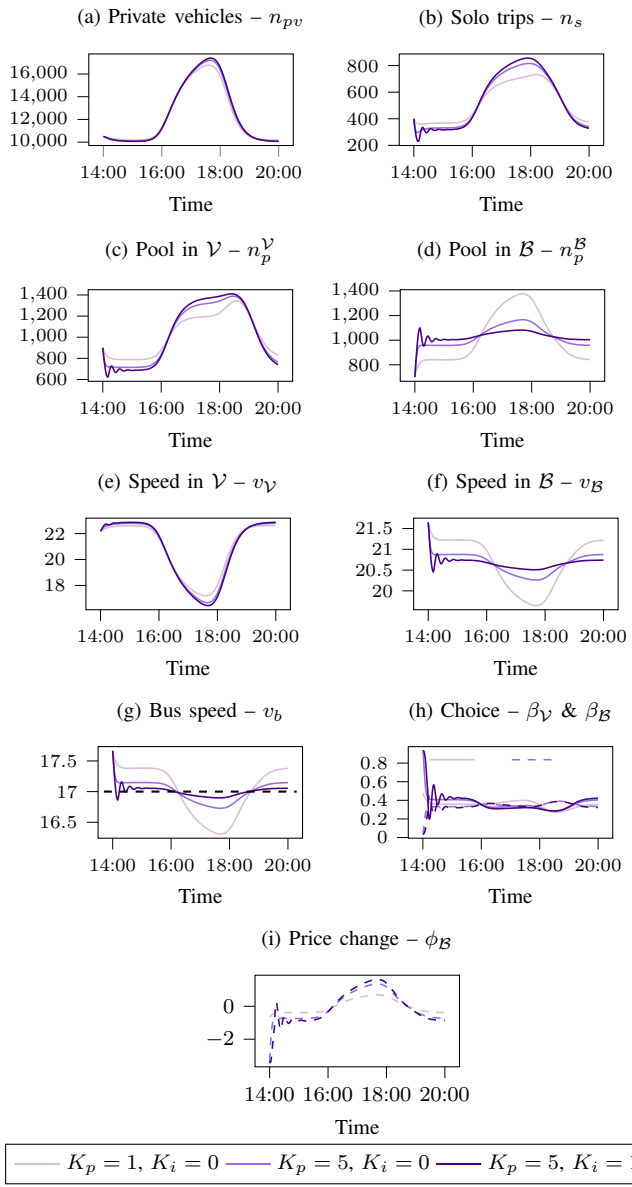

\centering 
\begin{tabular}{cc}
\hspace{-0.5cm}
\begin{tikzpicture}

\definecolor{color0}{rgb}{0.847058823529412,0.749019607843137,0.847058823529412}
\definecolor{color1}{rgb}{0.576470588235294,0.43921568627451,0.858823529411765}
\definecolor{color2}{rgb}{0.294117647058824,0,0.509803921568627}

\begin{axis}[
width=4.3cm,
height=2.8cm,
tick align=outside,
tick pos=left,
x grid style={white!69.0196078431373!black},
xlabel={\footnotesize Time},
xmin=13.5, xmax=20.5,
ymin=9702.05022850373, ymax=17766.3863713967,
ytick style={color=black},
scaled y ticks = false,
y tick label style={/pgf/number format/fixed, font = \scriptsize},
xlabel near ticks,
ylabel near ticks,
scaled x ticks = false,
x tick label style={/pgf/number format/fixed, font = \scriptsize},
x filter/.code={\pgfmathparse{#1/600+14}},
    xticklabel={ 
        \pgfmathsetmacro\hours{floor(\tick)}%
        \pgfmathsetmacro\minutes{(\tick-\hours)*0.6}%
        \pgfmathprintnumber{\hours}:\pgfmathprintnumber[fixed, fixed zerofill, skip 0.=true, dec sep={}]{\minutes}%
    },
title={\footnotesize (a) Private vehicles -- $n_{pv}$}
]
\addplot [semithick, color0]
table {%
0 10500
85 10396.66015625
115 10364.7919921875
143 10338.1689453125
171 10314.5830078125
199 10293.8984375
227 10275.896484375
257 10259.2998046875
288 10244.7392578125
321 10231.75390625
356 10220.4033203125
394 10210.4580078125
436 10201.8544921875
482 10194.80078125
532 10189.4736328125
584 10186.205078125
635 10185.21875
681 10186.478515625
722 10189.7939453125
757 10194.806640625
788 10201.4873046875
815 10209.546875
840 10219.390625
863 10230.9775390625
883 10243.4677734375
902 10257.8349609375
920 10274.1005859375
938 10293.3916015625
955 10314.8330078125
972 10339.87890625
989 10369.0458984375
1006 10402.8955078125
1023 10442.0302734375
1040 10487.083984375
1057 10538.708984375
1074 10597.552734375
1091 10664.2373046875
1109 10744.0166015625
1127 10833.7802734375
1146 10939.8046875
1166 11064.162109375
1187 11208.7177734375
1210 11382.794921875
1235 11589.072265625
1264 11847.1884765625
1301 12197.947265625
1378 12956.6943359375
1428 13435.787109375
1467 13789.0615234375
1503 14095.3076171875
1537 14365.7080078125
1570 14610.4462890625
1602 14831.46484375
1634 15037.09765625
1665 15222.4208984375
1696 15394.9208984375
1727 15555.4404296875
1758 15704.7900390625
1789 15843.71875
1820 15972.8916015625
1851 16092.87109375
1882 16204.09375
1912 16303.6640625
1941 16392.478515625
1969 16471.23046875
1995 16537.990234375
2019 16593.865234375
2041 16639.875
2062 16678.728515625
2081 16709.203125
2098 16732.31640625
2113 16749.134765625
2127 16761.51953125
2139 16769.37109375
2150 16774.14453125
2161 16776.427734375
2171 16776.19140625
2181 16773.6015625
2191 16768.501953125
2201 16760.73046875
2211 16750.1171875
2222 16734.94140625
2233 16715.861328125
2245 16690.296875
2257 16659.44921875
2270 16619.681640625
2283 16572.90234375
2297 16514.228515625
2312 16441.3046875
2327 16357.505859375
2343 16255.7060546875
2360 16133.1953125
2378 15987.2802734375
2398 15805.9560546875
2419 15594.8662109375
2443 15330.05859375
2470 15006.8359375
2505 14559.0673828125
2627 12972.572265625
2658 12611.248046875
2686 12309.70703125
2712 12052.5
2737 11826.4443359375
2761 11628.9443359375
2784 11457.1494140625
2807 11301.7587890625
2829 11167.6787109375
2851 11046.9619140625
2873 10938.7001953125
2894 10846.1220703125
2915 10763.2373046875
2936 10689.2529296875
2957 10623.40625
2978 10564.9697265625
2998 10515.572265625
3018 10471.7109375
3038 10432.8603515625
3059 10396.9267578125
3080 10365.455078125
3101 10337.9638671875
3122 10314.009765625
3144 10292.2705078125
3166 10273.5498046875
3189 10256.794921875
3213 10241.9716796875
3239 10228.5361328125
3266 10217.0224609375
3296 10206.6513671875
3329 10197.6396484375
3366 10189.90234375
3409 10183.3115234375
3459 10178.017578125
3520 10173.9208984375
3599 10171.0107421875
};
\addplot [semithick, color1]
table {%
0 10500
14 10481.173828125
32 10453.064453125
101 10341.341796875
125 10308.6611328125
148 10280.99609375
171 10256.6318359375
195 10234.365234375
220 10214.1943359375
246 10196.07421875
273 10179.9267578125
302 10165.18359375
333 10151.9580078125
367 10139.984375
404 10129.4560546875
444 10120.48828125
487 10113.1474609375
534 10107.400390625
584 10103.5712890625
633 10102.0302734375
679 10102.791015625
720 10105.7099609375
755 10110.3984375
786 10116.787109375
813 10124.578125
838 10134.1474609375
861 10145.4541015625
882 10158.3486328125
901 10172.5625
919 10188.671875
937 10207.794921875
954 10229.064453125
971 10253.9248046875
988 10282.892578125
1004 10314.41015625
1021 10352.9921875
1038 10397.447265625
1055 10448.4306640625
1072 10506.599609375
1089 10572.5888671875
1107 10651.630859375
1125 10740.68359375
1144 10846.029296875
1164 10969.8056640625
1185 11113.974609375
1208 11287.9931640625
1233 11494.7685546875
1261 11745.123046875
1296 12079.28515625
1423 13346.3623046875
1466 13751.5361328125
1504 14089.0302734375
1540 14389.2529296875
1575 14662.6474609375
1609 14911.1806640625
1643 15143.6845703125
1677 15361.123046875
1711 15564.5361328125
1745 15754.958984375
1780 15938.435546875
1815 16110.115234375
1850 16270.783203125
1885 16421.02734375
1919 16557.330078125
1952 16680.580078125
1983 16788.029296875
2012 16880.78515625
2039 16959.78125
2063 17023.359375
2085 17075.455078125
2104 17115.078125
2121 17145.791015625
2136 17168.751953125
2150 17186.3125
2162 17198.10546875
2173 17206.048828125
2183 17210.705078125
2193 17212.748046875
2202 17212.205078125
2211 17209.26171875
2220 17203.78125
2229 17195.615234375
2239 17183.208984375
2249 17167.091796875
2260 17144.822265625
2271 17117.51953125
2283 17081.662109375
2296 17035.271484375
2309 16980.62109375
2323 16912.09375
2338 16827.09375
2354 16722.818359375
2371 16596.35546875
2389 16444.8359375
2408 16265.6572265625
2429 16045.787109375
2452 15781.0615234375
2479 15443.251953125
2512 15000.9228515625
2622 13473.6318359375
2657 13023.615234375
2687 12665.478515625
2715 12357.279296875
2741 12094.8291015625
2766 11864.275390625
2790 11662.7578125
2813 11487.2626953125
2836 11328.234375
2859 11184.7548828125
2881 11061.1298828125
2903 10949.87890625
2925 10850.0615234375
2946 10764.59375
2967 10687.9111328125
2988 10619.2578125
3009 10557.9169921875
3030 10503.2099609375
3051 10454.5029296875
3072 10411.2080078125
3093 10372.7802734375
3114 10338.716796875
3136 10307.212890625
3158 10279.513671875
3181 10254.1533203125
3204 10232.0302734375
3228 10211.9677734375
3253 10193.9248046875
3279 10177.8330078125
3307 10163.1123046875
3337 10149.880859375
3370 10137.859375
3406 10127.2265625
3446 10117.853515625
3491 10109.716796875
3542 10102.8369140625
3599 10097.310546875
};
\addplot [semithick, color2]
table {%
0 10500
14 10481.1201171875
30 10455.64453125
49 10420.98046875
79 10360.771484375
109 10302.05859375
125 10275.2431640625
139 10255.87109375
153 10240.111328125
168 10226.416015625
189 10210.5615234375
249 10166.853515625
280 10144.6953125
300 10133.0166015625
320 10123.8818359375
344 10115.587890625
394 10101.6318359375
455 10085.7919921875
487 10080.1484375
526 10075.802734375
599 10070.2861328125
643 10068.9912109375
684 10070.126953125
725 10073.4765625
761 10078.58203125
791 10085.0712890625
817 10092.9296875
841 10102.5009765625
863 10113.65234375
883 10126.1611328125
902 10140.54296875
920 10156.826171875
938 10176.1484375
955 10197.64453125
972 10222.78125
989 10252.078125
1006 10286.1025390625
1023 10325.4560546875
1040 10370.7783203125
1057 10422.724609375
1074 10481.9560546875
1091 10549.107421875
1109 10629.4892578125
1127 10719.9921875
1146 10826.9736328125
1166 10952.572265625
1187 11098.732421875
1210 11274.9853515625
1235 11484.2041015625
1264 11746.6171875
1299 12084.6328125
1354 12640.4775390625
1426 13363.5302734375
1469 13772.78125
1508 14122.984375
1544 14426.9208984375
1579 14704.412109375
1614 14964.6337890625
1649 15208.451171875
1684 15436.92578125
1719 15651.1865234375
1755 15857.9111328125
1791 16051.888671875
1827 16234.0927734375
1863 16405.302734375
1899 16566.06640625
1934 16712.591796875
1968 16845.615234375
2000 16962.0078125
2029 17059.46484375
2055 17139.591796875
2079 17206.65234375
2100 17259.125
2119 17300.91796875
2136 17333.1328125
2151 17357.005859375
2164 17373.873046875
2176 17385.986328125
2187 17393.9296875
2197 17398.3203125
2206 17399.810546875
2215 17398.8203125
2224 17395.19921875
2232 17389.650390625
2241 17380.642578125
2250 17368.5546875
2260 17351.310546875
2271 17327.451171875
2282 17298.18359375
2294 17259.748046875
2307 17210.046875
2320 17151.5625
2334 17078.3359375
2349 16987.689453125
2365 16876.767578125
2382 16742.646484375
2400 16582.509765625
2420 16383.408203125
2441 16151.8798828125
2465 15862.1376953125
2493 15496.1796875
2529 14994.888671875
2643 13385.6650390625
2676 12962.9541015625
2705 12618.8125
2732 12323.5927734375
2758 12063.0771484375
2783 11834.5947265625
2807 11635.11328125
2830 11461.5146484375
2853 11304.2626953125
2876 11162.3935546875
2898 11040.1259765625
2920 10930.0380859375
2942 10831.1845703125
2964 10742.634765625
2985 10666.89453125
3006 10598.9765625
3027 10538.171875
3048 10483.8173828125
3069 10435.2939453125
3090 10392.0263671875
3112 10351.76171875
3134 10316.130859375
3156 10284.6279296875
3178 10256.796875
3201 10231.1806640625
3225 10207.7958984375
3250 10186.6181640625
3276 10167.58984375
3303 10150.6259765625
3331 10135.615234375
3361 10122.017578125
3394 10109.5556640625
3430 10098.431640625
3469 10088.744140625
3513 10080.1630859375
3563 10072.7509765625
3599 10068.611328125
};
\end{axis}

\end{tikzpicture} &
\hspace{-0.5cm}\input{n_s_PI_no_ab_initial_condition_vb=17.tikz} \\
\hspace{-0.4cm}\input{n_p_v_PI_no_ab_initial_condition_vb=17.tikz} &
\hspace{-0.7cm}
\begin{tikzpicture}

\definecolor{color0}{rgb}{0.847058823529412,0.749019607843137,0.847058823529412}
\definecolor{color1}{rgb}{0.576470588235294,0.43921568627451,0.858823529411765}
\definecolor{color2}{rgb}{0.294117647058824,0,0.509803921568627}

\begin{axis}[
width=4.3cm,
height=2.8cm,
tick align=outside,
tick pos=left,
x grid style={white!69.0196078431373!black},
xlabel={\footnotesize Time},
xmin=13.5, xmax=20.5,
xtick style={color=black},
y grid style={white!69.0196078431373!black},
ymin=666.187662110912, ymax=1410.05909567085,
ytick style={color=black},
scaled y ticks = false,
y tick label style={/pgf/number format/fixed, font = \scriptsize},
xlabel near ticks,
ylabel near ticks,
scaled x ticks = false,
x tick label style={/pgf/number format/fixed, font = \scriptsize},
x filter/.code={\pgfmathparse{#1/600+14}},
    xticklabel={ 
        \pgfmathsetmacro\hours{floor(\tick)}%
        \pgfmathsetmacro\minutes{(\tick-\hours)*0.6}%
        \pgfmathprintnumber{\hours}:\pgfmathprintnumber[fixed, fixed zerofill, skip 0.=true, dec sep={}]{\minutes}%
    },
title={\footnotesize (d) Pool in $\mc{B}$ -- $n_p^{\mc{B}}$}
]
\addplot [semithick, color0]
table {%
0 700
7 713.990356445312
14 726.738647460938
21 738.3251953125
28 748.834838867188
35 758.352661132812
43 768.121826171875
51 776.821166992188
59 784.559692382812
67 791.437561035156
75 797.545837402344
84 803.599792480469
93 808.888305664062
102 813.504028320312
112 817.942626953125
123 822.098815917969
134 825.602905273438
146 828.794555664062
159 831.6279296875
174 834.23291015625
190 836.378479003906
208 838.178039550781
229 839.651489257812
254 840.767639160156
285 841.502014160156
326 841.812316894531
390 841.599914550781
631 840.36328125
719 840.670227050781
784 841.487609863281
834 842.711547851562
874 844.27783203125
908 846.202087402344
938 848.513244628906
964 851.116516113281
988 854.134399414062
1010 857.524963378906
1030 861.214721679688
1049 865.333679199219
1067 869.854919433594
1084 874.734497070312
1101 880.256042480469
1117 886.078369140625
1133 892.538024902344
1149 899.656372070312
1165 907.444580078125
1181 915.901794433594
1198 925.604431152344
1215 936.012268066406
1233 947.744140625
1253 961.542541503906
1275 977.495056152344
1301 997.129272460938
1396 1070.89904785156
1427 1094.11315917969
1454 1113.56689453125
1480 1131.52355957031
1505 1148.02392578125
1529 1163.14282226562
1553 1177.55871582031
1578 1191.84436035156
1603 1205.41040039062
1628 1218.28845214844
1654 1230.98803710938
1681 1243.46838378906
1708 1255.26916503906
1736 1266.83166503906
1765 1278.12548828125
1795 1289.12243652344
1825 1299.46166992188
1856 1309.49169921875
1888 1319.18103027344
1920 1328.22045898438
1952 1336.62438964844
1984 1344.39331054688
2015 1351.29895019531
2045 1357.36865234375
2074 1362.61474609375
2101 1366.88952636719
2127 1370.37841796875
2151 1372.97521972656
2173 1374.75402832031
2194 1375.83984375
2214 1376.24328613281
2233 1375.98425292969
2251 1375.09411621094
2268 1373.61694335938
2284 1371.60961914062
2300 1368.95593261719
2315 1365.84045410156
2330 1362.08276367188
2345 1357.6533203125
2359 1352.89208984375
2373 1347.51232910156
2387 1341.5078125
2402 1334.38342285156
2417 1326.556640625
2432 1318.05187988281
2448 1308.27258300781
2465 1297.14404296875
2483 1284.62194824219
2503 1269.95239257812
2527 1251.53063964844
2637 1163.71948242188
2665 1142.31213378906
2691 1123.21899414062
2715 1106.34338378906
2739 1090.21374511719
2763 1074.82250976562
2788 1059.5380859375
2814 1044.38842773438
2841 1029.37426757812
2870 1013.95440673828
2902 997.661071777344
2936 981.061950683594
2970 965.138061523438
3003 950.34912109375
3034 937.133850097656
3062 925.855102539062
3088 916.022521972656
3113 907.213256835938
3137 899.394470214844
3161 892.226318359375
3185 885.716613769531
3209 879.857055664062
3233 874.624877929688
3258 869.804931640625
3284 865.423950195312
3311 861.492858886719
3340 857.894470214844
3371 854.669311523438
3405 851.758728027344
3443 849.146484375
3486 846.842346191406
3535 844.865539550781
3599 843.038696289062
};
\addplot [semithick, color1]
table {%
0 700
11 761.671813964844
18 797.721069335938
24 825.667114257812
29 846.501220703125
34 864.948974609375
39 881.001342773438
44 894.761962890625
49 906.414611816406
54 916.187316894531
59 924.321960449219
64 931.053527832031
69 936.598266601562
74 941.147827148438
79 944.868347167969
84 947.901306152344
90 950.800109863281
96 953.049743652344
103 955.034057617188
111 956.66162109375
121 958.007507324219
133 958.948486328125
149 959.522766113281
173 959.675842285156
353 958.062255859375
499 957.451965332031
658 957.390808105469
766 957.916137695312
837 958.842834472656
890 960.119873046875
933 961.749572753906
970 963.762939453125
1002 966.106506347656
1031 968.828857421875
1058 971.965393066406
1084 975.603271484375
1109 979.717529296875
1134 984.455688476562
1160 990.033996582031
1187 996.475708007812
1217 1004.28881835938
1254 1014.60803222656
1396 1054.91381835938
1435 1064.9482421875
1473 1074.09643554688
1512 1082.84765625
1552 1091.19104003906
1594 1099.31884765625
1638 1107.20532226562
1685 1114.99206542969
1734 1122.48327636719
1787 1129.94128417969
1842 1137.04187011719
1898 1143.65466308594
1955 1149.76635742188
2010 1155.0419921875
2060 1159.22570800781
2104 1162.30541992188
2143 1164.42626953125
2177 1165.67346191406
2208 1166.19409179688
2236 1166.04455566406
2262 1165.27905273438
2286 1163.95495605469
2309 1162.06311035156
2331 1159.63269042969
2353 1156.56018066406
2374 1153.01196289062
2396 1148.65344238281
2418 1143.66552734375
2442 1137.56750488281
2468 1130.29943847656
2499 1120.94787597656
2612 1085.03308105469
2649 1073.98059082031
2683 1064.47387695312
2716 1055.88696289062
2750 1047.68139648438
2786 1039.63195800781
2826 1031.32800292969
2874 1022.02563476562
2935 1010.87878417969
3003 999.081420898438
3055 990.64697265625
3097 984.431945800781
3135 979.423889160156
3171 975.298217773438
3207 971.7958984375
3244 968.817504882812
3284 966.230651855469
3328 964.02197265625
3378 962.148681640625
3437 960.577941894531
3509 959.299743652344
3599 958.31787109375
};
\addplot [semithick, color2]
table {%
0 700
28 859.613708496094
38 912.490417480469
46 951.841186523438
53 983.521423339844
59 1008.21844482422
65 1030.29577636719
70 1046.47680664062
75 1060.50427246094
79 1070.12854003906
83 1078.32885742188
87 1085.13391113281
91 1090.599609375
95 1094.80432128906
98 1097.18737792969
101 1098.95910644531
103 1099.265625
105 1098.63623046875
107 1097.15417480469
109 1094.91186523438
112 1090.32849121094
115 1084.54870605469
119 1075.4423828125
125 1059.82629394531
149 994.655090332031
154 983.399353027344
159 973.641906738281
163 967.063354492188
167 961.669067382812
171 957.514831542969
174 955.228759765625
177 953.652893066406
180 952.775695800781
183 952.57666015625
186 953.027282714844
189 954.092407226562
192 955.7314453125
196 958.732727050781
200 962.563354492188
205 968.341491699219
211 976.403015136719
220 989.794982910156
232 1007.52062988281
238 1015.34545898438
243 1020.98822021484
248 1025.66662597656
252 1028.63903808594
256 1030.8876953125
260 1032.39721679688
264 1033.17175292969
268 1033.23413085938
272 1032.62377929688
276 1031.39440917969
281 1029.08740234375
286 1026.06201171875
292 1021.7080078125
301 1014.29040527344
316 1001.82470703125
323 996.834228515625
329 993.28662109375
334 990.941467285156
339 989.201599121094
344 988.088562011719
349 987.6005859375
354 987.713623046875
359 988.383850097656
365 989.836853027344
372 992.259338378906
381 996.154357910156
405 1007.03314208984
413 1009.74273681641
420 1011.45245361328
427 1012.48101806641
434 1012.82104492188
441 1012.51287841797
449 1011.47229003906
458 1009.62921142578
472 1006.01013183594
488 1001.99322509766
498 1000.10961914062
507 999.027893066406
516 998.5810546875
525 998.743713378906
535 999.528930664062
548 1001.19348144531
578 1005.29467773438
590 1006.21661376953
602 1006.51361083984
615 1006.18048095703
632 1005.03918457031
666 1002.59136962891
682 1002.17877197266
699 1002.40209960938
723 1003.44891357422
754 1004.72320556641
776 1004.95587158203
805 1004.55749511719
842 1004.15692138672
869 1004.53485107422
920 1006.07922363281
1026 1009.45758056641
1061 1011.63623046875
1108 1015.22192382812
1156 1019.47552490234
1203 1024.29895019531
1334 1038.1298828125
1402 1044.53686523438
1463 1049.64440917969
1534 1054.94128417969
1606 1059.70837402344
1690 1064.61999511719
1780 1069.25439453125
1878 1073.67431640625
1974 1077.39807128906
2057 1080.01928710938
2122 1081.47338867188
2174 1082.03955078125
2218 1081.91247558594
2257 1081.17993164062
2292 1079.91467285156
2326 1078.06115722656
2359 1075.64636230469
2394 1072.45397949219
2434 1068.14611816406
2490 1061.40344238281
2586 1049.82727050781
2639 1044.138671875
2690 1039.29077148438
2745 1034.69116210938
2812 1029.74719238281
2913 1023.00250244141
3036 1015.33782958984
3101 1011.89575195312
3157 1009.54547119141
3215 1007.74114990234
3281 1006.32586669922
3363 1005.21960449219
3473 1004.39904785156
3599 1003.93341064453
};
\end{axis}
\end{tikzpicture} \\
\hspace{-0.1cm}\input{v_v_PI_no_ab_initial_condition_vb=17.tikz} &
\hspace{-0.2cm}\input{v_b_PI_no_ab_initial_condition_vb=17.tikz} \\
\input{bus_speed_PI_no_ab_initial_condition_vb=17.tikz} &
\hspace{-0.2cm}\input{betas_PI_no_ab_initial_condition_vb=17.tikz}
\end{tabular}
\input{phi_b_PI_no_ab_initial_condition_vb=17.tikz}
\begin{tikzpicture} 
    \begin{axis}[%
    hide axis,
    width=5cm,
    height=2cm,
    xmin=0,
    xmax=0.1,
    ymin=0,
    ymax=0.1,
    legend columns=3, 
    legend style={/tikz/every even column/.append style={column sep=0.0cm}}]
    ]
    \addlegendimage{thick, color0}
    \addlegendentry{\footnotesize $K_p = 1$, $K_i = 0$};
    \addlegendimage{thick, color1}
    \addlegendentry{\footnotesize $K_p = 5$, $K_i = 0$};
    \addlegendimage{thick, color2}
    \addlegendentry{\footnotesize $K_p = 5$, $K_i = 11$};
    \end{axis}
\end{tikzpicture}
\caption{Time-dependent model variables for the PI framework without abandonment for (a) private vehicle accumulation, (b) solo trip ride-hailing vehicles, (c) pool trip ride-hailing accumulation in~$\mc{V}$, (d) pool trip ride-hailing accumulation in~$\mc{B}$, (e) speed in the vehicle network~$\mc{V}$, (f) vehicle speed in the bus network~$\mc{B}$, (g) bus speed in the bus network~$\mc{B}$, (h) fraction of pool trip in $\mc{V}$ and $\mc{B}$ respectively, and (i) regulatory control fare for pooling in~$\mc{B}$.}
\label{fig:results_PI_no_ab}
\end{figure}

To guarantee that our allocation strategy does not worsen the situation for mainly bus users, we implement the PI control framework in our simulation and report the different state variables in Figure~\ref{fig:results_PI_no_ab}. The choice of the set point for the desired speed in the bus network, however, remains complex because bus users should ideally travel at the highest possible speed, and this speed is defined by the bus operator or the traffic regulators. In the results provided, we set $\bar{v}_b$ to $17$ km/hr while the default bus speed in the absence of cars is $v_b(0, n_b) = 19$ km/hr. This choice of set point ensures that the permissible decrease in bus speeds due to the bus lane usage remains within acceptable ranges.

\begin{figure}
    \centering
\begin{tikzpicture} %
        \begin{axis}[smooth , view={35}{20}, 
        grid, 
        width={8cm},
        height={5cm},
        xlabel={$\bar{v}_b$}, 
        ylabel={Multiplier}, 
        zlabel={PHT$+$WT},
        title={PHT$+$WT},
        scaled ticks = false,
      xticklabel={\pgfmathprintnumber[fixed, fixed zerofill, precision=0]{\tick}}, 
      zticklabel={\pgfmathprintnumber[fixed, fixed zerofill, precision=0]{\tick}}, 
       xtick ={12, 15, 18},
       ytick ={1, 1.5, 2},
       colormap/YlGnBu 
       ]%
        \addplot3[surf, shader=interp,samples=3, patch type=rectangle]  file {PHT_WT.txt};
        \end{axis} %
\end{tikzpicture}
\caption{Objective function values for different set points and different bus demand profiles under the PI control framework.}
\label{fig:of_PI}
\end{figure}
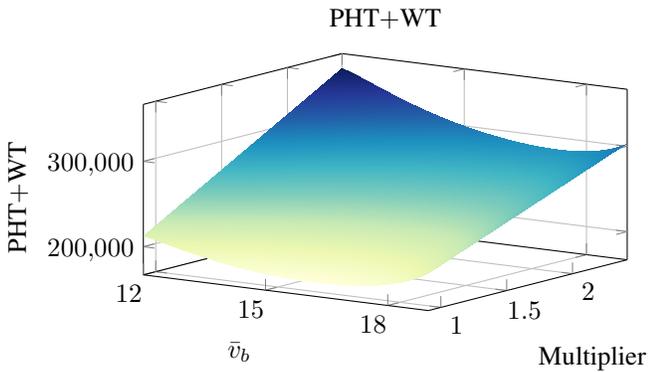

\begin{figure}
    \centering
\begin{tikzpicture} %
        \begin{axis}[smooth , view={55}{35}, 
        grid, 
        width={8cm},
        height={5cm},
        xlabel={$Q_{pv}$}, 
        ylabel={$Q_{b}$}, 
        zlabel={$\bar{v}_b$},
        title={Optimal set point},
        scaled ticks = false,
      xticklabel={\pgfmathprintnumber[fixed, fixed zerofill, precision=0]{\tick}}, 
      zticklabel={\pgfmathprintnumber[fixed, fixed zerofill, precision=0]{\tick}}, 
       ytick ={20000, 40000, 60000},
       xtick ={60000, 70000, 80000},
       ztick={12, 15, 18},
       colormap/YlGnBu 
       ]%
        \addplot3[surf, shader=interp,samples=3, patch type=rectangle]  file {set_points.txt};
        \end{axis} %
\end{tikzpicture}
\caption{Optimal PI bus speed set points for different private vehicles and bus demands.}
\label{fig:set_points}
\end{figure}
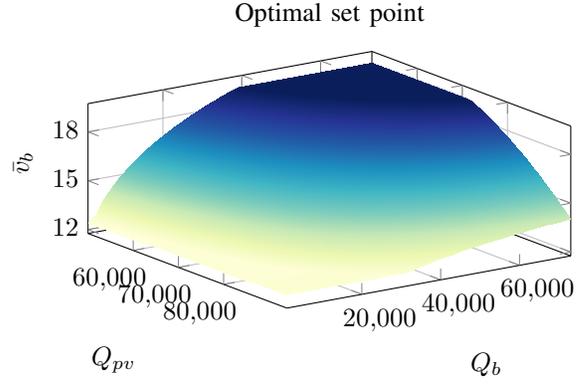

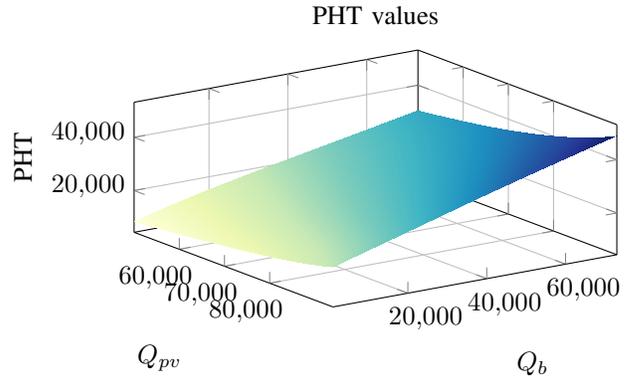
\begin{figure}
    \centering
\begin{tikzpicture} %
        \begin{axis}[smooth , view={55}{35}, 
        grid, 
        width={8cm},
        height={5cm},
        xlabel={$Q_{pv}$}, 
        ylabel={$Q_{b}$}, 
        zlabel={PHT},
        title={PHT values},
        scaled ticks = false,
      xticklabel={\pgfmathprintnumber[fixed, fixed zerofill, precision=0]{\tick}}, 
      zticklabel={\pgfmathprintnumber[fixed, fixed zerofill, precision=0]{\tick}}, 
       ytick ={20000, 40000, 60000},
       xtick ={60000, 70000, 80000},
       colormap/YlGnBu 
       ]%
        \addplot3[surf, shader=interp,samples=3, patch type=rectangle]  file {min_PHT.txt};
        \end{axis} %
\end{tikzpicture} 
\caption{PHT values for optimal set points.}
\label{fig:min_PHT}
\end{figure}

 The plots in Figure~\ref{fig:results_PI_no_ab} show the system dynamics for different proportional and integral gain values $K_p$ and $K_i$ respectively, all with $N_e = 100$ time steps. The variations of $n_{pv}$, $n_s$, $n_p^{\mc{B}}$, and $n_p^{\mc{V}}$ in Figures~\subfigref{fig:results_PI_no_ab}{a}-\subfigref{fig:results_PI_no_ab}{d} are almost similar to what is observed in Figures~\subfigref{fig:results_no_control_no_ab}{a}-\subfigref{fig:results_no_control_no_ab}{d}, except for the lower number of solo trips and higher number of pool trips in $\mc{B}$, especially during off-peak period. This is because the bus network is significantly underutilized during off-peak hours, and its capacity allows for some ride-hailing users to pool their trips in $\mc{B}$. This justifies the exorbitantly high discount $\phi_{\mc{B}}$ granted to users at the beginning of the simulation in Figure~\subfigref{fig:results_PI_no_ab}{i} to encourage them to pool their trips in $\mc{B}$. Note here that in the PI framework, we focus on controlling~$\phi_{\mc{B}}$ because it has a direct influence on the bus speed $v_b$. In other words, we set $\phi_{\mc{V}}$ to $0$ in our numerical experiment.

Moreover, we observe that picking reasonable values for $K_p$ and $K_i$ yields realistic pricing scenarios, all while achieving the desired objective of bridging the gap between the actual and target bus speeds as can be seen in Figure~\subfigref{fig:results_PI_no_ab}{i} compared to scenarios with no integral term. Due to the time-dependent nature of the demand in our simulations, the PI controller fails to achieve lower objective function values compared to the scenario with no control where the total delays are equal to $191383$ and $188954$ pax.hr, respectively. First, the dynamic nature of the problem implies that the choice of set points is not straightforward.

The approach we have adopted so far accounts for bus user delays without consideration of the overall network performance. Therefore, the previous choice of set point does not guarantee a convenient solution for all network users. In Figure~\ref{fig:of_PI}, we display the value of PHT and WT for various PI set points and different on-peak period bus demand multipliers. The displayed results reflect that a lower $\bar{v}_{b}$ is acceptable if the objective is to minimize multi-modal delays. A more practical way to determine this network operational point is to extend the static model developed in \cite{utilization_2023_fayed}, and compute the value of bus speed at optimality for the pair of time-invariant private vehicles and bus demands. The results of this approach are shown in Figure~\ref{fig:set_points}. From the figure, we can see that for the current setting, the set-point is demand-dependent. The corresponding minimum PHT for the different demand combinations are shown in Figure~\ref{fig:min_PHT}. Since bus lanes usually occupy a small fraction of the network infrastructure, increasing bus demand has the most significant impact on the PHT values. 

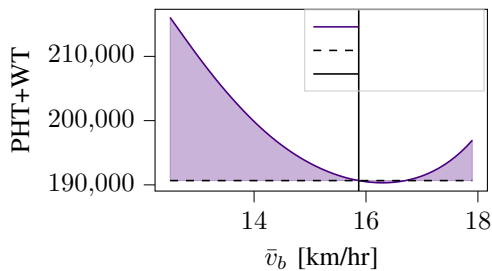
\begin{figure}
    \centering
\begin{tikzpicture}

\definecolor{color0}{rgb}{0.294117647058824,0,0.509803921568627}

\begin{axis}[
legend cell align={left},
legend style={at={(0.72,1)},anchor=north, fill opacity=0, draw opacity=1, text opacity=1, draw=white!80!black, fill = none},
width=6cm,
height=4cm,
tick align=outside,
tick pos=left,
x grid style={white!69.0196078431373!black},
xlabel={\(\displaystyle \bar{v}_b\) [km/hr]},
xmin=12.23, xmax=18.17,
xtick style={color=black},
y grid style={white!69.0196078431373!black},
ylabel={PHT+WT},
ymin=189034.95, ymax=217328.05,
ytick style={color=black},
xlabel near ticks,
ylabel near ticks,
scaled x ticks = false,
scaled y ticks = false,
x tick label style={/pgf/number format/fixed},
y tick label style={/pgf/number format/fixed},
]
\path [draw=color0, fill=color0, opacity=0.3]
(axis cs:12.5,190665)
--(axis cs:12.5,216042)
--(axis cs:12.6,214821)
--(axis cs:12.7,213597)
--(axis cs:12.8,212381)
--(axis cs:12.9,211178)
--(axis cs:13,209996)
--(axis cs:13.1,208839)
--(axis cs:13.2,207710)
--(axis cs:13.3,206613)
--(axis cs:13.4,205548)
--(axis cs:13.5,204518)
--(axis cs:13.6,203522)
--(axis cs:13.7,202558)
--(axis cs:13.8,201626)
--(axis cs:13.9,200727)
--(axis cs:14,199864)
--(axis cs:14.1,199038)
--(axis cs:14.2,198251)
--(axis cs:14.3,197505)
--(axis cs:14.4,196798)
--(axis cs:14.5,196131)
--(axis cs:14.6,195502)
--(axis cs:14.7,194912)
--(axis cs:14.8,194360)
--(axis cs:14.9,193844)
--(axis cs:15,193364)
--(axis cs:15.1,192920)
--(axis cs:15.2,192511)
--(axis cs:15.3,192136)
--(axis cs:15.4,191797)
--(axis cs:15.5,191492)
--(axis cs:15.6,191221)
--(axis cs:15.7,190985)
--(axis cs:15.8,190784)
--(axis cs:15.9,190619)
--(axis cs:16,190489)
--(axis cs:16.1,190395)
--(axis cs:16.2,190339)
--(axis cs:16.3,190321)
--(axis cs:16.4,190342)
--(axis cs:16.5,190404)
--(axis cs:16.6,190509)
--(axis cs:16.7,190657)
--(axis cs:16.8,190850)
--(axis cs:16.9,191091)
--(axis cs:17,191383)
--(axis cs:17.1,191729)
--(axis cs:17.2,192132)
--(axis cs:17.3,192596)
--(axis cs:17.4,193125)
--(axis cs:17.5,193725)
--(axis cs:17.6,194399)
--(axis cs:17.7,195155)
--(axis cs:17.8,195999)
--(axis cs:17.9,196938)
--(axis cs:17.9,190665)
--(axis cs:17.9,190665)
--(axis cs:17.8,190665)
--(axis cs:17.7,190665)
--(axis cs:17.6,190665)
--(axis cs:17.5,190665)
--(axis cs:17.4,190665)
--(axis cs:17.3,190665)
--(axis cs:17.2,190665)
--(axis cs:17.1,190665)
--(axis cs:17,190665)
--(axis cs:16.9,190665)
--(axis cs:16.8,190665)
--(axis cs:16.7,190665)
--(axis cs:16.6,190665)
--(axis cs:16.5,190665)
--(axis cs:16.4,190665)
--(axis cs:16.3,190665)
--(axis cs:16.2,190665)
--(axis cs:16.1,190665)
--(axis cs:16,190665)
--(axis cs:15.9,190665)
--(axis cs:15.8,190665)
--(axis cs:15.7,190665)
--(axis cs:15.6,190665)
--(axis cs:15.5,190665)
--(axis cs:15.4,190665)
--(axis cs:15.3,190665)
--(axis cs:15.2,190665)
--(axis cs:15.1,190665)
--(axis cs:15,190665)
--(axis cs:14.9,190665)
--(axis cs:14.8,190665)
--(axis cs:14.7,190665)
--(axis cs:14.6,190665)
--(axis cs:14.5,190665)
--(axis cs:14.4,190665)
--(axis cs:14.3,190665)
--(axis cs:14.2,190665)
--(axis cs:14.1,190665)
--(axis cs:14,190665)
--(axis cs:13.9,190665)
--(axis cs:13.8,190665)
--(axis cs:13.7,190665)
--(axis cs:13.6,190665)
--(axis cs:13.5,190665)
--(axis cs:13.4,190665)
--(axis cs:13.3,190665)
--(axis cs:13.2,190665)
--(axis cs:13.1,190665)
--(axis cs:13,190665)
--(axis cs:12.9,190665)
--(axis cs:12.8,190665)
--(axis cs:12.7,190665)
--(axis cs:12.6,190665)
--(axis cs:12.5,190665)
--cycle;

\addplot [semithick, color0]
table {%
12.5 216042
12.6 214821
12.7 213597
12.8 212381
12.9 211178
13 209996
13.1 208839
13.2 207710
13.3 206613
13.4 205548
13.5 204518
13.6 203522
13.7 202558
13.8 201626
13.9 200727
14 199864
14.1 199038
14.2 198251
14.3 197505
14.4 196798
14.5 196131
14.6 195502
14.7 194912
14.8 194360
14.9 193844
15 193364
15.1 192920
15.2 192511
15.3 192136
15.4 191797
15.5 191492
15.6 191221
15.7 190985
15.8 190784
15.9 190619
16 190489
16.1 190395
16.2 190339
16.3 190321
16.4 190342
16.5 190404
16.6 190509
16.7 190657
16.8 190850
16.9 191091
17 191383
17.1 191729
17.2 192132
17.3 192596
17.4 193125
17.5 193725
17.6 194399
17.7 195155
17.8 195999
17.9 196938
};
\addlegendentry{\tiny Dyanmic PHT+WT}
\addplot [semithick, black, dashed]
table {%
12.5 190665
12.6 190665
12.7 190665
12.8 190665
12.9 190665
13 190665
13.1 190665
13.2 190665
13.3 190665
13.4 190665
13.5 190665
13.6 190665
13.7 190665
13.8 190665
13.9 190665
14 190665
14.1 190665
14.2 190665
14.3 190665
14.4 190665
14.5 190665
14.6 190665
14.7 190665
14.8 190665
14.9 190665
15 190665
15.1 190665
15.2 190665
15.3 190665
15.4 190665
15.5 190665
15.6 190665
15.7 190665
15.8 190665
15.9 190665
16 190665
16.1 190665
16.2 190665
16.3 190665
16.4 190665
16.5 190665
16.6 190665
16.7 190665
16.8 190665
16.9 190665
17 190665
17.1 190665
17.2 190665
17.3 190665
17.4 190665
17.5 190665
17.6 190665
17.7 190665
17.8 190665
17.9 190665
};
\addlegendentry{\tiny Optimal PHT+WT}
\addplot [semithick, black]
table {%
15.87 189034.95
15.87 217328.05
};
\addlegendentry{ \tiny Optimal $\bar{v}_b$}
\end{axis}

\end{tikzpicture}
    \caption{Difference between the optimal static and dynamic bus speed set
points for the PI controller.}
    \label{fig:PI_different_set_points}
\end{figure}

Figure~\ref{fig:PI_different_set_points} shows the variation of the PHT results for different choices of the PI controller set points for the bus demand profile in Figure~\subfigref{fig:demand}{b}, and compare them with the optimal set point found by solving the static formulation. Clearly, the value of $\bar{v}_b$ minimizing delays does not fully coincide. Nevertheless, the static choice of $\bar{v}_b$ gives some insights into what value is potentially able to minimize delays. 

However, despite it being convenient from an implementation point of view, the formulation of the PI framework does not allow to explicitly achieve a multi-modal system optimum. Therefore, we broaden our formulation to include all network users by resorting to an MPC framework, and report the results for when the optimization framework runs with one control variable $\phi_{\mc{B}}$, and two control variables $\phi_{\mc{V}}$ and $\phi_{\mc{B}}$.

\subsubsection{MPC framework}
The results of the different MPC implementations in Table~\ref{tab:results_no_ab} show that when our regulatory prices dictate the amount of pooling in both the vehicle and the bus network, we get the lowest objective function value. This is because trip pooling, even if performed in network~$\mc{V}$, makes the ride-hailing vehicles available soon after for a new trip, hence reducing the waiting times of ride-hailing users. Note that even if we only consider one control variable $\phi_{\mc{B}}$, we already observe some improvements compared to the no control scenario. In fact, the results in Table~\ref{tab:results_no_ab} reveal that, in reference to the scenario with no control and no pooling in bus lanes, the MPC framework allows for $8\%$ improvement compared to the PI implementation. In addition to the previous results, we also report the outcomes of the MPC with a lower bound $\underbar{v}_b$ on the minimum bus speed allowed in the bus network $\mc{B}$. We do so to guarantee that the bus services do not significantly lose performance, even if this implies a better total PHT for the overall network. Clearly, when the value of $\underbar{v}_b$ is set to $17$ km/hr, the objective function value increases because the private vehicle and ride-hailing delays, that we denote by $\text{PHT}_{pv}$ and $\text{PHT}_{rs}$, also increase as reported in Table~\ref{tab:results_multi_delay}. However, when we numerically compute in Table~\ref{tab:results_bus_delay} the bus delays $\text{PHT}_b$ and the constraint violations of the minimum bus speed, we guarantee the public transit performance is minimally impacted. 

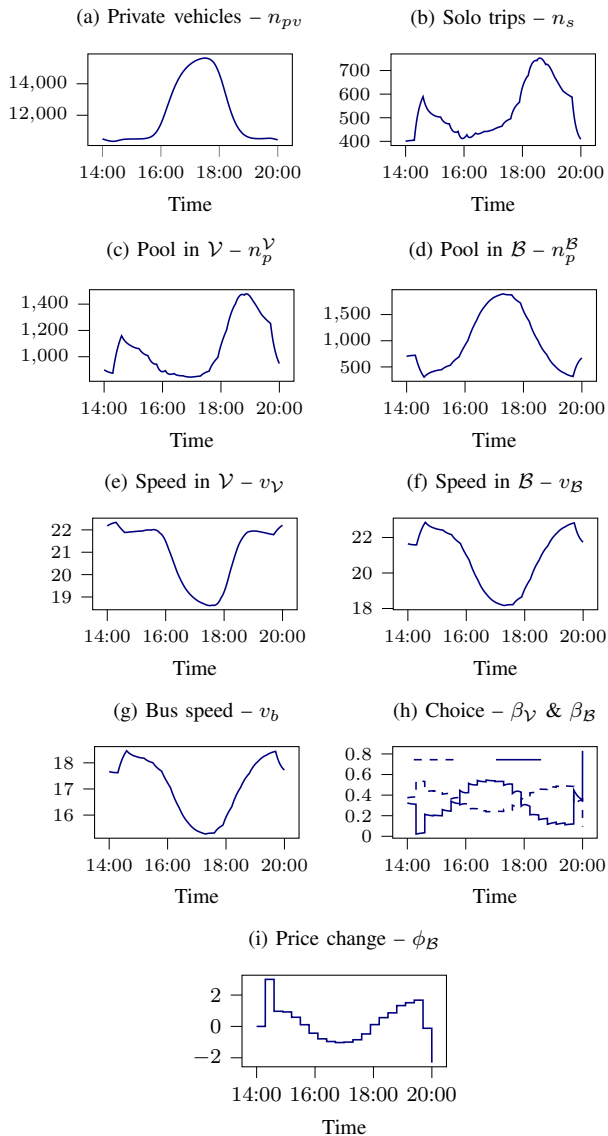
\begin{figure}
\centering 
\begin{tabular}{cc}
\hspace{-0.4cm}
\begin{tikzpicture}

\definecolor{color0}{rgb}{0.12156862745098,0.466666666666667,0.705882352941177}

\begin{axis}[
width=4.3cm,
height=2.8cm,
tick align=outside,
tick pos=left,
x grid style={white!69.0196078431373!black},
xlabel={\footnotesize Time},
xmin=13.5, xmax=20.5,
ytick style={color=black},
scaled y ticks = false,
y tick label style={/pgf/number format/fixed, font = \scriptsize},
ymin=10077.0787014165, ymax=15896.7328729128,
xlabel near ticks,
ylabel near ticks,
scaled x ticks = false,
x tick label style={/pgf/number format/fixed, font = \scriptsize},
x filter/.code={\pgfmathparse{#1/600+14}},
    xticklabel={ 
        \pgfmathsetmacro\hours{floor(\tick)}%
        \pgfmathsetmacro\minutes{(\tick-\hours)*0.6}%
        \pgfmathprintnumber{\hours}:\pgfmathprintnumber[fixed, fixed zerofill, skip 0.=true, dec sep={}]{\minutes}%
    },
title={\footnotesize (a) Private vehicles -- $n_{pv}$}
]

\addplot [semithick, blue!50.1960784313725!black]
table {%
0 10500
26 10469.5419921875
52 10442.4423828125
78 10418.47265625
105 10396.6015625
132 10377.5009765625
160 10360.30078125
186 10346.755859375
199 10342.9267578125
212 10341.6142578125
226 10342.642578125
241 10346.1484375
258 10352.6337890625
277 10362.44140625
299 10376.400390625
326 10396.2470703125
393 10447.0048828125
414 10457.935546875
436 10466.8720703125
461 10474.525390625
489 10480.650390625
522 10485.4462890625
562 10488.859375
616 10490.8662109375
706 10494.150390625
810 10502.513671875
836 10507.3984375
858 10513.7431640625
878 10521.80859375
896 10531.3681640625
918 10545.853515625
938 10561.4697265625
956 10578.4296875
972 10596.498046875
988 10617.9931640625
1003 10641.78515625
1018 10669.609375
1033 10701.9716796875
1049 10742.0576171875
1065 10788.4775390625
1081 10841.80078125
1100 10913.53125
1118 10990.0517578125
1137 11080.7919921875
1156 11182.4365234375
1176 11301.5693359375
1197 11439.9697265625
1220 11606.42578125
1246 11811.3427734375
1278 12082.302734375
1337 12603.728515625
1397 13127.435546875
1436 13448.48828125
1468 13694.3759765625
1497 13900.6865234375
1526 14091.2529296875
1554 14260.7021484375
1581 14411.138671875
1608 14549.5107421875
1635 14676.3896484375
1661 14787.82421875
1687 14889.4638671875
1713 14982.1162109375
1739 15066.5263671875
1765 15143.3623046875
1791 15213.21875
1817 15276.6923828125
1843 15334.466796875
1869 15386.83984375
1894 15432.2861328125
1918 15471.4853515625
1941 15505.009765625
1963 15533.34765625
1984 15556.9423828125
2009 15581.6181640625
2030 15599.4482421875
2049 15612.8134765625
2066 15622.1533203125
2081 15628.0439453125
2096 15631.451171875
2110 15632.1328125
2123 15630.375
2136 15626.0830078125
2148 15619.662109375
2160 15610.666015625
2175 15596.3837890625
2190 15578.89453125
2204 15559.0361328125
2218 15535.1142578125
2232 15506.48828125
2246 15472.5263671875
2260 15432.6171875
2274 15386.1728515625
2289 15328.53515625
2304 15262.1552734375
2320 15181.1279296875
2337 15082.9423828125
2357 14952.8134765625
2378 14801.4892578125
2400 14626.2607421875
2423 14425.1123046875
2449 14177.59375
2480 13859.90234375
2525 13372.8798828125
2575 12838.0341796875
2610 12486.703125
2640 12206.966796875
2668 11966.45703125
2694 11762.099609375
2718 11590.609375
2740 11448.73828125
2762 11320.8017578125
2784 11205.939453125
2805 11107.6611328125
2826 11019.662109375
2847 10941.1435546875
2867 10874.4609375
2887 10815.12890625
2905 10768.4111328125
2923 10727.59375
2941 10692.037109375
2959 10661.1611328125
2977 10634.4404296875
2996 10610.2255859375
3015 10589.6201171875
3034 10572.1826171875
3054 10556.8173828125
3074 10544.2451171875
3094 10534.3837890625
3116 10526.1923828125
3139 10520.1015625
3165 10515.681640625
3194 10513.177734375
3227 10512.662109375
3262 10514.271484375
3300 10518.458984375
3353 10526.7373046875
3433 10539.498046875
3446 10537.9765625
3461 10533.72265625
3478 10526.2919921875
3497 10515.4892578125
3521 10499.2294921875
3556 10472.5986328125
3599 10438.5732421875
};
\end{axis}

\end{tikzpicture} &
\hspace{-0.4cm}
\begin{tikzpicture}

\definecolor{color0}{rgb}{0.12156862745098,0.466666666666667,0.705882352941177}

\begin{axis}[
width=4.3cm,
height=2.8cm,
tick align=outside,
tick pos=left,
x grid style={white!69.0196078431373!black},
xlabel={\footnotesize Time},
xmin=13.5, xmax=20.5,
xtick style={color=black},
y grid style={white!69.0196078431373!black},
ymin=382.369994560726, ymax=770.230114224745,
ytick style={color=black},
scaled y ticks = false,
y tick label style={/pgf/number format/fixed, font = \scriptsize},
xlabel near ticks,
ylabel near ticks,
scaled x ticks = false,
x tick label style={/pgf/number format/fixed, font = \scriptsize},
x filter/.code={\pgfmathparse{#1/600+14}},
    xticklabel={ 
        \pgfmathsetmacro\hours{floor(\tick)}%
        \pgfmathsetmacro\minutes{(\tick-\hours)*0.6}%
        \pgfmathprintnumber{\hours}:\pgfmathprintnumber[fixed, fixed zerofill, skip 0.=true, dec sep={}]{\minutes}%
    },
title={\footnotesize (b) Solo trips -- $n_s$}
]
\addplot [semithick, blue!50.1960784313725!black]
table {%
0 400
11 400.79296875
26 401.512817382812
49 402.251037597656
97 403.399322509766
169 404.823089599609
179 404.990905761719
184 415.694427490234
190 427.6904296875
196 438.869995117188
202 449.326263427734
208 459.134765625
214 468.357574462891
220 477.046478271484
227 486.566772460938
234 495.478668212891
241 503.832336425781
248 511.671173095703
255 519.033142089844
262 525.952087402344
269 532.458374023438
277 539.423828125
285 545.923278808594
293 551.990600585938
301 557.656555175781
309 562.949462890625
318 568.490661621094
327 573.626892089844
336 578.389465332031
345 582.807250976562
355 587.343505859375
359 589.055419921875
365 583.475280761719
371 578.330688476562
377 573.570617675781
384 568.447631835938
391 563.736633300781
398 559.393005371094
406 554.831726074219
414 550.656494140625
422 546.828979492188
431 542.898132324219
440 539.326171875
449 536.078552246094
459 532.813842773438
469 529.875732421875
480 526.982482910156
492 524.187744140625
504 521.727966308594
517 519.396911621094
531 517.22705078125
545 515.070495605469
558 513.019897460938
572 511.132995605469
587 509.425354003906
604 507.821960449219
622 506.446960449219
642 505.24267578125
664 504.241851806641
689 503.439971923828
716 502.899780273438
719 502.857940673828
726 499.812042236328
734 496.683258056641
742 493.882507324219
751 491.074493408203
760 488.586090087891
770 486.151763916016
781 483.829040527344
792 481.835235595703
804 479.992218017578
817 478.342132568359
830 477.011871337891
844 475.897552490234
859 475.03125
875 474.443115234375
891 474.170745849609
899 474.145568847656
905 470.08447265625
911 466.393432617188
918 462.507293701172
925 459.028686523438
932 455.918212890625
939 453.142852783203
947 450.345581054688
955 447.913879394531
963 445.817535400391
971 444.030517578125
980 442.361907958984
989 441.028686523438
998 440.006927490234
1008 439.211944580078
1018 438.751770019531
1028 438.605560302734
1039 438.786407470703
1050 439.305816650391
1061 440.147094726562
1072 441.295928955078
1079 442.181549072266
1084 438.252532958984
1089 434.692352294922
1094 431.472503662109
1100 428.022979736328
1106 424.990142822266
1112 422.341461181641
1118 420.04833984375
1124 418.085571289062
1130 416.430541992188
1137 414.861724853516
1144 413.655181884766
1151 412.784637451172
1158 412.226135253906
1166 411.941802978516
1174 412.006011962891
1182 412.390716552734
1191 413.174194335938
1200 414.295562744141
1210 415.898315429688
1220 417.836151123047
1231 420.307861328125
1243 423.354461669922
1256 426.999725341797
1259 427.884521484375
1264 425.673065185547
1270 423.406646728516
1276 421.521514892578
1282 419.979339599609
1288 418.74658203125
1295 417.660003662109
1302 416.91259765625
1309 416.467498779297
1317 416.287109375
1325 416.415222167969
1334 416.879913330078
1344 417.737091064453
1355 419.026092529297
1367 420.767517089844
1381 423.141510009766
1398 426.378601074219
1421 431.127380371094
1439 434.972564697266
1446 433.813232421875
1454 432.820831298828
1463 432.061889648438
1472 431.619079589844
1482 431.433715820312
1494 431.557769775391
1507 432.029876708984
1522 432.907104492188
1540 434.299285888672
1563 436.427917480469
1598 440.038879394531
1620 442.218109130859
1632 441.801177978516
1646 441.672607421875
1662 441.86767578125
1681 442.433227539062
1705 443.494598388672
1736 445.212127685547
1785 448.290893554688
1801 449.464874267578
1818 451.543090820312
1839 453.774169921875
1864 456.096588134766
1893 458.461212158203
1927 460.90234375
1968 463.510070800781
1979 464.164123535156
1986 466.986877441406
1994 469.822875976562
2003 472.631958007812
2013 475.3837890625
2024 478.05224609375
2036 480.613891601562
2049 483.048583984375
2063 485.341857910156
2079 487.619720458984
2097 489.831634521484
2117 491.954376220703
2141 494.169067382812
2159 495.671752929688
2164 500.467987060547
2169 504.775817871094
2175 509.444458007812
2181 513.679443359375
2188 518.178466796875
2195 522.2822265625
2203 526.562622070312
2211 530.465026855469
2220 534.461364746094
2229 538.091613769531
2239 541.749755859375
2249 545.0634765625
2260 548.366027832031
2272 551.621948242188
2285 554.811218261719
2300 558.142883300781
2317 561.577392578125
2338 565.475158691406
2339 565.653930664062
2343 571.893127441406
2347 577.571105957031
2352 584.0615234375
2357 590.024353027344
2363 596.629516601562
2369 602.743469238281
2375 608.442260742188
2382 614.633666992188
2389 620.386962890625
2396 625.7451171875
2404 631.428466796875
2412 636.682922363281
2420 641.545349121094
2428 646.048278808594
2437 650.720458984375
2446 655.011291503906
2455 658.95361328125
2465 662.960266113281
2475 666.607116699219
2485 669.924194335938
2496 673.222778320312
2507 676.182495117188
2519 679.053161621094
2523 683.418884277344
2528 688.358276367188
2533 692.846130371094
2539 697.7587890625
2545 702.248168945312
2552 707.034362792969
2559 711.395874023438
2566 715.376403808594
2573 719.007202148438
2580 722.3125
2588 725.716857910156
2596 728.747314453125
2604 731.426086425781
2612 733.773071289062
2620 735.806335449219
2629 737.739135742188
2638 739.317138671875
2647 740.559936523438
2656 741.485229492188
2665 742.109313964844
2675 742.467407226562
2685 742.490234375
2695 742.194396972656
2699 741.990234375
2704 744.038146972656
2710 746.093505859375
2716 747.785400390625
2722 749.169494628906
2729 750.446533203125
2736 751.400634765625
2744 752.135131835938
2752 752.522155761719
2760 752.587463378906
2768 752.351867675781
2777 751.748596191406
2786 750.807800292969
2795 749.548461914062
2804 747.98779296875
2814 745.919494628906
2824 743.518859863281
2834 740.804870605469
2845 737.478698730469
2856 733.817626953125
2868 729.467529296875
2879 725.17626953125
2884 725.893249511719
2890 726.390075683594
2896 726.544982910156
2902 726.402893066406
2909 725.908142089844
2916 725.100341796875
2924 723.837890625
2932 722.253051757812
2941 720.126647949219
2950 717.675720214844
2960 714.614562988281
2971 710.8857421875
2983 706.444519042969
2996 701.262329101562
3010 695.330688476562
3027 687.755432128906
3048 678.024841308594
3060 672.538635253906
3072 668.781311035156
3087 663.712036132812
3107 656.582641601562
3172 633.156616210938
3193 626.053588867188
3212 619.954895019531
3231 614.203308105469
3239 611.890625
3263 608.282775878906
3320 599.624328613281
3349 595.603088378906
3376 592.185607910156
3403 589.097473144531
3419 587.421569824219
3424 575.295288085938
3429 563.950866699219
3434 553.326721191406
3439 543.36767578125
3444 534.024536132812
3450 523.564025878906
3456 513.859008789062
3462 504.849151611328
3468 496.480438232422
3474 488.704437255859
3480 481.477386474609
3486 474.759613037109
3492 468.514892578125
3498 462.710113525391
3504 457.314788818359
3510 452.300903320312
3517 446.898864746094
3524 441.942749023438
3531 437.397735595703
3538 433.231811523438
3545 429.415466308594
3553 425.447204589844
3561 421.862854003906
3569 418.628295898438
3578 415.368927001953
3587 412.47265625
3596 409.902954101562
3599 409.113159179688
};
\end{axis}

\end{tikzpicture} \\
\hspace{-0.25cm}
\begin{tikzpicture}

\definecolor{color0}{rgb}{0.12156862745098,0.466666666666667,0.705882352941177}

\begin{axis}[
width=4.3cm,
height=2.8cm,
tick align=outside,
tick pos=left,
x grid style={white!69.0196078431373!black},
xlabel={\footnotesize Time},
xmin=13.5, xmax=20.5,
xtick style={color=black},
y grid style={white!69.0196078431373!black},
ymin=813.210542324545, ymax=1509.73286027162,
ytick style={color=black},
scaled y ticks = false,
y tick label style={/pgf/number format/fixed, font = \scriptsize},
xlabel near ticks,
ylabel near ticks,
scaled x ticks = false,
x tick label style={/pgf/number format/fixed, font = \scriptsize},
x filter/.code={\pgfmathparse{#1/600+14}},
    xticklabel={ 
        \pgfmathsetmacro\hours{floor(\tick)}%
        \pgfmathsetmacro\minutes{(\tick-\hours)*0.6}%
        \pgfmathprintnumber{\hours}:\pgfmathprintnumber[fixed, fixed zerofill, skip 0.=true, dec sep={}]{\minutes}%
    },
title={\footnotesize (c) Pool in $\mc{V}$ -- $n_p^{\mc{V}}$}
]
\addplot [semithick, blue!50.1960784313725!black]
table {%
0 900
60 887.946166992188
82 884.462890625
105 881.418640136719
130 878.70751953125
158 876.279724121094
179 874.802856445312
187 895.410217285156
196 916.999145507812
206 939.352966308594
216 960.253723144531
226 979.891357421875
237 1000.19348144531
248 1019.26092529297
259 1037.19409179688
270 1054.07409667969
281 1069.96923828125
293 1086.2568359375
305 1101.5146484375
317 1115.80688476562
329 1129.193359375
341 1141.72998046875
353 1153.46923828125
359 1159.05541992188
368 1152.32043457031
378 1145.58874511719
389 1138.90759277344
401 1132.31628417969
414 1125.853515625
428 1119.56018066406
443 1113.47998046875
459 1107.65673828125
475 1102.43920898438
492 1097.48681640625
510 1092.8349609375
529 1088.51184082031
547 1084.42749023438
565 1080.29260253906
585 1076.30407714844
606 1072.70568847656
629 1069.36279296875
654 1066.34240722656
680 1063.78735351562
708 1061.61596679688
719 1060.91125488281
730 1054.6025390625
742 1048.45434570312
754 1042.95239257812
767 1037.62158203125
781 1032.525390625
796 1027.72595214844
811 1023.54205322266
827 1019.69561767578
843 1016.43255615234
860 1013.55615234375
878 1011.12921142578
896 1009.30267333984
899 1009.05474853516
908 1000.88421630859
917 993.502685546875
926 986.817260742188
936 980.120422363281
946 974.122375488281
956 968.765686035156
967 963.559692382812
978 959.025024414062
989 955.121215820312
1000 951.814453125
1011 949.076293945312
1023 946.709106445312
1035 944.962707519531
1047 943.814147949219
1059 943.243530273438
1071 943.233276367188
1079 943.52978515625
1087 935.031677246094
1095 927.456176757812
1103 920.713073730469
1111 914.729858398438
1119 909.447082519531
1127 904.814880371094
1135 900.790222167969
1144 896.941650390625
1153 893.765686035156
1162 891.218322753906
1171 889.258605957031
1180 887.847839355469
1190 886.879211425781
1200 886.49365234375
1211 886.686218261719
1222 887.465148925781
1234 888.912902832031
1247 891.099670410156
1259 893.609924316406
1267 888.464965820312
1275 884.060485839844
1283 880.308654785156
1292 876.77685546875
1301 873.8896484375
1311 871.345947265625
1321 869.414978027344
1332 867.904968261719
1344 866.887084960938
1357 866.407287597656
1371 866.484680175781
1387 867.169311523438
1406 868.596008300781
1430 871.025939941406
1439 872.050598144531
1449 868.921447753906
1460 866.086059570312
1472 863.581787109375
1486 861.284362792969
1501 859.41552734375
1518 857.875549316406
1538 856.66064453125
1562 855.818786621094
1591 855.403442382812
1621 855.057922363281
1637 852.730712890625
1656 850.603637695312
1678 848.759826660156
1704 847.199462890625
1734 846.002624511719
1770 845.178588867188
1800 844.939392089844
1830 846.683715820312
1879 848.878540039062
1980 853.078735351562
1990 857.042236328125
2002 861.140869140625
2016 865.296203613281
2032 869.445617675781
2051 873.742736816406
2072 877.878173828125
2097 882.173156738281
2128 886.859375
2159 891.170471191406
2166 898.617797851562
2174 906.162109375
2183 913.798217773438
2193 921.51904296875
2204 929.301147460938
2216 937.108703613281
2229 944.905334472656
2244 953.199829101562
2260 961.384643554688
2279 970.43701171875
2303 981.1923828125
2339 996.755249023438
2345 1008.21301269531
2352 1020.29132080078
2360 1032.98193359375
2370 1047.71850585938
2381 1062.88146972656
2393 1078.4443359375
2406 1094.34948730469
2420 1110.53637695312
2435 1126.95593261719
2451 1143.57470703125
2468 1160.37048339844
2486 1177.32275390625
2505 1194.40014648438
2519 1206.4892578125
2526 1219.1875
2534 1232.54943847656
2544 1248.08642578125
2555 1264.07824707031
2567 1280.45446777344
2579 1295.83776855469
2591 1310.29345703125
2603 1323.86157226562
2615 1336.56982421875
2627 1348.43957519531
2639 1359.48742675781
2651 1369.72668457031
2663 1379.16796875
2675 1387.81970214844
2686 1395.06262207031
2697 1401.65161132812
2699 1402.77954101562
2706 1410.51232910156
2714 1418.51818847656
2723 1426.65612792969
2732 1434.00183105469
2741 1440.63220214844
2750 1446.59362792969
2759 1451.916015625
2768 1456.62084960938
2777 1460.72436523438
2786 1464.24060058594
2795 1467.181640625
2804 1469.55944824219
2813 1471.38488769531
2822 1472.66943359375
2832 1473.4755859375
2842 1473.64306640625
2852 1473.18798828125
2862 1472.12683105469
2872 1470.47692871094
2879 1468.98132324219
2886 1471.93017578125
2893 1474.2099609375
2900 1475.90759277344
2908 1477.21533203125
2916 1477.91491699219
2924 1478.05908203125
2933 1477.60852050781
2942 1476.55505371094
2951 1474.93969726562
2961 1472.53039550781
2971 1469.51904296875
2982 1465.5634765625
2993 1460.9892578125
3005 1455.35864257812
3018 1448.58386230469
3032 1440.6015625
3047 1431.37756347656
3061 1422.68322753906
3074 1416.44812011719
3089 1408.56323242188
3107 1398.3955078125
3130 1384.67114257812
3169 1360.568359375
3211 1334.81787109375
3239 1318.34606933594
3319 1286.54479980469
3347 1276.29455566406
3374 1267.03686523438
3400 1258.73413085938
3419 1253.05139160156
3427 1228.14196777344
3435 1204.98779296875
3444 1180.80444335938
3453 1158.38708496094
3462 1137.55883789062
3471 1118.17309570312
3480 1100.10656738281
3489 1083.25329589844
3498 1067.52087402344
3508 1051.255859375
3518 1036.17370605469
3528 1022.18450927734
3538 1009.20684814453
3548 997.166381835938
3558 985.994873046875
3568 975.629516601562
3579 965.089477539062
3590 955.383422851562
3599 948.016418457031
};
\end{axis}
\end{tikzpicture} &
\hspace{-0.6cm}
\begin{tikzpicture}

\definecolor{color0}{rgb}{0.12156862745098,0.466666666666667,0.705882352941177}

\begin{axis}[
width=4.3cm,
height=2.8cm,
tick align=outside,
tick pos=left,
x grid style={white!69.0196078431373!black},
xlabel={\footnotesize Time},
xmin=13.5, xmax=20.5,
xtick style={color=black},
y grid style={white!69.0196078431373!black},
ymin=225.952704099035, ymax=1974.62587557585,
ytick style={color=black},
scaled y ticks = false,
y tick label style={/pgf/number format/fixed, font = \scriptsize},
xlabel near ticks,
ylabel near ticks,
scaled x ticks = false,
x tick label style={/pgf/number format/fixed, font = \scriptsize},
x filter/.code={\pgfmathparse{#1/600+14}},
    xticklabel={ 
        \pgfmathsetmacro\hours{floor(\tick)}%
        \pgfmathsetmacro\minutes{(\tick-\hours)*0.6}%
        \pgfmathprintnumber{\hours}:\pgfmathprintnumber[fixed, fixed zerofill, skip 0.=true, dec sep={}]{\minutes}%
    },
title={\footnotesize (d) Pool in $\mc{B}$ -- $n_p^{\mc{B}}$}
]
\addplot [semithick, blue!50.1960784313725!black]
table {%
0 700
16 704.943420410156
36 709.547607421875
61 713.72314453125
91 717.195068359375
127 719.853698730469
172 721.637939453125
179 721.804748535156
196 662.29052734375
213 607.917846679688
230 558.356994628906
247 513.269287109375
264 472.320190429688
281 435.185852050781
298 401.557189941406
315 371.141693115234
332 343.664367675781
349 318.868347167969
359 305.437835693359
378 323.861541748047
396 339.364990234375
414 353.026428222656
433 365.623443603516
453 377.093292236328
474 387.421813964844
496 396.628662109375
520 405.080230712891
581 423.885955810547
610 430.820617675781
642 436.923858642578
678 442.287872314453
721 448.619049072266
740 463.222320556641
759 475.981353759766
779 487.622222900391
800 498.140991210938
823 507.988555908203
848 517.090454101562
877 526.047302246094
899 532.00048828125
917 556.152099609375
934 576.624206542969
952 596.0341796875
971 614.34814453125
992 632.477355957031
1016 651.14208984375
1046 672.455871582031
1079 694.725830078125
1101 741.306579589844
1122 782.438049316406
1145 824.185485839844
1171 868.250915527344
1204 921.159362792969
1258 1004.56860351562
1260 1008.80267333984
1288 1082.34533691406
1312 1141.1796875
1335 1193.73852539062
1358 1242.77392578125
1382 1290.474609375
1406 1334.9052734375
1430 1376.29797363281
1441 1395.47644042969
1463 1442.0546875
1483 1481.12805175781
1503 1516.9892578125
1523 1549.80590820312
1543 1579.82897949219
1563 1607.32470703125
1584 1633.75231933594
1606 1659.04052734375
1628 1684.14611816406
1650 1709.92541503906
1671 1732.28112792969
1692 1752.53991699219
1714 1771.73181152344
1737 1789.8330078125
1761 1806.85205078125
1786 1822.81384277344
1803 1832.36938476562
1832 1845.1298828125
1869 1859.79809570312
1908 1873.70471191406
1947 1886.09436035156
1979 1895.14074707031
2027 1883.89575195312
2050 1880.65454101562
2077 1878.39587402344
2113 1876.97253417969
2159 1875.98181152344
2191 1834.25512695312
2207 1816.47338867188
2223 1800.89343261719
2240 1786.33630371094
2260 1771.21606445312
2284 1754.99865722656
2325 1729.40856933594
2339 1720.67456054688
2365 1650.75122070312
2382 1609.59545898438
2399 1572.19360351562
2417 1536.01428222656
2437 1499.09155273438
2459 1461.51232910156
2485 1420.08630371094
2515 1375.10595703125
2519 1369.28491210938
2541 1308.13952636719
2560 1259.734375
2579 1215.40222167969
2598 1174.70971679688
2618 1135.33215332031
2638 1099.07043457031
2659 1063.94653320312
2681 1030.00915527344
2701 999.787231445312
2720 959.92822265625
2739 923.507690429688
2758 890.322631835938
2777 860.046875
2797 830.959228515625
2817 804.377807617188
2838 778.821533203125
2860 754.294555664062
2880 732.775207519531
2898 699.335266113281
2916 668.888061523438
2934 641.220031738281
2953 614.74609375
2972 590.78564453125
2992 567.9814453125
3012 547.3828125
3033 527.864868164062
3055 509.482116699219
3062 503.155792236328
3082 483.141479492188
3102 465.202453613281
3123 448.390350341797
3144 433.442657470703
3166 419.571044921875
3189 406.809936523438
3213 395.173034667969
3238 384.654937744141
3242 382.405303955078
3264 369.760986328125
3287 358.272308349609
3311 347.945190429688
3336 338.754028320312
3363 330.366516113281
3392 322.863922119141
3419 317.057281494141
3435 373.403625488281
3450 420.729278564453
3464 460.116546630859
3478 495.16943359375
3492 526.263305664062
3506 553.795227050781
3520 578.148803710938
3534 599.6787109375
3548 618.705749511719
3563 636.639404296875
3578 652.340148925781
3594 666.933898925781
3599 671.084045410156
};
\end{axis}

\end{tikzpicture} \\
\hspace{0.2cm}
\begin{tikzpicture}

\definecolor{color0}{rgb}{0.12156862745098,0.466666666666667,0.705882352941177}

\begin{axis}[
width=4.3cm,
height=2.8cm,
tick align=outside,
tick pos=left,
x grid style={white!69.0196078431373!black},
xlabel={\footnotesize Time},
xmin=13.5, xmax=20.5,
xtick style={color=black},
y grid style={white!69.0196078431373!black},
ymin=18.4304726498738, ymax=22.5228654785651,
ytick style={color=black},
scaled y ticks = false,
y tick label style={/pgf/number format/fixed, font = \scriptsize},
xlabel near ticks,
ylabel near ticks,
scaled x ticks = false,
x tick label style={/pgf/number format/fixed, font = \scriptsize},
x filter/.code={\pgfmathparse{#1/600+14}},
    xticklabel={ 
        \pgfmathsetmacro\hours{floor(\tick)}%
        \pgfmathsetmacro\minutes{(\tick-\hours)*0.6}%
        \pgfmathprintnumber{\hours}:\pgfmathprintnumber[fixed, fixed zerofill, skip 0.=true, dec sep={}]{\minutes}%
    },
title={\footnotesize (e) Speed in $\mc{V}$ -- $v_{\mc{V}}$}
]
\addplot [semithick, blue!50.1960784313725!black]
table {%
0 22.1764698028564
18 22.2015686035156
39 22.2273292541504
62 22.2520122528076
87 22.2753105163574
113 22.2961921691895
142 22.3160209655762
173 22.3337745666504
179 22.3368473052979
219 22.2212810516357
246 22.146656036377
271 22.081018447876
294 22.0239734649658
317 21.9703254699707
340 21.9201507568359
359 21.8813190460205
402 21.8902740478516
530 21.9178676605225
569 21.9266242980957
624 21.9371128082275
673 21.9432201385498
724 21.9490661621094
753 21.9668769836426
780 21.9800796508789
806 21.9893741607666
831 21.9948463439941
854 21.9965744018555
876 21.9949150085449
897 21.9899349212646
900 21.9899978637695
917 22.0006141662598
933 22.0072898864746
948 22.0102787017822
963 22.0098342895508
977 22.0060997009277
991 21.9989376068115
1004 21.9890251159668
1017 21.9757747650146
1030 21.9589920043945
1042 21.9401893615723
1054 21.9180450439453
1066 21.8924007415771
1077 21.8656940460205
1080 21.8592700958252
1091 21.8439445495605
1102 21.8250732421875
1112 21.8046607971191
1122 21.781042098999
1132 21.7541561126709
1142 21.7239761352539
1152 21.6905097961426
1162 21.6537818908691
1172 21.6138477325439
1183 21.5663089752197
1194 21.5151176452637
1205 21.4604396820068
1216 21.4024677276611
1228 21.3357315063477
1240 21.2656593322754
1253 21.1863880157471
1261 21.1382083892822
1276 21.0572814941406
1292 20.9674587249756
1310 20.8627777099609
1332 20.7311248779297
1416 20.2244491577148
1436 20.1092720031738
1441 20.0820922851562
1459 19.9921131134033
1478 19.9007453918457
1496 19.8175296783447
1514 19.7376117706299
1532 19.6610450744629
1550 19.5878734588623
1568 19.518102645874
1586 19.4517154693604
1604 19.3886661529541
1620 19.3355617523193
1638 19.2822704315186
1657 19.2293796539307
1677 19.1770915985107
1698 19.1256885528564
1719 19.0776824951172
1741 19.0308628082275
1764 18.9855346679688
1787 18.9437236785889
1822 18.8844299316406
1846 18.8472766876221
1870 18.8135623931885
1894 18.7832126617432
1919 18.7551002502441
1943 18.7314472198486
1967 18.7110939025879
1984 18.696439743042
2009 18.6694622039795
2029 18.6512145996094
2048 18.637149810791
2067 18.6265182495117
2085 18.6198310852051
2102 18.6167697906494
2119 18.6171417236328
2135 18.620906829834
2150 18.6277275085449
2161 18.6326274871826
2178 18.6272315979004
2191 18.6266937255859
2203 18.6296997070312
2214 18.6356792449951
2225 18.6449298858643
2236 18.6576061248779
2247 18.673864364624
2257 18.6918792724609
2267 18.7130947113037
2277 18.7376251220703
2287 18.7655773162842
2297 18.7970542907715
2307 18.8321495056152
2317 18.870943069458
2327 18.9135036468506
2336 18.9550685882568
2340 18.9726753234863
2351 19.0083255767822
2360 19.0408420562744
2369 19.0767612457275
2378 19.1161918640137
2387 19.1591320037842
2396 19.2055263519287
2405 19.2552757263184
2414 19.3082580566406
2424 19.3707389831543
2434 19.4368057250977
2444 19.5062065124512
2455 19.5860691070557
2466 19.6692485809326
2478 19.7632846832275
2492 19.8766174316406
2508 20.0098190307617
2520 20.1102447509766
2550 20.3292427062988
2591 20.6288089752197
2610 20.7639122009277
2626 20.8744258880615
2641 20.9747047424316
2655 21.0649948120117
2669 21.1518020629883
2682 21.229097366333
2695 21.3030700683594
2700 21.329870223999
2713 21.3894729614258
2727 21.450065612793
2741 21.50705909729
2755 21.5605335235596
2769 21.6105556488037
2783 21.657205581665
2797 21.7005729675293
2812 21.7435131072998
2827 21.7829418182373
2842 21.8190059661865
2858 21.8539409637451
2874 21.8854236602783
2880 21.8954925537109
2898 21.9101028442383
2918 21.9229106903076
2940 21.9334716796875
2964 21.9414119720459
2990 21.9464740753174
3019 21.9485626220703
3051 21.9474353790283
3064 21.9447898864746
3120 21.9214305877686
3257 21.8602466583252
3304 21.8343830108643
3351 21.8119125366211
3399 21.7923622131348
3419 21.7851657867432
3442 21.8581848144531
3460 21.9117240905762
3477 21.9588260650635
3494 22.0025272369385
3512 22.0452213287354
3530 22.0844459533691
3549 22.1223411560059
3568 22.1568984985352
3588 22.1899509429932
3599 22.2067794799805
};
\end{axis}

\end{tikzpicture} &
\hspace{-0.2cm}
\begin{tikzpicture}

\definecolor{color0}{rgb}{0.12156862745098,0.466666666666667,0.705882352941177}

\begin{axis}[
width=4.3cm,
height=2.8cm,
tick align=outside,
tick pos=left,
x grid style={white!69.0196078431373!black},
xlabel={\footnotesize Time},
xmin=13.5, xmax=20.5,
xtick style={color=black},
y grid style={white!69.0196078431373!black},
ymin=17.9302450067802, ymax=23.0919079412466,
ytick style={color=black},
scaled y ticks = false,
y tick label style={/pgf/number format/fixed, font = \scriptsize},
xlabel near ticks,
ylabel near ticks,
scaled x ticks = false,
x tick label style={/pgf/number format/fixed, font = \scriptsize},
x filter/.code={\pgfmathparse{#1/600+14}},
    xticklabel={ 
        \pgfmathsetmacro\hours{floor(\tick)}%
        \pgfmathsetmacro\minutes{(\tick-\hours)*0.6}%
        \pgfmathprintnumber{\hours}:\pgfmathprintnumber[fixed, fixed zerofill, skip 0.=true, dec sep={}]{\minutes}%
    },
title={\footnotesize (f) Speed in $\mc{B}$ -- $v_{\mc{B}}$}
]
\addplot [semithick, blue!50.1960784313725!black]
table {%
0 21.649486541748
16 21.6345348358154
36 21.6206130981445
60 21.6084175109863
89 21.5980739593506
124 21.5899848937988
168 21.5843963623047
179 21.5835704803467
188 21.6808433532715
197 21.7738094329834
206 21.8625831604004
215 21.9472923278809
224 22.0280647277832
234 22.1133613586426
244 22.1941528320312
254 22.2706260681152
264 22.3429641723633
275 22.4179801940918
286 22.4884548187256
297 22.5546207427979
308 22.6167030334473
320 22.6800270080566
332 22.7390251159668
345 22.7983589172363
358 22.8532409667969
359 22.8572864532471
375 22.8088264465332
390 22.7676372528076
406 22.7281837463379
422 22.6930541992188
439 22.6600303649902
457 22.6293640136719
476 22.6012020111084
497 22.5744285583496
520 22.54958152771
580 22.4925327301025
607 22.4723854064941
637 22.4542751312256
671 22.4380283355713
712 22.4229526519775
719 22.4207630157471
735 22.3820991516113
752 22.3455257415771
769 22.3132381439209
788 22.281644821167
808 22.2527980804443
830 22.2254333496094
855 22.1988143920898
884 22.1724720001221
899 22.1602687835693
912 22.1062049865723
925 22.0562953948975
939 22.0069885253906
954 21.9588069915771
970 21.912036895752
987 21.8667449951172
1006 21.8204288482666
1029 21.768856048584
1060 21.7040596008301
1079 21.6654434204102
1093 21.5745830535889
1107 21.4882564544678
1121 21.4064159393311
1136 21.3232154846191
1152 21.2387924194336
1171 21.1431159973145
1193 21.0368232727051
1222 20.9013023376465
1260 20.7240791320801
1276 20.5977687835693
1290 20.4913120269775
1304 20.3891448974609
1318 20.2913284301758
1332 20.1977462768555
1347 20.1019687652588
1362 20.0105934143066
1378 19.9177093505859
1394 19.8292942047119
1410 19.7451133728027
1427 19.6600780487061
1441 19.5899181365967
1456 19.4968643188477
1470 19.414587020874
1483 19.3423938751221
1497 19.2691535949707
1511 19.2004146575928
1525 19.135929107666
1540 19.0712432861328
1555 19.010778427124
1571 18.950569152832
1588 18.8910331726074
1606 18.8325347900391
1626 18.7680511474609
1644 18.7070693969727
1661 18.6537570953369
1679 18.601770401001
1697 18.5540618896484
1716 18.5079154968262
1736 18.4635219573975
1758 18.4190788269043
1781 18.3769435882568
1801 18.344030380249
1829 18.3089866638184
1865 18.2684211730957
1903 18.2298336029053
1942 18.1944847106934
1979 18.1648654937744
2027 18.1964340209961
2050 18.2055377960205
2078 18.2120494842529
2116 18.21608543396
2159 18.2186660766602
2188 18.3259296417236
2201 18.3681659698486
2214 18.406135559082
2228 18.4428005218506
2244 18.4802665710449
2262 18.5180549621582
2285 18.5617694854736
2324 18.6307182312012
2339 18.6572551727295
2359 18.8120288848877
2370 18.8918762207031
2381 18.9670448303223
2392 19.037784576416
2404 19.1104259490967
2417 19.1844844818115
2431 19.2596740722656
2446 19.3358917236328
2463 19.4178924560547
2482 19.5052185058594
2504 19.6019878387451
2519 19.6658782958984
2533 19.7806453704834
2545 19.8744316101074
2556 19.9562015533447
2568 20.0409183502197
2580 20.1212253570557
2592 20.197437286377
2605 20.2757320404053
2619 20.355489730835
2633 20.4309120178223
2648 20.5073127746582
2664 20.5842094421387
2680 20.6567687988281
2697 20.7295265197754
2699 20.7378101348877
2712 20.820686340332
2724 20.8930587768555
2737 20.9669742584229
2750 21.0364265441895
2763 21.1017036437988
2777 21.1676864624023
2792 21.2338275909424
2807 21.2956829071045
2823 21.3573894500732
2840 21.4186038970947
2858 21.4790534973145
2877 21.5385341644287
2879 21.5445594787598
2891 21.6132736206055
2904 21.6831893920898
2917 21.7485618591309
2930 21.8096408843994
2944 21.8709354400635
2958 21.9279270172119
2973 21.9845962524414
2988 22.0370979309082
3004 22.0888957977295
3021 22.1395950317383
3039 22.1888446807861
3058 22.2363357543945
3061 22.2452487945557
3077 22.2950763702393
3094 22.3435134887695
3111 22.3876476287842
3129 22.4300765991211
3148 22.47047996521
3168 22.5086059570312
3189 22.5442562103271
3211 22.5772933959961
3235 22.6088619232178
3242 22.6194438934326
3262 22.6551113128662
3283 22.6880970001221
3305 22.7182750701904
3329 22.7467041015625
3354 22.7719898223877
3381 22.7950534820557
3411 22.8163166046143
3419 22.8213119506836
3427 22.7319183349609
3435 22.6472053527832
3443 22.5673332214355
3451 22.4923038482666
3459 22.4220085144043
3467 22.356273651123
3475 22.2948875427246
3483 22.2376155853271
3492 22.1778049468994
3501 22.1225566864014
3510 22.0715427398682
3520 22.0194473266602
3530 21.9717864990234
3541 21.9240417480469
3552 21.8807563781738
3564 21.8381404876709
3576 21.7998561859131
3589 21.7627658843994
3599 21.7370338439941
};
\end{axis}

\end{tikzpicture} \\
\hspace{0.25cm}
\begin{tikzpicture}

\definecolor{color0}{rgb}{0.12156862745098,0.466666666666667,0.705882352941177}

\begin{axis}[
width=4.3cm,
height=2.8cm,
tick align=outside,
tick pos=left,
x grid style={white!69.0196078431373!black},
xlabel={\footnotesize Time},
xmin=13.5, xmax=20.5,
xtick style={color=black},
y grid style={white!69.0196078431373!black},
ymin=15.1152895616132, ymax=18.620982089781,
ytick style={color=black},
scaled y ticks = false,
y tick label style={/pgf/number format/fixed, font = \scriptsize},
xlabel near ticks,
ylabel near ticks,
scaled x ticks = false,
x tick label style={/pgf/number format/fixed, font = \scriptsize},
x filter/.code={\pgfmathparse{#1/600+14}},
    xticklabel={ 
        \pgfmathsetmacro\hours{floor(\tick)}%
        \pgfmathsetmacro\minutes{(\tick-\hours)*0.6}%
        \pgfmathprintnumber{\hours}:\pgfmathprintnumber[fixed, fixed zerofill, skip 0.=true, dec sep={}]{\minutes}%
    },
title={\footnotesize (g) Bus speed -- $v_b$},
]
\addplot [semithick, blue!50.1960784313725!black]
table {%
0 17.6656169891357
16 17.6556606292725
36 17.6463871002197
60 17.6382617950439
90 17.631175994873
126 17.6257457733154
170 17.6221446990967
179 17.6217021942139
188 17.6864891052246
197 17.7483062744141
206 17.8072452545166
215 17.8634033203125
225 17.9226570129395
235 17.9787330627441
245 18.0317707061768
255 18.0819034576416
265 18.1292629241943
276 18.1783123016357
287 18.2243328094482
298 18.2674884796143
310 18.3114814758301
322 18.3524475097656
334 18.3905696868896
347 18.4288692474365
359 18.4616317749023
375 18.430004119873
391 18.4014053344727
407 18.3758029937744
424 18.3516578674316
442 18.3291988372803
461 18.3085517883301
481 18.2897777557373
503 18.2721405029297
527 18.2559490203857
598 18.2138366699219
628 18.2011222839355
662 18.1897315979004
701 18.1796436309814
720 18.1741409301758
736 18.1488933563232
753 18.1250076293945
771 18.1027526855469
790 18.0822486877441
811 18.0626449584961
834 18.0442295074463
860 18.0264759063721
891 18.0084705352783
899 18.0042400360107
913 17.9659023284912
927 17.9307098388672
941 17.8985023498535
956 17.8669948577881
972 17.8363628387451
990 17.8049736022949
1010 17.7731113433838
1034 17.7378845214844
1069 17.6897468566895
1079 17.6762390136719
1093 17.6157131195068
1107 17.5581169128418
1121 17.503438949585
1136 17.4477729797363
1153 17.3877544403076
1172 17.3236961364746
1195 17.2491722106934
1226 17.1518325805664
1259 17.0499973297119
1276 16.9590358734131
1291 16.8817501068115
1305 16.8125076293945
1319 16.7461051940918
1333 16.6824741363525
1348 16.617244720459
1363 16.55491065979
1379 16.4914436340332
1395 16.4309368133545
1411 16.3732433319092
1428 16.3148765563965
1440 16.2742538452148
1455 16.2098064422607
1469 16.1527004241943
1482 16.1025085449219
1496 16.0515174865723
1510 16.0035972595215
1524 15.958589553833
1539 15.9133920669556
1554 15.8711023330688
1570 15.8289470672607
1587 15.7872247695923
1604 15.7483901977539
1627 15.6964540481567
1644 15.6562166213989
1661 15.618857383728
1678 15.58434009552
1696 15.5506896972656
1715 15.5181169509888
1735 15.4867610931396
1756 15.4567127227783
1779 15.4267873764038
1800 15.4020671844482
1827 15.378098487854
1862 15.3500719070435
1900 15.3225975036621
1938 15.2980012893677
1977 15.2757024765015
1980 15.2751350402832
2024 15.2958974838257
2047 15.3027038574219
2073 15.3073711395264
2107 15.3103790283203
2159 15.3126630783081
2187 15.3859462738037
2200 15.4159564971924
2213 15.4429225921631
2227 15.468936920166
2242 15.4939060211182
2260 15.5207862854004
2282 15.5504570007324
2316 15.5928773880005
2339 15.6213102340698
2358 15.7244310379028
2368 15.7755575180054
2378 15.8239183425903
2389 15.8740434646606
2400 15.9212427139282
2412 15.9697971343994
2425 16.0194244384766
2439 16.0699577331543
2455 16.1246795654297
2473 16.1831569671631
2493 16.2451553344727
2516 16.3135013580322
2519 16.3222236633301
2533 16.4012050628662
2544 16.4603900909424
2555 16.5167331695557
2566 16.5702896118164
2578 16.6257152557373
2590 16.6782321929932
2603 16.7321014404297
2616 16.7830772399902
2630 16.8350086212158
2644 16.8841209411621
2659 16.933874130249
2675 16.9839534759521
2691 17.0312099456787
2700 17.0582733154297
2713 17.1140384674072
2725 17.1626529693604
2738 17.2122325897217
2751 17.2587604522705
2764 17.3024463653564
2778 17.3465557098389
2793 17.3907299041748
2808 17.4320011138916
2824 17.4731349945068
2841 17.5139064788818
2859 17.5541343688965
2878 17.5936851501465
2879 17.5956897735596
2891 17.6414966583252
2903 17.6845798492432
2916 17.7282886505127
2929 17.7690830230713
2943 17.8099803924561
2957 17.8479652404785
2972 17.8856964111328
2988 17.9228515625
3004 17.9570999145508
3021 17.9905910491943
3039 18.0230979919434
3058 18.0544185638428
3061 18.0602931976318
3077 18.0931224822998
3094 18.1250095367432
3112 18.1556644439697
3130 18.1833992004395
3149 18.209789276123
3169 18.2346744537354
3190 18.257926940918
3213 18.2803783416748
3237 18.3008270263672
3242 18.3061599731445
3262 18.3295154571533
3283 18.3511028289795
3306 18.3716735839844
3330 18.3901386260986
3356 18.4071578979492
3384 18.4225273132324
3414 18.4361381530762
3419 18.4381561279297
3427 18.3797588348389
3435 18.3243389129639
3443 18.2720146179199
3451 18.2227973937988
3459 18.176628112793
3467 18.1334056854248
3475 18.0929985046387
3483 18.0552616119385
3492 18.015811920166
3501 17.9793376922607
3510 17.945629119873
3520 17.9111747741699
3530 17.8796272277832
3541 17.8479976654053
3552 17.8193016052246
3564 17.7910270690918
3576 17.7656097412109
3589 17.7409687042236
3599 17.7238655090332
};
\end{axis}
\end{tikzpicture} &
\hspace{-0.3cm}
\begin{tikzpicture}

\definecolor{color0}{rgb}{0.12156862745098,0.466666666666667,0.705882352941177}
\definecolor{color1}{rgb}{1,0.498039215686275,0.0549019607843137}

\begin{axis}[
legend cell align={left},
legend style={at={(0.5,1.05)},anchor=north, fill opacity=0, draw opacity=1, text opacity=1, draw=white!80!black, legend columns=2, fill = none, legend style={draw=none}},
tick align=outside,
tick pos=left,
x grid style={white!69.0196078431373!black},
width=4.3cm,
height=2.8cm,
x grid style={white!69.0196078431373!black},
xlabel={\footnotesize Time},
xmin=13.5, xmax=20.5,
xtick style={color=black},
y grid style={white!69.0196078431373!black},
ymin=-0.018467284715793, ymax=0.870989824297609,
ytick style={color=black},
scaled y ticks = false,
y tick label style={/pgf/number format/fixed, font = \scriptsize},
xlabel near ticks,
ylabel near ticks,
scaled x ticks = false,
x tick label style={/pgf/number format/fixed, font = \scriptsize},
x filter/.code={\pgfmathparse{#1/600+14}},
    xticklabel={ 
        \pgfmathsetmacro\hours{floor(\tick)}%
        \pgfmathsetmacro\minutes{(\tick-\hours)*0.6}%
        \pgfmathprintnumber{\hours}:\pgfmathprintnumber[fixed, fixed zerofill, skip 0.=true, dec sep={}]{\minutes}%
    },
title={\footnotesize (h) Choice -- $\beta_{\mc{V}}$ \& $\beta_{\mc{B}}$ }
]
\addplot [semithick, blue!50.1960784313725!black, dashed]
table {%
0 0.374650955200195
28 0.376857042312622
65 0.379004955291748
111 0.380914092063904
168 0.382501840591431
178 0.382714629173279
179 0.54316520690918
265 0.537420988082886
332 0.533697724342346
358 0.5324547290802
359 0.433356761932373
403 0.43651556968689
447 0.43891429901123
498 0.440935254096985
538 0.442095637321472
539 0.438360929489136
599 0.440041780471802
671 0.441287517547607
718 0.441762685775757
719 0.408677816390991
754 0.41163957118988
791 0.414010286331177
831 0.415823340415955
874 0.417031049728394
898 0.417399406433105
899 0.363581538200378
923 0.367374181747437
947 0.370418190956116
972 0.372827410697937
998 0.374568104743958
1025 0.375605344772339
1052 0.375895142555237
1078 0.375487089157104
1079 0.305509090423584
1099 0.309272289276123
1118 0.31208610534668
1137 0.314147472381592
1157 0.315547585487366
1178 0.316239953041077
1200 0.316203951835632
1224 0.315395951271057
1250 0.313768148422241
1258 0.313135266304016
1259 0.264427423477173
1284 0.266894102096558
1307 0.268409967422485
1331 0.269255638122559
1358 0.269459247589111
1390 0.268928289413452
1430 0.267482042312622
1438 0.267123222351074
1439 0.244261503219604
1479 0.2462317943573
1514 0.247182130813599
1554 0.247509002685547
1607 0.247145056724548
1618 0.246999979019165
1619 0.238305568695068
1680 0.23960542678833
1740 0.240037560462952
1798 0.240029811859131
1799 0.243775010108948
1898 0.242139220237732
1969 0.241807103157043
1978 0.241820454597473
1979 0.261371850967407
2024 0.25735068321228
2053 0.255624890327454
2083 0.254577994346619
2115 0.254218697547913
2147 0.254615187644958
2158 0.254929423332214
2159 0.302288889884949
2187 0.296703100204468
2204 0.29415225982666
2221 0.292355537414551
2238 0.291284918785095
2256 0.29089879989624
2274 0.291239023208618
2293 0.292344808578491
2312 0.294177174568176
2332 0.296845436096191
2338 0.297785997390747
2339 0.368813276290894
2358 0.365673303604126
2372 0.364091515541077
2386 0.363218307495117
2401 0.363025426864624
2417 0.363577246665955
2434 0.364900827407837
2453 0.367119073867798
2475 0.370435476303101
2503 0.37542986869812
2518 0.3783118724823
2519 0.422700643539429
2548 0.423238515853882
2577 0.424598932266235
2618 0.427354574203491
2698 0.43299126625061
2699 0.457924604415894
2802 0.456655025482178
2878 0.455596089363098
2879 0.487017512321472
2940 0.482787370681763
3004 0.479124665260315
3058 0.476532816886902
3059 0.487608313560486
3121 0.484336853027344
3187 0.481613516807556
3238 0.4799644947052
3239 0.489542245864868
3305 0.487197518348694
3380 0.485299110412598
3418 0.484578371047974
3419 0.302664518356323
3436 0.311342120170593
3452 0.318748831748962
3468 0.325388073921204
3484 0.331295490264893
3501 0.336838364601135
3519 0.341972827911377
3539 0.346903800964355
3560 0.351327657699585
3583 0.355423212051392
3598 0.35773241519928
3599 0.0939099788665771
};
\addlegendentry{\tiny $\beta_{\mathcal{V}}$}
\addplot [semithick, blue!50.1960784313725!black]
table {%
0 0.323701977729797
21 0.320918560028076
48 0.318125367164612
80 0.315573692321777
118 0.313299179077148
163 0.311370253562927
178 0.310868620872498
179 0.0219626426696777
253 0.0277913808822632
314 0.0318582057952881
358 0.034321665763855
359 0.214023947715759
391 0.209798455238342
423 0.206336855888367
458 0.203325510025024
497 0.200731158256531
538 0.198668837547302
539 0.205440521240234
584 0.203173637390137
636 0.201325535774231
696 0.199953556060791
718 0.199613571166992
719 0.259556889533997
746 0.255507230758667
773 0.252198696136475
802 0.24937629699707
834 0.24701714515686
868 0.245261788368225
898 0.244290828704834
899 0.34172534942627
917 0.33656919002533
935 0.332135200500488
954 0.32821798324585
973 0.325036644935608
993 0.322426080703735
1014 0.320443868637085
1035 0.31919264793396
1057 0.318629026412964
1078 0.318779587745667
1079 0.445712208747864
1094 0.440261483192444
1109 0.435570001602173
1124 0.431666135787964
1139 0.42853045463562
1154 0.426121592521667
1170 0.424296021461487
1187 0.423125028610229
1205 0.42266047000885
1224 0.422931551933289
1245 0.42401134967804
1258 0.425029516220093
1259 0.514411330223083
1278 0.509937763214111
1296 0.506460666656494
1314 0.503750920295715
1333 0.501657724380493
1353 0.500192880630493
1376 0.499280571937561
1402 0.499029159545898
1433 0.499518156051636
1438 0.499657988548279
1439 0.542428016662598
1469 0.538113117218018
1494 0.535297989845276
1520 0.533144950866699
1549 0.531528830528259
1582 0.530457973480225
1618 0.529940366744995
1619 0.546453833580017
1662 0.543278694152832
1698 0.541416883468628
1741 0.539980888366699
1798 0.538873434066772
1799 0.531660318374634
1896 0.533206105232239
1954 0.533215761184692
1978 0.532993078231812
1979 0.495226621627808
2018 0.501627922058105
2039 0.504193663597107
2061 0.506137251853943
2084 0.507418870925903
2108 0.507999897003174
2132 0.507843255996704
2156 0.506953477859497
2158 0.506845712661743
2159 0.415239930152893
2183 0.424580097198486
2196 0.428757429122925
2209 0.43213951587677
2222 0.434732675552368
2235 0.436570167541504
2248 0.437689781188965
2262 0.438134431838989
2276 0.437830567359924
2290 0.436816096305847
2305 0.434982061386108
2320 0.432418346405029
2336 0.428932189941406
2338 0.428444743156433
2339 0.292198538780212
2355 0.298253893852234
2366 0.301648736000061
2377 0.304279804229736
2388 0.306155204772949
2399 0.307320237159729
2411 0.307851076126099
2424 0.307648539543152
2438 0.306646466255188
2453 0.304816126823425
2470 0.3019859790802
2490 0.29789924621582
2517 0.291587948799133
2518 0.291344046592712
2519 0.208329796791077
2540 0.210164308547974
2559 0.211067676544189
2581 0.211331009864807
2608 0.210861802101135
2649 0.209306359291077
2698 0.20724093914032
2699 0.161670684814453
2739 0.165219664573669
2784 0.168391704559326
2857 0.172667503356934
2878 0.173842549324036
2879 0.11689031124115
2918 0.122359991073608
2958 0.127201199531555
3002 0.131761908531189
3051 0.136074304580688
3058 0.136633038520813
3059 0.116565346717834
3104 0.120739102363586
3151 0.124347329139709
3202 0.127515077590942
3238 0.129351139068604
3239 0.111971497535706
3289 0.114896416664124
3345 0.117406487464905
3409 0.119503259658813
3418 0.119744658470154
3419 0.450195074081421
3432 0.438396811485291
3444 0.428271293640137
3456 0.418925762176514
3468 0.410353183746338
3480 0.402519226074219
3493 0.394810676574707
3506 0.387846112251282
3520 0.38110089302063
3535 0.374654054641724
3551 0.368567705154419
3568 0.362887382507324
3586 0.357643127441406
3598 0.354534983634949
3599 0.830559968948364
};
\addlegendentry{\tiny $\beta_{\mathcal{B}}$}
\end{axis}

\end{tikzpicture}
\end{tabular}
\begin{tikzpicture}

\definecolor{color0}{rgb}{0.12156862745098,0.466666666666667,0.705882352941177}

\begin{axis}[
width=4.3cm,
height=2.8cm,
tick align=outside,
tick pos=left,
x grid style={white!69.0196078431373!black},
xlabel={\footnotesize Time},
xmin=13.5, xmax=20.5,
xtick style={color=black},
y grid style={white!69.0196078431373!black},
ymin=-2.57482447282108, ymax=3.2654678320391,
ytick style={color=black},
scaled y ticks = false,
y tick label style={/pgf/number format/fixed, font = \footnotesize},
xlabel near ticks,
ylabel near ticks,
scaled x ticks = false,
x tick label style={/pgf/number format/fixed, font = \footnotesize},
x filter/.code={\pgfmathparse{#1/600+14}},
    xticklabel={ 
        \pgfmathsetmacro\hours{floor(\tick)}%
        \pgfmathsetmacro\minutes{(\tick-\hours)*0.6}%
        \pgfmathprintnumber{\hours}:\pgfmathprintnumber[fixed, fixed zerofill, skip 0.=true, dec sep={}]{\minutes}%
    },
title={\footnotesize (i) Price change -- $\phi_{\mc{B}}$}
]

\addplot [semithick, blue!50.1960784313725!black]
table {%
0 -0
178 -0
179 3
358 3
359 0.965336203575134
538 0.965336203575134
539 0.923060655593872
718 0.923060655593872
719 0.582537531852722
898 0.582537531852722
899 0.108728051185608
1078 0.108728051185608
1079 -0.432538986206055
1258 -0.432538986206055
1259 -0.791999578475952
1438 -0.791999578475952
1439 -0.963369965553284
1618 -0.963369965553284
1619 -1.02984654903412
1798 -1.02984654903412
1799 -1.00090777873993
1978 -1.00090777873993
1979 -0.849699258804321
2158 -0.849699258804321
2159 -0.480193257331848
2338 -0.480193257331848
2339 0.115325212478638
2518 0.115325212478638
2519 0.560298681259155
2698 0.560298681259155
2699 0.86427903175354
2878 0.86427903175354
2879 1.32824671268463
3058 1.32824671268463
3059 1.51073801517487
3238 1.51073801517487
3239 1.67520451545715
3418 1.67520451545715
3419 -0.119515895843506
3598 -0.119515895843506
3599 -2.30935668945312
};
\end{axis}

\end{tikzpicture}
\caption{Time-dependent model variables for the MPC framework without abandonment for (a) private vehicle accumulation, (b) solo trip ride-hailing vehicles, (c) pool trip ride-hailing accumulation in~$\mc{V}$, (d) pool trip ride-hailing accumulation in~$\mc{B}$, (e) speed in the vehicle network~$\mc{V}$, (f) vehicle speed in the bus network~$\mc{B}$, (g) bus speed in the bus network~$\mc{B}$, (h) fraction of pool trip in $\mc{V}$ and $\mc{B}$ respectively, and (i) regulatory control fare for pooling in~$\mc{B}$.}
\label{fig:results_mpc_no_ab}
\end{figure}

To understand how the MPC framework influences the results, we display the variations of the main simulation variables in Figure~\ref{fig:results_mpc_no_ab}. Note that for this simulation, we set $\xi_{\text{min}}$ and $\xi_{\text{max}}$ to be equal to $e^{-3}$ and $e^{3}$ respectively. Moreover, for practicality reasons, we only allow the control variable to be updated every $180$ time steps such that $N_u = 180$. As opposed to the PI controller, the MPC shows a more demand-responsive behavior where, during off-peak periods, the speed in the vehicle network in Figure~\subfigref{fig:results_mpc_no_ab}{e} is lower than that in the bus network observed in Figure~\subfigref{fig:results_mpc_no_ab}{f}. This substantiates the relatively high values of $\phi_{\mc{B}}$ as the use of the bus network is privileged in this case. During the on-peak period, however, $\phi_{\mc{B}}$ becomes negative, implying that ride-hailing users in the vehicle networks are causing significant delays compared to bus users, and therefore encouraging people to pool their rides in bus lanes is necessary to reduce the total delays for multi-modal users.

\subsection{Multi-modal network delays with abandonment}
\label{subsec:with}
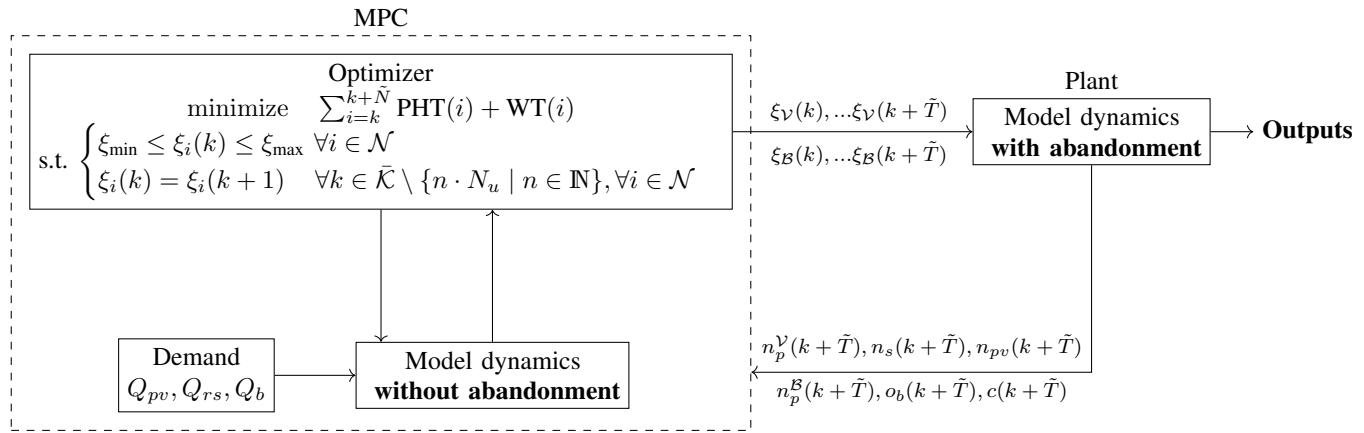
\begin{figure*}
    \centering
    \resizebox{0.99\textwidth}{!}{%
   \begin{tikzpicture}[scale=0.95]
     \node[align = center] (empty) at (16.3, 0) {\textbf{Outputs}};
     \node[draw,align=center] (optimizer) at (3,0) { Optimizer\\ $ \minimize \quad \sum_{i=k}^{k+\tilde{N}} \text{PHT}(i) + \text{WT}(i) $ \\ s.t. $ \begin{cases} 
        \xi_{\text{min}}\leq\xi_i(k)\leq \xi_{\text{max}} \, \, \forall i\in\mc{N} \\  \xi_i(k) = \xi_i(k+1) \, \quad \forall k \in \bar{\mc{K}} \setminus \{n \cdot N_u \mid n \in \N \}, \forall i\in \mc{N} \end{cases}$ };
              \node[draw,align=center, label = Plant] (plant) at (13.2, 0) { Model dynamics \\ \textbf{ with abandonment}
              };
      \node[draw,align=center] (demand) at (0.35, -3.5) {Demand \\ $Q_{pv}, Q_{rs}, Q_b$};
        \node[draw,align=center] (model) at (4.6, -3.5) { Model dynamics \\ \textbf{ without abandonment}
  };
      \node[draw, dashed, rectangle, label = MPC, fit=(demand) (model) (optimizer), inner xsep=7pt, inner ysep=7pt] (mpc) {};
      \draw[->] (optimizer) -- node[right, align = center, pos = 0.12] {\footnotesize $\xi_{\mc{V}}(k),...\xi_{\mc{V}}(k+\tilde{T})$ \\[0.5em] \footnotesize $\xi_{\mc{B}}(k),...\xi_{\mc{B}}(k+\tilde{T})$} (plant);
      \draw[->] (demand) -- (model);
      \draw[->] (model.north) --  (model.north |- optimizer.south);
      \draw[->] (optimizer.south) --  (optimizer.south |- model.north);
      \draw[->] (plant.south) |- node[left, align = center, pos = 0.48] { \\  \footnotesize $n_p^{\mc{V}}(k+\tilde{T}), n_s(k+\tilde{T}), n_{pv} (k+\tilde{T})$ \\[0.51em]  \footnotesize $n_p^{\mc{B}}(k+\tilde{T}), o_b(k+\tilde{T}), c(k+\tilde{T})$}($(mpc.east)-(0,2)$);
      \draw[->] (plant) -- (empty);
    \end{tikzpicture}}
    \caption{Implementation of the MPC framework with abandonment. In this framework, the MPC dynamics run without abandonment with a prediction horizon of $\tilde{N}$ whereas the plant dynamics run with abandonment. After every update step $\tilde{T}$, we reinitialize the MPC with some updated initial states retrieved from the plant dynamics.}
        \label{fig:mpc_framework}

\end{figure*}
The results we reported so far display scenarios where the ride-hailing users never abandon their requested trip, regardless of the waiting time. Table~\ref{tab:results_ab} shows the delays and abandonment values for when the ride-hailing users' waiting tolerance $w_{\text{max}}$ is set to $15$ min. Irrespective of the scenario under consideration, the values of the objective function when abandonment is considered in the dynamic model are lower than the values with no abandonment in Table~\ref{tab:results_no_ab}. Again, for this case, the lowest objective function values are observed for the MPC implementation. In fact, when looking at the individual delays of multi-modal users in Table~\ref{tab:delays_ab}, we notice that the MPC framework is capable of returning a solution that is convenient for all multi-modal users. We note here that due to the complexity of the abandonment function, the MPC dynamics are run without any abandoning requests, and the output control variables are then applied in the real settings dynamics with abandonment. Figure~\ref{fig:mpc_framework} describes the approach we follow when implementing the MPC. For each optimization run, we consider a prediction horizon of $\tilde{N} = 650$ time steps, i.e., we assume that the optimizer is aware of the private vehicle, bus, and ride-hailing demand for the upcoming $650\tau$ duration. However, after every update step of $\tilde{T} = 200$ time steps, we reinitialize the MPC dynamics with the actual state variables for the model dynamics with abandonment to bridge the gap between the MPC predictions and the actual network dynamics. 

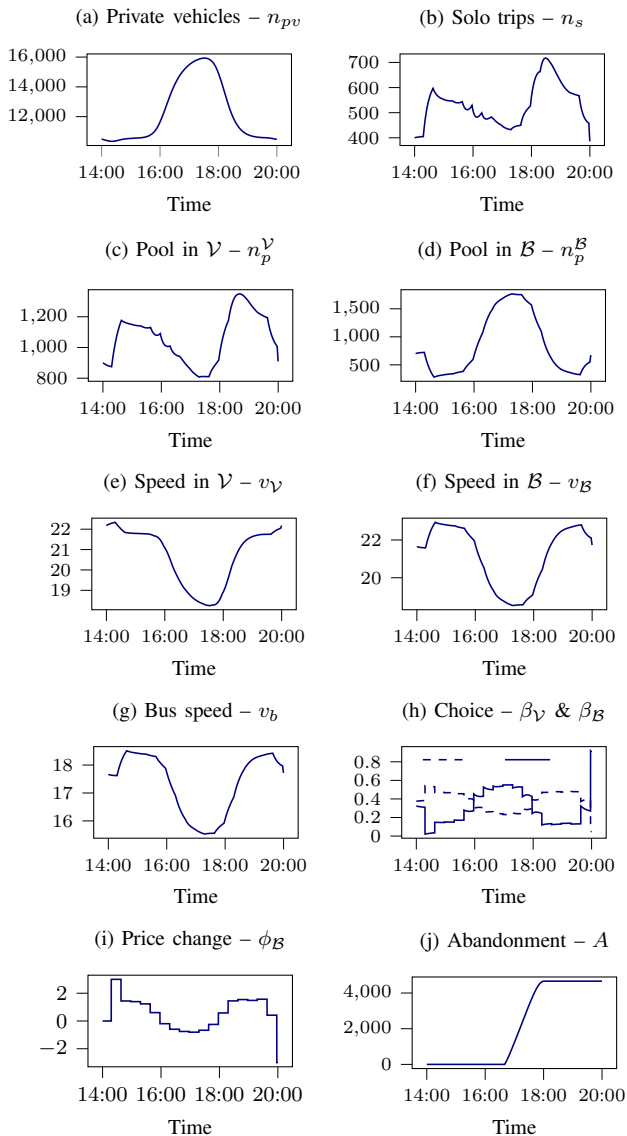
\begin{figure}
\centering 
\begin{tabular}{cc}
\hspace{-0.4cm}
\begin{tikzpicture}

\definecolor{color0}{rgb}{0.12156862745098,0.466666666666667,0.705882352941177}

\begin{axis}[
width=4.3cm,
height=2.8cm,
tick align=outside,
tick pos=left,
x grid style={white!69.0196078431373!black},
xlabel={\footnotesize Time},
xmin=13.5, xmax=20.5,
ytick style={color=black},
scaled y ticks = false,
y tick label style={/pgf/number format/fixed, font = \scriptsize},
ymin=10062.1415391577, ymax=16210.4132775058,
xlabel near ticks,
ylabel near ticks,
scaled x ticks = false,
x tick label style={/pgf/number format/fixed, font = \scriptsize},
x filter/.code={\pgfmathparse{#1/600+14}},
    xticklabel={ 
        \pgfmathsetmacro\hours{floor(\tick)}%
        \pgfmathsetmacro\minutes{(\tick-\hours)*0.6}%
        \pgfmathprintnumber{\hours}:\pgfmathprintnumber[fixed, fixed zerofill, skip 0.=true, dec sep={}]{\minutes}%
    },
title={\footnotesize (a) Private vehicles -- $n_{pv}$}
]

\addplot [semithick, blue!50.1960784313725!black]
table {%
0 10500
26 10469.5419921875
52 10442.4423828125
78 10418.47265625
105 10396.6015625
132 10377.5009765625
160 10360.30078125
186 10346.755859375
199 10342.9267578125
212 10341.6142578125
226 10342.642578125
241 10346.1484375
258 10352.6337890625
277 10362.44140625
299 10376.400390625
326 10396.2470703125
364 10427.1826171875
404 10459.5029296875
428 10475.4091796875
454 10489.9287109375
482 10502.9599609375
513 10514.8525390625
548 10525.7998046875
590 10536.4296875
639 10546.3408203125
751 10568.1240234375
787 10577.986328125
821 10589.515625
847 10600.5966796875
869 10612.25390625
889 10625.232421875
908 10640.1708984375
926 10657.1533203125
943 10676.1787109375
960 10698.5849609375
976 10723.224609375
995 10756.9560546875
1013 10793.173828125
1030 10832.1640625
1047 10876.73046875
1064 10927.751953125
1081 10986.0625
1098 11052.4208984375
1115 11127.490234375
1133 11217.05859375
1152 11323.318359375
1172 11448.40234375
1195 11607.556640625
1222 11809.744140625
1252 12052.04296875
1287 12353.5859375
1342 12849.619140625
1402 13385.564453125
1435 13659.5888671875
1466 13899.392578125
1496 14114.7578125
1525 14307.408203125
1554 14485.19140625
1582 14643.306640625
1607 14772.7333984375
1633 14895.9765625
1660 15013.466796875
1687 15121.5
1715 15224.48046875
1743 15319
1771 15405.6962890625
1798 15482.24609375
1824 15549.41796875
1850 15610.56640625
1876 15666.0517578125
1901 15714.3193359375
1925 15756.1162109375
1949 15793.51953125
1971 15823.91796875
1994 15851.8623046875
2017 15876.5498046875
2037 15895.0322265625
2055 15908.908203125
2072 15919.26171875
2087 15925.8857421875
2101 15929.6904296875
2114 15930.9462890625
2127 15929.7890625
2139 15926.373046875
2151 15920.498046875
2163 15911.9482421875
2175 15900.4892578125
2188 15884.791015625
2202 15864.720703125
2216 15840.87890625
2230 15812.634765625
2244 15779.3349609375
2258 15740.3271484375
2272 15694.97265625
2287 15638.650390625
2302 15573.666015625
2318 15494.126953125
2334 15403.466796875
2351 15294.5
2369 15164.6865234375
2389 15003.83203125
2414 14784.9931640625
2441 14528.6240234375
2472 14212.232421875
2514 13758.47265625
2594 12890.33984375
2623 12602.873046875
2650 12356.3720703125
2676 12139.02734375
2701 11948.6865234375
2725 11782.869140625
2748 11638.9853515625
2771 11509.123046875
2793 11397.35546875
2814 11301.5
2835 11215.4658203125
2856 11138.4794921875
2877 11069.783203125
2898 11008.6435546875
2919 10954.36328125
2940 10906.2822265625
2961 10863.783203125
2982 10826.2939453125
3003 10793.189453125
3025 10762.650390625
3047 10735.890625
3069 10712.4873046875
3092 10691.1904296875
3116 10672.0048828125
3141 10654.892578125
3167 10639.783203125
3194 10626.630859375
3222 10615.5732421875
3253 10605.880859375
3287 10597.7177734375
3325 10590.966796875
3370 10585.396484375
3396 10581.81640625
3415 10576.2939453125
3438 10566.98828125
3467 10552.5419921875
3512 10527.216796875
3594 10478.7216796875
3599 10474.1015625
};
\end{axis}

\end{tikzpicture} &
\hspace{-0.4cm}
\begin{tikzpicture}

\definecolor{color0}{rgb}{0.12156862745098,0.466666666666667,0.705882352941177}

\begin{axis}[
width=4.3cm,
height=2.8cm,
tick align=outside,
tick pos=left,
x grid style={white!69.0196078431373!black},
xlabel={\footnotesize Time},
xmin=13.5, xmax=20.5,
xtick style={color=black},
y grid style={white!69.0196078431373!black},
ymin=369.340197310781, ymax=733.902456092716,
ytick style={color=black},
scaled y ticks = false,
y tick label style={/pgf/number format/fixed, font = \scriptsize},
xlabel near ticks,
ylabel near ticks,
scaled x ticks = false,
x tick label style={/pgf/number format/fixed, font = \scriptsize},
x filter/.code={\pgfmathparse{#1/600+14}},
    xticklabel={ 
        \pgfmathsetmacro\hours{floor(\tick)}%
        \pgfmathsetmacro\minutes{(\tick-\hours)*0.6}%
        \pgfmathprintnumber{\hours}:\pgfmathprintnumber[fixed, fixed zerofill, skip 0.=true, dec sep={}]{\minutes}%
    },
title={\footnotesize (b) Solo trips -- $n_s$}
]
\addplot [semithick, blue!50.1960784313725!black]
table {%
0 400
11 400.79296875
25 401.47314453125
47 402.195831298828
91 403.266204833984
162 404.700866699219
179 404.990905761719
184 415.694427490234
189 425.750610351562
195 437.059234619141
201 447.63037109375
207 457.542205810547
213 466.858795166016
219 475.633453369141
225 483.911163330078
231 491.730377197266
238 500.317626953125
245 508.372161865234
252 515.934143066406
259 523.0390625
266 529.718688964844
273 536.001586914062
280 541.913879394531
288 548.247436523438
296 554.160766601562
304 559.683654785156
312 564.843566894531
320 569.665893554688
329 574.716491699219
338 579.400024414062
347 583.744812011719
356 587.776794433594
366 591.918823242188
376 595.734252929688
379 596.819396972656
386 593.275268554688
393 590.052795410156
401 586.70458984375
410 583.301696777344
419 580.228881835938
429 577.148681640625
439 574.375061035156
450 571.634460449219
462 568.970825195312
475 566.422241210938
488 564.181335449219
502 562.069519042969
517 560.110168457031
533 558.320373535156
551 556.625122070312
570 555.146728515625
583 554.112487792969
599 552.577453613281
617 551.158386230469
637 549.894592285156
659 548.817321777344
683 547.950927734375
709 547.315490722656
737 546.931030273438
767 546.824890136719
780 546.704528808594
791 545.083923339844
803 543.611511230469
816 542.311157226562
830 541.207214355469
845 540.325561523438
861 539.694641113281
877 539.3583984375
894 539.306091308594
911 539.558471679688
928 540.113647460938
945 540.977355957031
961 542.0830078125
977 543.488037109375
979 543.685607910156
985 539.468811035156
991 535.668334960938
997 532.2451171875
1003 529.166687011719
1009 526.405700683594
1015 523.938781738281
1022 521.406066894531
1029 519.219421386719
1036 517.355773925781
1043 515.794982910156
1050 514.51953125
1058 513.391418457031
1066 512.595458984375
1074 512.113586425781
1082 511.929748535156
1090 512.029357910156
1099 512.463928222656
1108 513.223388671875
1117 514.291809082031
1126 515.654235839844
1136 517.495666503906
1146 519.663635253906
1156 522.139709472656
1167 525.197021484375
1178 528.579650878906
1179 528.902404785156
1184 523.137145996094
1189 517.912353515625
1194 513.183898925781
1199 508.912841796875
1204 505.064636230469
1209 501.608306884766
1214 498.516021728516
1219 495.762542724609
1224 493.324890136719
1229 491.181945800781
1235 488.972290039062
1241 487.128326416016
1247 485.621765136719
1253 484.426483154297
1259 483.518280029297
1266 482.791534423828
1273 482.391662597656
1280 482.287902832031
1288 482.495330810547
1296 483.012756347656
1305 483.920196533203
1314 485.126495361328
1324 486.764739990234
1335 488.867034912109
1348 491.672515869141
1363 495.227661132812
1379 499.265869140625
1384 496.016479492188
1389 493.107940673828
1394 490.508697509766
1400 487.758728027344
1406 485.371551513672
1412 483.310028076172
1418 481.541473388672
1425 479.81005859375
1432 478.397644042969
1440 477.128631591797
1448 476.183258056641
1457 475.456085205078
1466 475.036437988281
1476 474.877471923828
1487 475.016265869141
1499 475.474761962891
1513 476.330627441406
1529 477.628051757812
1549 479.581298828125
1576 482.557067871094
1579 482.901153564453
1586 481.169952392578
1594 479.536376953125
1605 477.697296142578
1619 475.156433105469
1676 464.453521728516
1695 461.350494384766
1714 458.551330566406
1734 455.915924072266
1755 453.45947265625
1777 451.186401367188
1780 450.785766601562
1792 448.466369628906
1805 446.269073486328
1819 444.205902099609
1835 442.164886474609
1852 440.300903320312
1871 438.521728515625
1892 436.860626220703
1916 435.281341552734
1942 433.882446289062
1971 432.634460449219
1979 432.340515136719
1987 434.378051757812
1996 436.357421875
2006 438.233520507812
2017 439.974060058594
2029 441.558624267578
2042 442.977386474609
2057 444.309997558594
2074 445.516967773438
2095 446.684387207031
2121 447.807037353516
2168 449.463562011719
2179 449.853332519531
2184 454.706481933594
2189 459.175506591797
2195 464.090759277344
2201 468.575805664062
2207 472.682952880859
2213 476.456787109375
2220 480.488891601562
2227 484.170593261719
2235 488.005096435547
2243 491.494354248047
2252 495.065185546875
2261 498.315002441406
2271 501.607879638672
2282 504.908294677734
2294 508.196838378906
2308 511.712829589844
2325 515.64501953125
2347 520.390014648438
2379 527.011962890625
2384 536.719055175781
2389 545.666015625
2394 553.948120117188
2399 561.64453125
2404 568.822082519531
2409 575.537414550781
2415 583.05322265625
2421 590.042785644531
2427 596.561767578125
2433 602.644836425781
2439 608.319580078125
2445 613.610717773438
2451 618.540222167969
2457 623.128112792969
2464 628.073120117188
2471 632.604309082031
2478 636.745544433594
2485 640.518676757812
2492 643.943603515625
2499 647.038757324219
2506 649.820922851562
2513 652.3056640625
2521 654.799438476562
2529 656.943115234375
2537 658.754699707031
2545 660.250793457031
2553 661.446594238281
2561 662.356323242188
2570 663.054077148438
2579 663.423522949219
2583 668.327514648438
2588 673.917114257812
2593 678.976318359375
2598 683.56591796875
2603 687.735168457031
2608 691.524230957031
2613 694.966247558594
2618 698.088806152344
2624 701.44677734375
2630 704.412170410156
2636 707.013549804688
2642 709.275512695312
2648 711.219482421875
2654 712.864379882812
2660 714.227416992188
2667 715.482116699219
2674 716.396911621094
2681 716.992797851562
2688 717.289367675781
2695 717.304748535156
2703 717.000366210938
2711 716.3759765625
2719 715.454772949219
2728 714.091186523438
2737 712.410034179688
2747 710.204284667969
2757 707.678955078125
2768 704.574279785156
2783 700.244567871094
2793 697.704345703125
2804 694.590454101562
2816 690.882507324219
2830 686.233581542969
2847 680.24365234375
2869 672.136047363281
2943 644.573852539062
2964 637.210632324219
2985 629.876892089844
3000 624.404357910156
3015 619.263366699219
3030 614.434997558594
3046 609.613098144531
3062 605.115539550781
3078 600.927795410156
3095 596.801635742188
3112 592.991149902344
3130 589.280395507812
3148 585.881713867188
3167 582.609130859375
3180 580.621520996094
3269 573.9404296875
3306 571.587890625
3343 569.5400390625
3379 567.825073242188
3384 560.6875
3389 554.015380859375
3395 546.567016601562
3401 539.671630859375
3407 533.278076171875
3413 527.342102050781
3419 521.825317382812
3425 516.693908691406
3431 511.917907714844
3438 506.759674072266
3445 502.011077880859
3452 497.638366699219
3459 493.611297607422
3466 489.902435302734
3474 486.021636962891
3482 482.490447998047
3490 479.278381347656
3499 476.011962890625
3508 473.078857421875
3517 470.446899414062
3527 467.840423583984
3537 465.534027099609
3548 463.305236816406
3560 461.198760986328
3573 459.250305175781
3579 458.454772949219
3585 435.099578857422
3591 413.144256591797
3597 392.509155273438
3599 385.911193847656
};
\end{axis}

\end{tikzpicture} \\
\hspace{-0.25cm}
\begin{tikzpicture}

\definecolor{color0}{rgb}{0.12156862745098,0.466666666666667,0.705882352941177}

\begin{axis}[
width=4.3cm,
height=2.8cm,
tick align=outside,
tick pos=left,
x grid style={white!69.0196078431373!black},
xlabel={\footnotesize Time},
xmin=13.5, xmax=20.5,
xtick style={color=black},
y grid style={white!69.0196078431373!black},
ymin=780.026084408287, ymax=1375.99765167158,
ytick style={color=black},
scaled y ticks = false,
y tick label style={/pgf/number format/fixed, font = \scriptsize},
xlabel near ticks,
ylabel near ticks,
scaled x ticks = false,
x tick label style={/pgf/number format/fixed, font = \scriptsize},
x filter/.code={\pgfmathparse{#1/600+14}},
    xticklabel={ 
        \pgfmathsetmacro\hours{floor(\tick)}%
        \pgfmathsetmacro\minutes{(\tick-\hours)*0.6}%
        \pgfmathprintnumber{\hours}:\pgfmathprintnumber[fixed, fixed zerofill, skip 0.=true, dec sep={}]{\minutes}%
    },
title={\footnotesize (c) Pool in $\mc{V}$ -- $n_p^{\mc{V}}$}
]
\addplot [semithick, blue!50.1960784313725!black]
table {%
0 900
59 888.11865234375
79 884.903503417969
100 882.032409667969
123 879.410034179688
148 877.081298828125
176 874.998046875
179 874.802856445312
187 895.410217285156
195 914.674133300781
204 935.005798339844
213 954.123474121094
223 974.123840332031
233 992.959838867188
243 1010.73999023438
253 1027.54772949219
263 1043.45031738281
274 1059.96472167969
285 1075.51708984375
296 1090.16528320312
307 1103.96203613281
318 1116.95617675781
329 1129.193359375
340 1140.71643066406
351 1151.56591796875
363 1162.67858886719
375 1173.08239746094
379 1176.40051269531
390 1173.69189453125
404 1170.79040527344
421 1167.79992675781
442 1164.63391113281
466 1161.51965332031
493 1158.51538085938
523 1155.69421386719
555 1153.19995117188
587 1150.72998046875
612 1148.23937988281
640 1145.95874023438
670 1144.01599121094
702 1142.44458007812
735 1141.31811523438
769 1140.66040039062
780 1140.3525390625
795 1137.4462890625
811 1134.84204101562
829 1132.43359375
847 1130.521484375
866 1129.00732421875
885 1127.99487304688
904 1127.48132324219
923 1127.47583007812
941 1127.95703125
959 1128.93688964844
976 1130.349609375
979 1130.65063476562
987 1123.66711425781
995 1117.37255859375
1003 1111.69458007812
1011 1106.57836914062
1020 1101.44177246094
1029 1096.91564941406
1038 1092.96301269531
1047 1089.55419921875
1056 1086.66467285156
1065 1084.2734375
1074 1082.36206054688
1083 1080.91442871094
1093 1079.83142089844
1103 1079.28283691406
1113 1079.25024414062
1123 1079.71508789062
1133 1080.65869140625
1144 1082.22692871094
1155 1084.32458496094
1166 1086.9228515625
1178 1090.29235839844
1179 1090.59704589844
1185 1081.83618164062
1192 1072.544921875
1199 1064.15966796875
1206 1056.59643554688
1213 1049.78405761719
1220 1043.66076660156
1227 1038.17297363281
1234 1033.27270507812
1241 1028.91662597656
1249 1024.55395507812
1257 1020.79602050781
1265 1017.59265136719
1273 1014.89715576172
1282 1012.41717529297
1291 1010.46594238281
1301 1008.85168457031
1311 1007.75256347656
1322 1007.06121826172
1334 1006.83453369141
1347 1007.10388183594
1362 1007.94299316406
1379 1009.40338134766
1386 1002.88421630859
1393 996.990417480469
1401 990.928588867188
1409 985.500244140625
1417 980.630798339844
1426 975.742492675781
1435 971.405578613281
1445 967.154846191406
1455 963.428527832031
1466 959.857177734375
1478 956.50634765625
1491 953.424377441406
1505 950.64208984375
1520 948.173522949219
1537 945.896667480469
1556 943.870544433594
1578 942.046325683594
1579 941.974243164062
1588 937.791381835938
1598 933.679626464844
1617 926.123596191406
1648 912.969421386719
1678 900.468322753906
1702 890.990966796875
1725 882.421936035156
1749 874.020690917969
1773 866.151000976562
1781 863.402648925781
1797 856.845458984375
1814 850.437622070312
1832 844.183532714844
1852 837.784912109375
1873 831.610473632812
1895 825.670593261719
1918 819.978942871094
1942 814.552001953125
1967 809.40966796875
1979 807.11572265625
1991 808.53173828125
2005 809.676330566406
2022 810.531433105469
2042 811.015563964844
2068 811.109008789062
2108 810.682556152344
2164 810.153991699219
2179 810.210021972656
2186 817.098999023438
2194 824.240356445312
2202 830.730773925781
2211 837.387512207031
2221 844.123901367188
2232 850.881530761719
2244 857.629516601562
2257 864.363220214844
2272 871.570068359375
2291 880.099670410156
2318 891.606140136719
2361 909.928161621094
2379 918.017578125
2386 934.281188964844
2393 949.346252441406
2401 965.367614746094
2410 982.153503417969
2419 997.877319335938
2429 1014.32867431641
2439 1029.83703613281
2449 1044.48498535156
2460 1059.68298339844
2471 1073.99182128906
2482 1087.47058105469
2493 1100.16760253906
2504 1112.12292480469
2515 1123.36950683594
2526 1133.93395996094
2537 1143.83813476562
2548 1153.09887695312
2559 1161.72985839844
2570 1169.74145507812
2579 1175.84143066406
2586 1189.90637207031
2593 1202.8837890625
2601 1216.591796875
2609 1229.26501464844
2617 1241.02416992188
2625 1251.95520019531
2633 1262.12097167969
2641 1271.56921386719
2649 1280.33703613281
2658 1289.42504882812
2666 1296.84204101562
2674 1303.66015625
2682 1309.89965820312
2690 1315.57995605469
2698 1320.71948242188
2707 1325.87744140625
2716 1330.39892578125
2725 1334.30895996094
2734 1337.63269042969
2743 1340.39501953125
2752 1342.62121582031
2761 1344.33642578125
2770 1345.56591796875
2801 1348.72241210938
2811 1348.90295410156
2821 1348.58728027344
2832 1347.7177734375
2843 1346.35205078125
2855 1344.35375976562
2868 1341.65747070312
2882 1338.21533203125
2897 1333.99963378906
2914 1328.67883300781
2933 1322.19738769531
2956 1313.8125
2980 1304.60900878906
3016 1288.56555175781
3046 1275.71423339844
3073 1264.67749023438
3098 1254.98266601562
3122 1246.19274902344
3146 1237.93200683594
3170 1230.20849609375
3181 1227.03637695312
3229 1217.35534667969
3263 1211.03405761719
3296 1205.41296386719
3329 1200.3046875
3363 1195.5625
3379 1193.50634765625
3387 1179.08020019531
3395 1165.69421386719
3403 1153.22338867188
3412 1140.16552734375
3421 1128.02966308594
3430 1116.72619628906
3439 1106.18127441406
3449 1095.279296875
3459 1085.16442871094
3469 1075.77380371094
3479 1067.05151367188
3489 1058.94763183594
3500 1050.69372558594
3511 1043.07861328125
3522 1036.05249023438
3534 1029.00549316406
3546 1022.55065917969
3558 1016.63836669922
3571 1010.79302978516
3579 1007.46276855469
3588 961.971435546875
3597 918.988525390625
3599 909.763977050781
};
\end{axis}
\end{tikzpicture} &
\hspace{-0.6cm}
\begin{tikzpicture}

\definecolor{color0}{rgb}{0.12156862745098,0.466666666666667,0.705882352941177}

\begin{axis}[
width=4.3cm,
height=2.8cm,
tick align=outside,
tick pos=left,
x grid style={white!69.0196078431373!black},
xlabel={\footnotesize Time},
xmin=13.5, xmax=20.5,
xtick style={color=black},
y grid style={white!69.0196078431373!black},
ymin=206.641310734423, ymax=1840.98919469827,
ytick style={color=black},
scaled y ticks = false,
y tick label style={/pgf/number format/fixed, font = \scriptsize},
xlabel near ticks,
ylabel near ticks,
scaled x ticks = false,
x tick label style={/pgf/number format/fixed, font = \scriptsize},
x filter/.code={\pgfmathparse{#1/600+14}},
    xticklabel={ 
        \pgfmathsetmacro\hours{floor(\tick)}%
        \pgfmathsetmacro\minutes{(\tick-\hours)*0.6}%
        \pgfmathprintnumber{\hours}:\pgfmathprintnumber[fixed, fixed zerofill, skip 0.=true, dec sep={}]{\minutes}%
    },
title={\footnotesize (d) Pool in $\mc{B}$ -- $n_p^{\mc{B}}$}
]
\addplot [semithick, blue!50.1960784313725!black]
table {%
0 700
16 704.943420410156
35 709.348571777344
59 713.438598632812
88 716.907165527344
122 719.561889648438
164 721.415710449219
179 721.804748535156
196 662.29052734375
213 607.917846679688
229 561.145202636719
245 518.3525390625
261 479.26025390625
277 443.596771240234
293 411.101684570312
309 381.527282714844
325 354.639617919922
342 328.769500732422
359 305.437866210938
376 284.416015625
379 280.929840087891
410 290.825531005859
439 298.636108398438
470 305.53857421875
504 311.666687011719
542 317.087127685547
617 327.17822265625
658 332.136413574219
706 336.533782958984
768 340.757293701172
781 342.067810058594
808 350.356231689453
838 358.118225097656
873 365.705841064453
919 374.154541015625
979 384.352081298828
997 410.90380859375
1015 434.959075927734
1034 457.963165283203
1054 480.001220703125
1077 503.246398925781
1105 529.528991699219
1179 596.099060058594
1200 663.651062011719
1219 720.403991699219
1238 773.049926757812
1258 824.50244140625
1279 874.771606445312
1301 923.944763183594
1324 972.114929199219
1349 1021.24468994141
1374 1067.40991210938
1380 1079.47045898438
1401 1143.48986816406
1419 1194.21520996094
1437 1240.85192871094
1455 1283.55786132812
1473 1322.63073730469
1491 1358.40515136719
1509 1391.20849609375
1528 1422.9453125
1547 1452.02966308594
1567 1480.09130859375
1582 1501.220703125
1602 1535.82849121094
1613 1552.35266113281
1625 1567.91247558594
1638 1582.58837890625
1653 1597.42407226562
1670 1612.22497558594
1689 1626.8896484375
1710 1641.36059570312
1734 1656.17651367188
1760 1670.58203125
1789 1686.55444335938
1815 1701.99572753906
1840 1715.29711914062
1866 1727.59057617188
1893 1738.86682128906
1922 1749.50354003906
1953 1759.41271972656
1979 1766.70068359375
2032 1758.95129394531
2062 1756.28063964844
2098 1754.55456542969
2179 1752.43591308594
2203 1720.04895019531
2221 1698.05419921875
2238 1679.36486816406
2256 1661.62854003906
2275 1644.83447265625
2297 1627.25634765625
2326 1605.97131347656
2379 1568.20104980469
2399 1502.46789550781
2417 1448.01599121094
2434 1400.78137207031
2452 1354.72155761719
2471 1309.8193359375
2491 1265.95971679688
2513 1221.00183105469
2537 1175.13854980469
2563 1128.50439453125
2579 1101.15087890625
2597 1037.71154785156
2615 979.363342285156
2632 928.738586425781
2649 882.163940429688
2667 836.901611328125
2685 795.43603515625
2703 757.41552734375
2721 722.520629882812
2739 690.464416503906
2758 659.424255371094
2777 630.99462890625
2789 612.727844238281
2808 585.399475097656
2827 560.462341308594
2846 537.71240234375
2866 515.921325683594
2886 496.146179199219
2907 477.355377197266
2928 460.401550292969
2950 444.426727294922
2972 430.10693359375
2985 422.9072265625
3011 410.990570068359
3039 399.696929931641
3068 389.4931640625
3099 380.06884765625
3132 371.509124755859
3167 363.863159179688
3185 359.668884277344
3214 351.487121582031
3245 344.173675537109
3279 337.584075927734
3316 331.808532714844
3358 326.661224365234
3379 324.546295166016
3394 356.625396728516
3408 383.529846191406
3422 407.620849609375
3436 429.107269287109
3450 448.227752685547
3465 466.359558105469
3480 482.325439453125
3495 496.376495361328
3511 509.506286621094
3528 521.618103027344
3545 532.082153320312
3563 541.608337402344
3579 548.912658691406
3599 674.036743164062
};
\end{axis}
\end{tikzpicture} \\
\hspace{0.2cm}
\begin{tikzpicture}

\definecolor{color0}{rgb}{0.12156862745098,0.466666666666667,0.705882352941177}

\begin{axis}[
width=4.3cm,
height=2.8cm,
tick align=outside,
tick pos=left,
x grid style={white!69.0196078431373!black},
xlabel={\footnotesize Time},
xmin=13.5, xmax=20.5,
xtick style={color=black},
y grid style={white!69.0196078431373!black},
ymin=18.0474929098696, ymax=22.5411026090415,,
ytick style={color=black},
scaled y ticks = false,
y tick label style={/pgf/number format/fixed, font = \scriptsize},
xlabel near ticks,
ylabel near ticks,
scaled x ticks = false,
x tick label style={/pgf/number format/fixed, font = \scriptsize},
x filter/.code={\pgfmathparse{#1/600+14}},
    xticklabel={ 
        \pgfmathsetmacro\hours{floor(\tick)}%
        \pgfmathsetmacro\minutes{(\tick-\hours)*0.6}%
        \pgfmathprintnumber{\hours}:\pgfmathprintnumber[fixed, fixed zerofill, skip 0.=true, dec sep={}]{\minutes}%
    },
title={\footnotesize (e) Speed in $\mc{V}$ -- $v_{\mc{V}}$}
]
\addplot [semithick, blue!50.1960784313725!black]
table {%
0 22.1764698028564
19 22.2028751373291
41 22.2296123504639
65 22.2549915313721
91 22.2787342071533
119 22.3005771636963
150 22.3209209442139
179 22.3368473052979
221 22.2156372070312
249 22.1385917663574
275 22.0708599090576
300 22.0096454620361
324 21.9546852111816
348 21.9035129547119
372 21.8560943603516
380 21.8424854278564
412 21.8293113708496
451 21.8170719146729
498 21.8061466217041
557 21.7962627410889
600 21.7920608520508
711 21.7840518951416
763 21.7765655517578
789 21.7735538482666
821 21.7719879150391
850 21.7668571472168
876 21.7585010528564
900 21.7469253540039
921 21.7331485748291
941 21.7162818908691
959 21.697473526001
976 21.6760902404785
980 21.6717624664307
995 21.6673393249512
1009 21.6593685150146
1022 21.6483211517334
1035 21.6335029602051
1047 21.6162700653076
1059 21.5954475402832
1071 21.5708656311035
1083 21.5423679351807
1094 21.5126838684082
1105 21.4794979095459
1116 21.4427280426025
1127 21.4023189544678
1138 21.3582305908203
1149 21.3104572296143
1160 21.2590141296387
1171 21.203950881958
1180 21.1584548950195
1191 21.1210880279541
1202 21.0800609588623
1213 21.0353221893311
1224 20.9869537353516
1236 20.9302444458008
1248 20.8697147369385
1261 20.800235748291
1275 20.7214241027832
1290 20.6331119537354
1307 20.5291862487793
1328 20.3968334197998
1363 20.1716594696045
1380 20.0635242462158
1408 19.920877456665
1437 19.7768707275391
1462 19.6564617156982
1484 19.5541248321533
1505 19.4601078033447
1525 19.3742122650146
1545 19.2920665740967
1565 19.213773727417
1580 19.1580867767334
1598 19.1024017333984
1668 18.8944797515869
1692 18.8284664154053
1715 18.7688407897949
1738 18.7128505706787
1761 18.6604442596436
1782 18.6160507202148
1806 18.5716686248779
1831 18.5291614532471
1857 18.4886512756348
1884 18.4503803253174
1911 18.4159030914307
1938 18.3852310180664
1964 18.3593902587891
1988 18.3355751037598
2013 18.3084011077881
2035 18.2880878448486
2056 18.2724132537842
2076 18.2612705230713
2095 18.2544765472412
2113 18.2518081665039
2130 18.2530117034912
2146 18.2577953338623
2161 18.2658233642578
2176 18.2776184082031
2180 18.2802791595459
2195 18.2794933319092
2208 18.2824096679688
2220 18.2886295318604
2232 18.2986316680908
2243 18.3114051818848
2254 18.3278541564941
2265 18.3481864929199
2275 18.3701972961426
2285 18.3957004547119
2295 18.4248199462891
2305 18.4576625823975
2315 18.4943237304688
2325 18.5348815917969
2335 18.5793972015381
2345 18.6279144287109
2355 18.6804466247559
2365 18.7369899749756
2375 18.7975044250488
2379 18.822811126709
2389 18.8654098510742
2399 18.911771774292
2409 18.9619960784912
2419 19.0160827636719
2429 19.0739440917969
2439 19.135404586792
2450 19.2068881988525
2461 19.2820930480957
2473 19.3679275512695
2486 19.464786529541
2500 19.5728130340576
2517 19.7079086303711
2540 19.8948421478271
2579 20.2130489349365
2601 20.3495960235596
2621 20.4699802398682
2639 20.5745067596436
2656 20.6693267822266
2672 20.7547130584717
2687 20.8311176300049
2702 20.9038257598877
2717 20.9727344512939
2732 21.0377883911133
2747 21.0989875793457
2762 21.1563625335693
2777 21.2099838256836
2786 21.2393798828125
2802 21.288013458252
2818 21.3328227996826
2834 21.374002456665
2851 21.4139976501465
2868 21.4503574371338
2886 21.4851684570312
2905 21.5181045532227
2925 21.5489120483398
2946 21.5773906707764
2968 21.6034145355225
3002 21.6388683319092
3027 21.6610240936279
3054 21.6812038421631
3084 21.6997241973877
3117 21.7161350250244
3153 21.7301788330078
3184 21.7390689849854
3233 21.7448139190674
3296 21.7483863830566
3380 21.7516212463379
3404 21.8015880584717
3426 21.8437213897705
3448 21.8821582794189
3471 21.918550491333
3495 21.9526767730713
3520 21.9843997955322
3546 22.0136375427246
3574 22.0412788391113
3579 22.0458278656006
3592 22.1300754547119
3599 22.1764087677002
};
\end{axis}

\end{tikzpicture} &
\hspace{-0.2cm}
\begin{tikzpicture}

\definecolor{color0}{rgb}{0.12156862745098,0.466666666666667,0.705882352941177}

\begin{axis}[
width=4.3cm,
height=2.8cm,
tick align=outside,
tick pos=left,
x grid style={white!69.0196078431373!black},
xlabel={\footnotesize Time},
xmin=13.5, xmax=20.5,
xtick style={color=black},
y grid style={white!69.0196078431373!black},
ymin=18.3064954468224, ymax=23.1535700529202,
ytick style={color=black},
scaled y ticks = false,
y tick label style={/pgf/number format/fixed, font = \scriptsize},
xlabel near ticks,
ylabel near ticks,
scaled x ticks = false,
x tick label style={/pgf/number format/fixed, font = \scriptsize},
x filter/.code={\pgfmathparse{#1/600+14}},
    xticklabel={ 
        \pgfmathsetmacro\hours{floor(\tick)}%
        \pgfmathsetmacro\minutes{(\tick-\hours)*0.6}%
        \pgfmathprintnumber{\hours}:\pgfmathprintnumber[fixed, fixed zerofill, skip 0.=true, dec sep={}]{\minutes}%
    },
title={\footnotesize (f) Speed in $\mc{B}$ -- $v_{\mc{B}}$}
]
\addplot [semithick, blue!50.1960784313725!black]
table {%
0 21.649486541748
15 21.6353511810303
34 21.621826171875
57 21.609733581543
85 21.5992736816406
119 21.5909080505371
159 21.5852203369141
179 21.5835704803467
188 21.6808433532715
197 21.7738094329834
206 21.8625831604004
215 21.9472923278809
224 22.0280647277832
233 22.105037689209
243 22.1862716674805
253 22.2631664276123
263 22.3359107971191
273 22.4046859741211
283 22.4696712493896
294 22.5369892120361
305 22.6001625061035
316 22.6594123840332
328 22.7198238372803
340 22.7760829925537
352 22.8284454345703
365 22.8810539245605
378 22.929666519165
379 22.9332485198975
408 22.9043960571289
436 22.8806629180908
466 22.8594970703125
498 22.8410472869873
535 22.8240070343018
576 22.8093204498291
588 22.8036098480225
624 22.7870807647705
665 22.7724227905273
713 22.7593803405762
777 22.7463836669922
781 22.7439594268799
807 22.7192249298096
836 22.6958503723145
869 22.6734371185303
911 22.6491756439209
979 22.6134414672852
992 22.5535621643066
1005 22.4978523254395
1018 22.4460334777832
1032 22.3941268920898
1048 22.3390579223633
1065 22.2846202850342
1085 22.224681854248
1111 22.1511688232422
1179 21.9647750854492
1190 21.8553676605225
1201 21.750186920166
1211 21.658483505249
1222 21.561897277832
1233 21.4696044921875
1244 21.3813247680664
1256 21.2892322540283
1268 21.2011451721191
1281 21.1098022460938
1295 21.0157146453857
1309 20.9256267547607
1324 20.8331108093262
1340 20.738582611084
1357 20.6424503326416
1374 20.5504188537598
1379 20.5240898132324
1391 20.4140090942383
1402 20.3173122406006
1412 20.2333240509033
1422 20.1531944274902
1432 20.076904296875
1442 20.0043506622314
1453 19.9286842346191
1464 19.8571357727051
1475 19.7894668579102
1487 19.7197856903076
1499 19.6541309356689
1512 19.5872173309326
1525 19.5243549346924
1539 19.4608287811279
1553 19.4012794494629
1568 19.3415222167969
1581 19.2897167205811
1595 19.2187824249268
1605 19.1713523864746
1613 19.1375846862793
1622 19.1035308837891
1632 19.0696258544922
1644 19.0332584381104
1657 18.998010635376
1672 18.9614849090576
1689 18.924243927002
1708 18.8866844177246
1730 18.8474197387695
1754 18.808744430542
1793 18.7471523284912
1817 18.7071189880371
1841 18.6710987091064
1866 18.6376304626465
1893 18.6056518554688
1922 18.5755081176758
1953 18.5474452972412
1979 18.5268173217773
2032 18.5487518310547
2063 18.5565032958984
2100 18.5613803863525
2179 18.5672016143799
2199 18.6443157196045
2214 18.6979637145996
2228 18.7440719604492
2243 18.7892513275146
2259 18.8330593109131
2276 18.8754234313965
2296 18.9210109710693
2322 18.9758415222168
2379 19.0921382904053
2390 19.1977081298828
2400 19.2899894714355
2410 19.3784027099609
2420 19.4628410339355
2430 19.5433731079102
2441 19.6277656555176
2452 19.7081680297852
2464 19.7917976379395
2477 19.8781318664551
2491 19.9667587280273
2506 20.0573692321777
2522 20.1497364044189
2539 20.2436923980713
2557 20.3390960693359
2576 20.4358005523682
2579 20.4507236480713
2589 20.5567321777344
2599 20.6582946777344
2609 20.7553977966309
2619 20.8481216430664
2629 20.9365978240967
2639 21.0209903717041
2649 21.1014747619629
2660 21.1857089996338
2671 21.2656688690186
2682 21.3415908813477
2693 21.4136924743652
2705 21.4882431030273
2717 21.5587577819824
2730 21.6308727264404
2743 21.6988182067871
2756 21.7628574371338
2770 21.8277320861816
2812 22.0139331817627
2827 22.073371887207
2842 22.1287155151367
2858 22.1835460662842
2874 22.2343349456787
2891 22.2841911315918
2909 22.3327102661133
2927 22.3771705627441
2946 22.4200477600098
2966 22.461051940918
2982 22.4901943206787
3006 22.52467918396
3032 22.5578937530518
3059 22.5883045196533
3088 22.6168403625488
3119 22.6431655883789
3152 22.6670436859131
3190 22.6942577362061
3218 22.7180137634277
3249 22.7400703430176
3282 22.7594203948975
3319 22.7769527435303
3360 22.7922554016113
3379 22.7981376647949
3388 22.7373733520508
3397 22.6804332733154
3406 22.6272830963135
3416 22.57253074646
3426 22.5220775604248
3436 22.4756546020508
3447 22.4289112091064
3458 22.3863334655762
3470 22.3442249298096
3482 22.306224822998
3495 22.2692432403564
3509 22.2337799072266
3523 22.2023429870605
3538 22.1726226806641
3554 22.1449012756348
3571 22.1193790435791
3579 22.1086158752441
3597 21.7654571533203
3599 21.7280864715576
};
\end{axis}

\end{tikzpicture} \\
\hspace{0.25cm}
\begin{tikzpicture}

\definecolor{color0}{rgb}{0.12156862745098,0.466666666666667,0.705882352941177}

\begin{axis}[
width=4.3cm,
height=2.8cm,
tick align=outside,
tick pos=left,
x grid style={white!69.0196078431373!black},
xlabel={\footnotesize Time},
xmin=13.5, xmax=20.5,
xtick style={color=black},
y grid style={white!69.0196078431373!black},
ymin=15.3806943703235, ymax=18.6602250541736,
ytick style={color=black},
scaled y ticks = false,
y tick label style={/pgf/number format/fixed, font = \scriptsize},
xlabel near ticks,
ylabel near ticks,
scaled x ticks = false,
x tick label style={/pgf/number format/fixed, font = \scriptsize},
x filter/.code={\pgfmathparse{#1/600+14}},
    xticklabel={ 
        \pgfmathsetmacro\hours{floor(\tick)}%
        \pgfmathsetmacro\minutes{(\tick-\hours)*0.6}%
        \pgfmathprintnumber{\hours}:\pgfmathprintnumber[fixed, fixed zerofill, skip 0.=true, dec sep={}]{\minutes}%
    },
title={\footnotesize (g) Bus speed -- $v_b$},
]
\addplot [semithick, blue!50.1960784313725!black]
table {%
0 17.6656169891357
15 17.6562042236328
34 17.6471939086914
57 17.639139175415
85 17.6321697235107
119 17.626594543457
159 17.622802734375
179 17.6217021942139
188 17.6864891052246
197 17.7483062744141
206 17.8072452545166
215 17.8634033203125
224 17.9168758392334
233 17.9677658081055
243 18.0214004516602
253 18.0721015930176
263 18.1200065612793
273 18.1652431488037
284 18.2120723724365
295 18.2559928894043
306 18.2971630096436
317 18.3357334136963
329 18.3750190734863
341 18.4115657806396
354 18.4482727050781
367 18.4821758270264
379 18.5111560821533
409 18.4917545318604
437 18.476375579834
467 18.4626598358154
500 18.4503650665283
537 18.4393882751465
578 18.4299201965332
588 18.4265995025635
625 18.4155406951904
666 18.40602684021
716 18.3972511291504
784 18.3856315612793
811 18.3691921234131
841 18.3537864685059
876 18.3386859893799
924 18.3210906982422
979 18.3022289276123
992 18.2629852294922
1005 18.2264385223389
1019 18.1898937225342
1034 18.153621673584
1050 18.1177310943604
1068 18.0801486968994
1089 18.0391025543213
1117 17.9873218536377
1179 17.8749847412109
1191 17.795991897583
1202 17.7264003753662
1213 17.6596946716309
1224 17.595853805542
1235 17.5347480773926
1247 17.4709911346436
1259 17.4100151062012
1272 17.3468036651611
1285 17.2862644195557
1299 17.2237567901611
1314 17.159574508667
1330 17.0939865112305
1346 17.0310974121094
1363 16.9669990539551
1379 16.9090595245361
1391 16.8342704772949
1402 16.7684574127197
1412 16.7112064361572
1422 16.6565093994141
1432 16.6043605804443
1443 16.5498714447021
1454 16.4982757568359
1465 16.4494247436523
1476 16.4031658172607
1488 16.3554725646973
1500 16.3104801177979
1513 16.2645683288574
1526 16.2213840484619
1540 16.1776924133301
1554 16.1366882324219
1569 16.095495223999
1580 16.0658664703369
1595 16.0130405426025
1605 15.9801006317139
1613 15.9566326141357
1622 15.9329509735107
1632 15.9093599319458
1643 15.886043548584
1656 15.8612823486328
1671 15.8356370925903
1688 15.8094987869263
1707 15.7831430435181
1728 15.756781578064
1751 15.7306032180786
1777 15.7038021087646
1783 15.6967477798462
1808 15.6664371490479
1832 15.6401538848877
1856 15.616605758667
1882 15.5939168930054
1909 15.5730962753296
1938 15.5534420013428
1969 15.5351314544678
1980 15.5300798416138
2030 15.5447206497192
2060 15.5502099990845
2095 15.5537176132202
2179 15.5581302642822
2198 15.6096267700195
2213 15.6474256515503
2227 15.6799097061157
2241 15.7096920013428
2256 15.738844871521
2272 15.767237663269
2291 15.7980794906616
2314 15.8325223922729
2357 15.8934364318848
2379 15.9250249862671
2390 15.9984092712402
2400 16.0624446868896
2410 16.1236991882324
2420 16.1821117401123
2430 16.2377452850342
2440 16.2907905578613
2451 16.3464183807373
2463 16.4041843414307
2475 16.4592361450195
2488 16.5161743164062
2502 16.5747375488281
2517 16.6347217559814
2533 16.6959648132324
2550 16.7583389282227
2568 16.8217182159424
2579 16.8592300415039
2588 16.9241504669189
2597 16.9865303039551
2606 17.0463638305664
2615 17.1036930084229
2625 17.1645393371582
2635 17.2224960327148
2645 17.2776927947998
2655 17.3302593231201
2666 17.3852024078369
2677 17.4372940063477
2688 17.4866943359375
2700 17.5377006530762
2712 17.5858783721924
2724 17.6314029693604
2737 17.6779155731201
2750 17.7216949462891
2764 17.7659912109375
2778 17.8075256347656
2783 17.822998046875
2797 17.8652725219727
2811 17.9048042297363
2826 17.9443054199219
2841 17.981050491333
2857 18.0174198150635
2873 18.0510807037354
2890 18.0840950012207
2907 18.1144866943359
2925 18.1440258026123
2944 18.1724948883057
2964 18.1997032165527
2981 18.2204132080078
3006 18.2440414428711
3032 18.2658252716064
3059 18.2857608795166
3089 18.3050537109375
3120 18.322208404541
3154 18.338191986084
3189 18.354528427124
3217 18.3701572418213
3248 18.3846645355225
3281 18.397388458252
3318 18.4089126586914
3360 18.4191837310791
3379 18.4230251312256
3389 18.3790664672852
3399 18.3381462097168
3409 18.3002300262451
3419 18.265209197998
3429 18.2329273223877
3439 18.2032108306885
3450 18.1732711791992
3461 18.1459846496582
3473 18.1189823150635
3485 18.0946006774902
3498 18.0708599090576
3512 18.0480785369873
3527 18.0265235900879
3543 18.0064010620117
3560 17.9878597259521
3578 17.9709911346436
3579 17.9701290130615
3599 17.7179164886475
};
\end{axis}
\end{tikzpicture} &
\hspace{-0.3cm}
\begin{tikzpicture}

\definecolor{color0}{rgb}{0.12156862745098,0.466666666666667,0.705882352941177}
\definecolor{color1}{rgb}{1,0.498039215686275,0.0549019607843137}

\begin{axis}[
legend cell align={left},
legend style={at={(0.5,1.05)},anchor=north, fill opacity=0, draw opacity=1, text opacity=1, draw=white!80!black, legend columns=2, fill = none, legend style={draw=none}},
tick align=outside,
tick pos=left,
x grid style={white!69.0196078431373!black},
width=4.3cm,
height=2.8cm,
x grid style={white!69.0196078431373!black},
xlabel={\footnotesize Time},
xmin=13.5, xmax=20.5,
xtick style={color=black},
y grid style={white!69.0196078431373!black},
ymin=-0.0228729039920502, ymax=0.963507829099009,
ytick style={color=black},
scaled y ticks = false,
y tick label style={/pgf/number format/fixed, font = \scriptsize},
xlabel near ticks,
ylabel near ticks,
scaled x ticks = false,
x tick label style={/pgf/number format/fixed, font = \scriptsize},
x filter/.code={\pgfmathparse{#1/600+14}},
    xticklabel={ 
        \pgfmathsetmacro\hours{floor(\tick)}%
        \pgfmathsetmacro\minutes{(\tick-\hours)*0.6}%
        \pgfmathprintnumber{\hours}:\pgfmathprintnumber[fixed, fixed zerofill, skip 0.=true, dec sep={}]{\minutes}%
    },
title={\footnotesize (h) Choice -- $\beta_{\mc{V}}$ \& $\beta_{\mc{B}}$ }
]
\addplot [semithick, blue!50.1960784313725!black, dashed]
table {%
0 0.374650955200195
22 0.376436948776245
49 0.378158569335938
82 0.379792094230652
121 0.381245851516724
167 0.382479667663574
178 0.382714629173279
179 0.54316520690918
249 0.538417339324951
303 0.535217523574829
355 0.532592535018921
378 0.531574249267578
379 0.469301700592041
541 0.470118045806885
578 0.47020947933197
579 0.466897130012512
671 0.467435836791992
764 0.467533946037292
778 0.46750545501709
779 0.455462336540222
825 0.456254601478577
871 0.456588506698608
915 0.45644998550415
955 0.455871224403381
978 0.455304861068726
979 0.395837306976318
997 0.397995114326477
1015 0.399703025817871
1033 0.400964260101318
1052 0.401824712753296
1071 0.402217030525208
1090 0.402154684066772
1109 0.401648998260498
1129 0.400648951530457
1149 0.399182081222534
1169 0.397265076637268
1178 0.396261096000671
1179 0.297216296195984
1193 0.300345897674561
1206 0.30282187461853
1219 0.304863452911377
1233 0.306588053703308
1247 0.307851433753967
1262 0.308738946914673
1278 0.309213399887085
1295 0.30925452709198
1314 0.308825850486755
1335 0.307879686355591
1359 0.306332349777222
1378 0.304847478866577
1379 0.251543641090393
1402 0.254837512969971
1421 0.257087469100952
1440 0.258864045143127
1460 0.260257720947266
1481 0.26126492023468
1505 0.261946320533752
1532 0.262246489524841
1564 0.262138485908508
1578 0.26197612285614
1579 0.241249084472656
1607 0.243403077125549
1620 0.243872404098511
1636 0.243990421295166
1657 0.243669271469116
1687 0.242718577384949
1746 0.240291237831116
1778 0.238951921463013
1779 0.232446551322937
1978 0.230031490325928
1979 0.247502088546753
2017 0.244430065155029
2044 0.242730855941772
2070 0.241560578346252
2096 0.240861296653748
2121 0.240641713142395
2146 0.240884184837341
2170 0.241581439971924
2178 0.241921782493591
2179 0.295314788818359
2197 0.291912794113159
2212 0.289543867111206
2226 0.287798285484314
2240 0.286534428596497
2253 0.285801768302917
2267 0.285487174987793
2281 0.28565788269043
2295 0.286302804946899
2309 0.287408351898193
2323 0.288958430290222
2338 0.291091442108154
2353 0.293686628341675
2369 0.296928763389587
2378 0.298951387405396
2379 0.393073916435242
2392 0.391144752502441
2404 0.38980233669281
2416 0.388903617858887
2428 0.388444900512695
2441 0.38841187953949
2455 0.388857841491699
2470 0.389813423156738
2487 0.391387701034546
2506 0.39362359046936
2530 0.396935701370239
2570 0.403002500534058
2578 0.40422797203064
2579 0.471096515655518
2607 0.470564246177673
2638 0.470444798469543
2679 0.470782995223999
2778 0.472207427024841
2779 0.478063106536865
2888 0.478101253509521
2978 0.477791309356689
2979 0.474023342132568
3085 0.473989963531494
3178 0.473766565322876
3179 0.47920298576355
3282 0.478224992752075
3378 0.477645516395569
3379 0.371814370155334
3395 0.37634539604187
3411 0.380417466163635
3428 0.384261608123779
3445 0.387651920318604
3464 0.390964865684509
3484 0.393979430198669
3506 0.39681613445282
3530 0.399427652359009
3556 0.401783466339111
3578 0.403456449508667
3579 0.0449600219726562
3593 0.0492123365402222
3599 0.051150918006897
};
\addlegendentry{\tiny $\beta_{\mathcal{V}}$}
\addplot [semithick, blue!50.1960784313725!black]
table {%
0 0.323701977729797
22 0.320801496505737
50 0.317944765090942
85 0.315232872962952
126 0.312903165817261
175 0.310964107513428
178 0.310868620872498
179 0.0219626426696777
257 0.0280805826187134
321 0.0322765111923218
378 0.0353134870529175
379 0.148298025131226
494 0.146610260009766
578 0.145956993103027
579 0.151971459388733
670 0.150922536849976
771 0.150582790374756
778 0.150591135025024
779 0.172469019889832
828 0.170934200286865
878 0.170192122459412
927 0.170286536216736
972 0.171193838119507
978 0.171383738517761
979 0.279592633247375
998 0.275399208068848
1017 0.272032737731934
1036 0.269472479820251
1056 0.26760721206665
1076 0.266550779342651
1097 0.266271829605103
1118 0.26680850982666
1140 0.268211483955383
1162 0.270437955856323
1178 0.272550344467163
1179 0.454323291778564
1194 0.447697043418884
1208 0.442358255386353
1222 0.437872171401978
1236 0.4342120885849
1251 0.431142210960388
1267 0.428744196891785
1284 0.427067637443542
1302 0.42613160610199
1322 0.42593514919281
1344 0.426543712615967
1370 0.428110361099243
1378 0.428735733032227
1379 0.528562784194946
1402 0.521250605583191
1421 0.516077637672424
1439 0.511989593505859
1458 0.508486270904541
1479 0.505472421646118
1502 0.503042459487915
1527 0.501234292984009
1556 0.499973773956299
1578 0.499470114707947
1579 0.539032697677612
1607 0.534010410308838
1620 0.532691717147827
1636 0.531946897506714
1656 0.531878590583801
1684 0.532678842544556
1729 0.53490138053894
1778 0.537655115127563
1779 0.550219535827637
1881 0.550845265388489
1978 0.55186653137207
1979 0.517821192741394
2018 0.523432016372681
2045 0.526432275772095
2071 0.528484106063843
2097 0.52969765663147
2123 0.530065298080444
2148 0.529596328735352
2172 0.528342485427856
2178 0.527898788452148
2179 0.423718452453613
2197 0.430346131324768
2212 0.435031771659851
2226 0.438566446304321
2240 0.441235780715942
2254 0.44302499294281
2268 0.443941593170166
2282 0.444005250930786
2296 0.443243861198425
2310 0.441689610481262
2325 0.439186811447144
2340 0.435867786407471
2356 0.431492924690247
2373 0.425988435745239
2378 0.424216628074646
2379 0.243049621582031
2392 0.247756481170654
2405 0.251605987548828
2417 0.254355669021606
2430 0.256475806236267
2443 0.257769346237183
2457 0.258345246315002
2472 0.258156895637512
2489 0.257119297981262
2509 0.255043029785156
2534 0.251577138900757
2577 0.244627118110657
2578 0.244462490081787
2579 0.119619607925415
2604 0.123209714889526
2630 0.126119613647461
2660 0.128627777099609
2696 0.13077986240387
2744 0.132765412330627
2778 0.133836507797241
2779 0.123148322105408
2847 0.125854969024658
2938 0.128605365753174
2978 0.129598498344421
2979 0.136477470397949
3178 0.138709783554077
3179 0.128830671310425
3270 0.130570650100708
3378 0.13182270526886
3379 0.324183225631714
3395 0.31632924079895
3411 0.309296131134033
3428 0.30268132686615
3446 0.296551704406738
3465 0.290940403938293
3485 0.285857439041138
3507 0.281097173690796
3531 0.276737809181213
3558 0.272692203521729
3578 0.270182371139526
3579 0.918672323226929
3593 0.91110897064209
3599 0.907665371894836
};
\addlegendentry{\tiny $\beta_{\mathcal{B}}$}
\end{axis}

\end{tikzpicture} \\
\hspace{-0.1cm} 
\begin{tikzpicture}

\definecolor{color0}{rgb}{0.12156862745098,0.466666666666667,0.705882352941177}

\begin{axis}[
width=4.3cm,
height=2.8cm,
tick align=outside,
tick pos=left,
x grid style={white!69.0196078431373!black},
xlabel={\footnotesize Time},
xmin=13.5, xmax=20.5,
xtick style={color=black},
y grid style={white!69.0196078431373!black},
ymin=-3.29999983437911, ymax=3.29999999211329,
ytick style={color=black},
scaled y ticks = false,
y tick label style={/pgf/number format/fixed, font = \footnotesize},
xlabel near ticks,
ylabel near ticks,
scaled x ticks = false,
x tick label style={/pgf/number format/fixed, font = \footnotesize},
x filter/.code={\pgfmathparse{#1/600+14}},
    xticklabel={ 
        \pgfmathsetmacro\hours{floor(\tick)}%
        \pgfmathsetmacro\minutes{(\tick-\hours)*0.6}%
        \pgfmathprintnumber{\hours}:\pgfmathprintnumber[fixed, fixed zerofill, skip 0.=true, dec sep={}]{\minutes}%
    },
title={\footnotesize (i) Price change -- $\phi_{\mc{B}}$}
]

\addplot [semithick, blue!50.1960784313725!black]
table {%
0 -0
178 -0
179 3
378 3
379 1.4418705701828
578 1.4418705701828
579 1.39437830448151
778 1.39437830448151
779 1.23264706134796
978 1.23264706134796
979 0.60351300239563
1178 0.60351300239563
1179 -0.194258213043213
1378 -0.194258213043213
1379 -0.595309495925903
1578 -0.595309495925903
1579 -0.753939867019653
1778 -0.753939867019653
1779 -0.804366946220398
1978 -0.804366946220398
1979 -0.667446136474609
2178 -0.667446136474609
2179 -0.248525738716125
2378 -0.248525738716125
2379 0.580536007881165
2578 0.580536007881165
2579 1.44731485843658
2778 1.44731485843658
2779 1.54305839538574
2978 1.54305839538574
2979 1.48361325263977
3178 1.48361325263977
3179 1.56897842884064
3378 1.56897842884064
3379 0.418728828430176
3578 0.418728828430176
3579 -2.99999976158142
3599 -2.99999976158142
};
\end{axis}

\end{tikzpicture} & \hspace{-0.3cm}
\begin{tikzpicture}

\definecolor{color0}{rgb}{0.12156862745098,0.466666666666667,0.705882352941177}

\begin{axis}[
width=4.3cm,
height=2.8cm,
tick align=outside,
tick pos=left,
x grid style={white!69.0196078431373!black},
xlabel={\footnotesize Time},
xmin=13.5, xmax=20.5,
xtick style={color=black},
y grid style={white!69.0196078431373!black},
ymin=-233.259557358528, ymax=4898.45070452908,
ytick style={color=black},
scaled y ticks = false,
y tick label style={/pgf/number format/fixed, font = \scriptsize},
xlabel near ticks,
ylabel near ticks,
scaled x ticks = false,
x tick label style={/pgf/number format/fixed, font = \scriptsize},
x filter/.code={\pgfmathparse{#1/600+14}},
    xticklabel={ 
        \pgfmathsetmacro\hours{floor(\tick)}%
        \pgfmathsetmacro\minutes{(\tick-\hours)*0.6}%
        \pgfmathprintnumber{\hours}:\pgfmathprintnumber[fixed, fixed zerofill, skip 0.=true, dec sep={}]{\minutes}%
    },
title={\footnotesize (j) Abandonment -- $A$}
]
\addplot [semithick, blue!50.1960784313725!black]
table {%
0 0
1599 0
1602 11.243350982666
1616 75.1509780883789
1634 168.460174560547
1658 305.321197509766
1691 507.222839355469
1736 797.052795410156
1794 1185.43103027344
1885 1811.2919921875
2022 2773.95947265625
2094 3275.9619140625
2140 3581.48022460938
2178 3819.6669921875
2211 4012.79809570312
2241 4181.4228515625
2264 4298.4423828125
2284 4389.14404296875
2302 4460.8994140625
2318 4516.19482421875
2333 4560.345703125
2346 4592.318359375
2358 4616.46533203125
2369 4633.95556640625
2379 4645.92578125
2384 4649.09130859375
2409 4663.06884765625
2418 4665.08935546875
2441 4665.19091796875
3599 4665.19091796875
};
\end{axis}
\end{tikzpicture}
\end{tabular}
\caption{Time-dependent model variables for the MPC framework with abandonment for (a) private vehicle accumulation, (b) solo trip ride-hailing vehicles, (c) pool trip ride-hailing accumulation in~$\mc{V}$, (d) pool trip ride-hailing accumulation in~$\mc{B}$, (e) speed in the vehicle network~$\mc{V}$, (f) vehicle speed in the bus network~$\mc{B}$, (g) bus speed in the bus network~$\mc{B}$, (h) fraction of pool trip in $\mc{V}$ and $\mc{B}$ respectively, (i) regulatory control fare for pooling in~$\mc{B}$, and (j) cumulative abandoning requests.}
\label{fig:results_mpc_with_ab}
\end{figure}

With slightly higher objective function value of $184884$ pax.hr, the MPC framework with two decision variables $\phi_{\mc{V}}$ and $\phi_{\mc{B}}$ generally performs better with significantly fewer abandonment compared to the MPC framework with one decision variable $\phi_{\mc{V}}$ where the total delays yield a value of $184691$ pax.hr but the abandoning requests are significantly larger. Again, this accentuates the importance of pooling, even in the vehicle network, which significantly reduces waiting times, and improves empty vehicle availability. 

As in Figure~\ref{fig:results_mpc_no_ab}, the results of the MPC with one control variable are shown in Figure~\ref{fig:results_mpc_with_ab}, in addition to the variation of the cumulative abandoning requests in Figure~\subfigref{fig:results_mpc_with_ab}{j}. Clearly, the number of abandoning ride-hailing requests increases during peak hours, yet the variation of the decision variable $\phi_{\mc{V}}$ in Figures~\subfigref{fig:results_mpc_no_ab}{i} and~\subfigref{fig:results_mpc_with_ab}{i} are almost similar. To conclude, this strategy therefore has a positive influence on the service level of ride-hailing platforms, even if the question here arises on which party should bear the incentivizing fare for pooling.  
\begin{table}
    \centering
    \caption{Macro-simulation results with abandonment}
    \begin{tabular}{ccc}
    \hline
         \multirow{2}{*}{\textbf{Scenario}}& \textbf{PHT+WT}& \textbf{Abandonment} \\
         & \textbf{[pax.hr]} & \textbf{[pax]} \\ \hline
         $\beta_{\mc{V}}=0$ \& $\beta_{\mc{B}}= 0$&$211898$ & $17816$ \\
         $\beta_{\mc{V}}\in[0,1]$ \& $\beta_{\mc{B}}= 0$&$210505$ & $15970$ \\
         $\beta_{\mc{V}}=0$ \& $\beta_{\mc{B}}\in[0,1]$ &$185664$ & $4812$\\
         $\beta_{\mc{V}}=0$ \& $\beta_{\mc{B}} = 1$ & $196565$ & $12630$\\
        $\beta_{\mc{V}}\in[0,1]$ \& $\beta_{\mc{B}} \in[0,1]$ & $185157$ & $4732$\\
        PI control -- $\phi_{\mc{B}}$ &$186471$ & $5398$ \\
        MPC -- $\phi_{\mc{B}}$& $184691$& $4665$\\
        MPC -- $\phi_{\mc{V}}$ and $\phi_{\mc{B}}$& $184884$ & $3979$\\
         \hline
    \end{tabular}
    \label{tab:results_ab}
\end{table}

\begin{table}
    \centering
    \caption{Multi-modal delays}
    \begin{tabular}{cccc}
    \hline
         \multirow{2}{*}{\textbf{Scenario}}& $\text{\textbf{PHT}}_\mathbf{pv}$& $\text{\textbf{PHT}}_\mathbf{rs}$ &  $\text{\textbf{PHT}}_\mathbf{b}$\\
         & \textbf{[pax.hr]} & \textbf{[pax.hr]} & \textbf{[pax.hr]} \\ \hline
         $\beta_{\mc{V}}=0$ \& $\beta_{\mc{B}}= 0$&$114856$ & $17815$ & $65578$\\
         $\beta_{\mc{V}}\in[0,1]$ \& $\beta_{\mc{B}}= 0$&$114855$ & $21367$ & $65058$\\
         $\beta_{\mc{V}}=0$ \& $\beta_{\mc{B}}\in[0,1]$ &$84902$ & $19811$ & $72407$\\
         $\beta_{\mc{V}}=0$ \& $\beta_{\mc{B}} = 1$ & $77072$ & $24039$ & $87451$\\
        $\beta_{\mc{V}}\in[0,1]$ \& $\beta_{\mc{B}} \in[0,1]$ & $88282$ & $20504$ & $69169$\\
        PI control -- $\phi_{\mc{B}}$ &$89558$ & $20587$ & $69086$\\
        MPC -- $\phi_{\mc{B}}$& $88494$& $20479$ & $68444$\\
        MPC -- $\phi_{\mc{V}}$ and $\phi_{\mc{B}}$& $88409$ & $20637$ & $68335$\\
         \hline
    \end{tabular}
    \label{tab:delays_ab}
\end{table}
\section{Conclusions}
\label{sec:conclusion}
In this work, we develop an aggregate dynamic for a multi-modal network with private vehicles, ride-hailing services, and public transportation. Our modeling approach aims at evaluating an occupancy-dependent space allocation policy where a fraction of the pool ride-hailing users choose to utilize dedicated bus lanes. Despite ameliorating the total user delays and the ride-hailing service levels, our allocation strategy is not capable by itself of modifying the selfish user choices that do not necessarily align with the network-level optimum. The need for a more elaborate pricing scheme to steer the user's choice towards a more convenient solution for multi-modal users is therefore substantiated. Consequently, we build both a PI and an MPC control framework, and analyze what the additional pooling discount or fare should be that improves the overall travel times for all network users, and the waiting time for ride-hailing users in specific. Our results show that pricing indeed influences the preferences of ride-hailing requests in a manner that reduces the PHT for all modes. We performed the same analysis for the cases with and without request abandonment, and demonstrated how the complexity of our abandonment function is circumvented in the MPC solution.  

In future work, we plan to give further attention to demand-dependent trip detours. This is because the detour distance itself for a pool trip is lower when more passengers opt for pooling. Therefore, the efficiency of our proposed policy becomes more accentuated if this factor is accounted for in the modeling formulation. Moreover, we have considered so far pool trips with only two passengers sharing their rides. A potential direction for this work is to extend this work to incorporate a high-capacity on-demand micro-transit service utilizing the bus network along with buses, and observe the additional occupancy-dependent improvements that we achieve if these services progressively gain more and more momentum.

\bibliographystyle{IEEEtran}
\bibliography{references.bib} 

\end{document}